\documentclass[11pt]{article}

\DeclareMathAlphabet{\scr}{U}{rsfs}{m}{n}

\usepackage{latexsym}
\usepackage{epsfig}
\usepackage[mathscr]{eucal}
\usepackage{amsfonts}
\usepackage{amscd}
\usepackage{cite}
\usepackage{array}   
\usepackage{amssymb}
\usepackage{colordvi}
\usepackage[centertags]{amsmath}
\usepackage{enumerate}
\usepackage{graphicx}
\usepackage{booktabs}
\usepackage{theorem}
\usepackage[footnotesize]{caption}
\usepackage{soul}
\usepackage{mcite}
\usepackage{slashed}
\usepackage{color}
\usepackage{ulem}
\usepackage{bbm}
\usepackage[utf8]{inputenc}
\newcommand{\newc}{\newcommand}
\newc{\be}{\begin{equation}}
\newc{\ee}{\end{equation}}
\newc{\bea}{\begin{eqnarray}}
\newc{\eea}{\end{eqnarray}}
\newc{\ol}{\overline}
\newc{\wt}{\widetilde}
\newc{\bs}{\boldsymbol}
\newc{\m}{\mathcal}
\newc{\la}{\langle}
\newc{\ra}{\rangle}

\newcommand{\beq}{\begin{eqnarray}} 
\newcommand{\eeq}{\end{eqnarray}} 
\newcommand{\bpmatrix}{\begin{pmatrix}}
\newcommand{\epmatrix}{\end{pmatrix}}
\newcommand{\ba}{\begin{array}}
\newcommand{\ea}{\end{array}}

\renewcommand{\ol}{\text{1l}}

\renewcommand{\eqref}[1]{Eq.~(\ref{#1})}

\newcommand{\bc}{\begin{center}}
\newcommand{\ec}{\end{center}}

\newcommand{\gsim}{\raisebox{-0.13cm}{~\shortstack{$>$ \\[-0.07cm]
      $\sim$}}~}

\newcommand{\s}{\newline \vspace*{-3.5mm}}

\begin{document}

\title{
\vspace*{-3cm}
\phantom{h} \hfill\mbox{\small KA-TP-15-2016}
\\[1cm]
\textbf{Gauge-independent Renormalization of the \\
 2-Higgs-Doublet Model \\[4mm]}}

\date{}
\author{
Marcel Krause$^{1\,}$\footnote{E-mail:
  \texttt{marcel.krause@mail.de}} ,
Robin Lorenz$^{1\,}$\footnote{E-mail:
  \texttt{rwllorenz@gmail.com}} , 
Margarete M\"{u}hlleitner$^{1\,}$\footnote{E-mail:
  \texttt{margarete.muehlleitner@kit.edu}} ,
Rui Santos$^{2\,, \, 3\,}$\footnote{E-mail:
  \texttt{rasantos@fc.ul.pt}} ,
Hanna Ziesche$^{1\,}$\footnote{E-mail: \texttt{hanna.ziesche@kit.edu}}
\\[9mm]
{\small\it
$^1$Institute for Theoretical Physics, Karlsruhe Institute of Technology,} \\
{\small\it 76128 Karlsruhe, Germany.}\\[3mm]
{\small\it
$^2$ISEL -
 Instituto Superior de Engenharia de Lisboa,} \\
{\small \it  Instituto Polit\'ecnico de Lisboa,  1959-007 Lisboa, Portugal}\\[3mm]
{\small\it
$^3$Centro de F\'{\i}sica Te\'{o}rica e Computacional,
    Faculdade de Ci\^{e}ncias,} \\
{\small \it    Universidade de Lisboa, Campo Grande, Edif\'{\i}cio C8
  1749-016 Lisboa, Portugal} 
}

\maketitle

\begin{abstract}
\noindent
The 2-Higgs-Doublet Model (2HDM) belongs to the simplest extensions of the
Standard Model (SM) Higgs sector that are in accordance with theoretical
and experimental constraints. In order to be able to properly
investigate the experimental Higgs data and, in the long term to
distinguish between possible models beyond the SM, precise predictions
for the Higgs boson observables have to be made available on the
theory side. This requires the inclusion of the higher order 
corrections. In this work, we investigate in detail the renormalization
of the 2HDM, a pre-requisite for the computation of higher order
corrections. We pay particular attention to the renormalization of the
mixing angles $\alpha$ and $\beta$, which diagonalize the Higgs mass
matrices and which enter all Higgs observables. The implications of various
renormalization schemes in next-to-leading order corrections to the sample
processes $H^\pm \to W^\pm 
h/H$ and $H \to ZZ$ are investigated. Based on our findings, we will present a
renormalization scheme that is at the same time process independent,
gauge independent and numerically stable. 
\end{abstract}
\thispagestyle{empty}
\vfill
\newpage
\setcounter{page}{1}

\section{Introduction}
The discovery of a new scalar particle by the LHC experiments ATLAS
\cite{Aad:2012tfa} and CMS \cite{Chatrchyan:2012ufa} in 2012 and its
subsequent confirmation as being the Higgs boson
\cite{Khachatryan:2014kca,Aad:2015mxa,Khachatryan:2014jba,Aad:2015gba}
marked a milestone for particle physics. At the same time, it 
triggered a change of paradigm. The Higgs particle, which formerly was
the object of experimental searches, has itself become a tool in the
search for New Physics (NP). Although the Standard Model (SM) of
particle physics has been tested in previous and present experiments at
the highest accuracy, there remain many open questions that cannot be
answered within this model. The SM is therefore regarded as the
low-energy description of some more fundamental theory that becomes
effective at higher energy scales. A plethora of NP models have been
discussed, among them {\it e.g.} supersymmetry (SUSY) as one of the
most popular and most intensely studied Beyond the SM (BSM)
extensions. Supersymmetry requires the introduction of at least two
complex Higgs doublets. The Higgs sector of the Minimal Supersymmetric
extension of the SM (MSSM)
\cite{Gunion:1989we,Martin:1997ns,Dawson:1997tz,Djouadi:2005gj} is a
special case of the 2-Higgs-Doublet Model (2HDM)
\cite{Gunion:1989we,Lee:1973iz,Branco:2011iw} 
type II. While the parameters of the SUSY Higgs potential are
restricted due to SUSY relations, general 2HDMs allow for much
more freedom in the choice of the parameters. They are therefore an
ideal framework to study the implications of an extended Higgs sector
for Higgs phenomenology at the LHC. This is reflected in the
experimental analyses that interpret the results in various benchmark
models, among them the 2HDM. The precise investigation of the
Higgs sector aims at getting insights into the nature of electroweak
symmetry breaking (EWSB) and at clarifying the question whether it is based 
on weakly or strongly interacting dynamics. Deviations in the properties of the discovered
SM-like Higgs boson are hints towards NP. In particular, the higher
precision in the Higgs couplings measurements at the LHC run 2 and in
the high-luminosity option allows to search indirectly for BSM
effects. This becomes increasingly important in view of the null
results of direct searches for NP so far.\footnote{Recent hints like
  the diphoton excess at 750~GeV \cite{Atlas750,CMS-PAS-EXO-15-004}
  need further data for more conclusive interpretations.} The precise
measurements on the experimental side, 
however, call for precise predictions of parameters and observables
from theory. Accurate theory predictions are indispensable not only for the
proper interpretation of the experimental data, but also for the
correct determination of the parameter space that is still allowed in
the various models, and, finally, for the distinction between 
different BSM extensions. \s

With this paper we contribute to the effort of
providing precise predictions for parameters and observables relevant
for the phenomenology at the LHC and future $e^+e^-$ colliders. 
We investigate higher order corrections in the
framework of the 2HDM. While 2HDMs are interesting because they 
contain the MSSM Higgs sector as a special case, they 
also belong to the simplest SM extensions respecting basic
experimental and theoretical constraints that are testable at the
LHC. After EWSB they feature five physical Higgs
  bosons, two neutral CP-even, one neutral CP-odd and two charged
  Higgs bosons.
They represent an ideal benchmark framework to
investigate the various possible NP effects to be expected at the LHC
in multi-Higgs boson sectors. Finally, specific 2HDM versions also allow
for a Dark Matter (DM) candidate~\cite{Barbieri:2006dq, LopezHonorez:2006gr,
Cao:2007rm, Dolle:2009fn, Gustafsson:2012aj, Arhrib:2013ela, Ilnicka:2015jba}. 
In the past, 
numerous papers have provided higher order corrections to the 2HDM
parameters, production cross sections and decay
widths. Several papers have dealt with the 
renormalization of the 2HDM (see {\it
  e.g.}~\cite{Santos:1996vt,Kanemura:2004mg,Kanemura:2015mxa}).
In particular, the renormalization of the mixing angles $\alpha$ and
$\beta$ is of interest. While $\alpha$ is introduced to
  diagonalize the mass matrix of the neutral CP-even Higgs sector, the
  angle $\beta$ appears in the diagonalization of the CP-odd and the
  charged Higgs sector, respectively.
These angles define the Higgs couplings to the
  SM particles and thus enter all Higgs observables like {\it
    e.g.}~production cross sections and decay widths. 
For the MSSM it was stated in \cite{Freitas:2002um} that a
renormalization scheme for the only mixing angle taken as an independent
parameter from the scalar sector, $\beta$,
cannot be simultaneously gauge independent, process independent and
numerically stable. In the 2HDM also $\alpha$ needs to be
renormalized, which has important consequences for the choice of the
renormalization scheme. If the tadpoles are treated in the usual way,
which we call the standard approach ({\it cf.}~\ref{sec:standard}), a
process-independent definition of the angular counterterms is prone to lead to
gauge-dependent amplitudes and consequently to gauge-dependent physical observables. This is the
case {\it e.g.}~in the scheme presented in \cite{Kanemura:2004mg}. There are
essentially two possibilities to circumvent the emergence of this gauge
dependence. Either one gives up the requirement of process
independence and fixes $\alpha$ and $\beta$ in terms of a physical
observable or one changes the treatment of the tadpoles.
As we will see, this will decouple the issue of gauge dependence from
the definition of $\delta\alpha$ and $\delta\beta$ and allow for 
process- and gauge-independent angular counterterms leading to
manifestly gauge-independent amplitudes. 
\s

In this paper we study in detail the renormalization of the 2HDM Higgs
sector with the main focus on the investigation of the 
gauge dependence of the renormalization of the mixings angles $\alpha$
and $\beta$. We propose several schemes and compare them both to the
ones in the literature and amongst each other. In sample decay processes 
we investigate the numerical differences and in particular the
numerical stability of the various renormalization prescriptions. Our
results presented here will serve as basis for the further computation
of the one-loop electroweak corrections to all 2HDM Higgs boson
decays. \s

The organization of the paper is as follows. In sec.~\ref{sec:model}
we introduce the model and set up our notation. The following
sec.~\ref{sec:renorm} is devoted to the detailed presentation and
discussion of the various renormalization prescriptions that will be
applied. Section~\ref{sec:oneloopdec} deals with the computation of
the electroweak (EW) one-loop corrections to various decay processes and the
discussion of the gauge dependence of the angular counterterms. 
In sec.~\ref{sec:numerics} we present our numerical
results. We finish with the conclusions in sec.~\ref{sec:concl}. The
paper is accompanied by an extensive Appendix to serve as starting
point for further investigations of the 2HDM renormalization.

\section{Description of the Model \label{sec:model}}
\setcounter{equation}{0}
We work in the framework of a general 2HDM with a global discrete $\mathbb{Z}_2$
symmetry that is softly broken. The kinetic term of the two $SU(2)$ Higgs
doublets $\Phi_1$ and $\Phi_2$ is given by
\beq
{\cal L}_{\text{kin}} = \sum_{i=1}^2 (D_\mu \Phi_i)^\dagger (D^\mu \Phi_i)
\eeq
in terms of the covariant derivative
\beq
D_\mu = \partial_\mu + \frac{i}{2} g \sum_{a=1}^3 \tau^a W_\mu^a +
\frac{i}{2} g' B_\mu \;, \label{eq:covdiv}
\eeq
where $\tau^a$ denote the Pauli matrices, $W_\mu^a$ and $B_\mu$ the
$SU(2)_L$ and $U(1)_Y$ gauge bosons, respectively, and $g$ and $g'$ the
corresponding gauge couplings. The scalar potential that can be built from the
two $SU(2)$ Higgs doublets can be written as
\beq
V &=& m_{11}^2 |\Phi_1|^2 + m_{22}^2 |\Phi_2|^2 - m_{12}^2 (\Phi_1^\dagger
\Phi_2 + h.c.) + \frac{\lambda_1}{2} (\Phi_1^\dagger \Phi_1)^2 +
\frac{\lambda_2}{2} (\Phi_2^\dagger \Phi_2)^2 \nonumber \\
&& + \lambda_3
(\Phi_1^\dagger \Phi_1) (\Phi_2^\dagger \Phi_2) + \lambda_4 
(\Phi_1^\dagger \Phi_2) (\Phi_2^\dagger \Phi_1) + \frac{\lambda_5}{2}
[(\Phi_1^\dagger \Phi_2)^2 + h.c.] \;.
\eeq
The discrete $\mathbb{Z}_2$ symmetry ($\Phi_1 \to -\Phi_1, \Phi_2 \to
\Phi_2$) ensures the absence of tree-level 
Flavour Changing Neutral Currents. Assuming CP conservation, the
2HDM potential depends on eight real parameters, three mass parameters,
$m_{11}$, $m_{22}$ and $m_{12}$, and five coupling parameters
$\lambda_1$-$\lambda_5$. Through the term proportional to $m_{12}^2$ the
discrete $\mathbb{Z}_2$ symmetry is softly broken. 
After EWSB the neutral components of the Higgs doublets develop
vacuum expectation values (VEVs), which are real in the CP-conserving
case. Expanding about the VEVs $v_{1}$ and $v_{2}$ and expressing each
doublet $\Phi_i$ $(i=1,2)$ in terms of the charged complex field
$\phi_i^+$ and the real neutral CP-even and CP-odd fields $\rho_i$ and
$\eta_i$, respectively,  
\beq
\Phi_1 = \left(
\begin{array}{c}
\phi_1^+ \\
\frac{\rho_1 + i \eta_1 + v_1}{\sqrt{2}}
\end{array}
\right) \qquad \mbox{and} \qquad
\Phi_2 = \left(
\begin{array}{c}
\phi_2^+ \\
\frac{\rho_2 + i \eta_2 + v_2}{\sqrt{2}}
\end{array}
\right) \;, \label{eq:vevexpansion}
\eeq
leads to the mass matrices, which are obtained from the terms bilinear in the Higgs
fields in the potential. Due to charge and CP conservation they decompose into $2
\times 2$ matrices ${\cal M}_S$, ${\cal M}_P$ and ${\cal M}_C$ for the
neutral CP-even, neutral CP-odd and charged Higgs sector. They are
diagonalized by the following orthogonal transformations
\beq
\left( \begin{array}{c} \rho_1 \\ \rho_2 \end{array} \right) &=&
R(\alpha) \left( \begin{array}{c} H \\ h \end{array} \right)  \; , \label{eq:diagHh} \\
\left( \begin{array}{c} \eta_1 \\ \eta_2 \end{array} \right) &=&
R(\beta) \left( \begin{array}{c} G^0 \\ A \end{array} \right)  \;
, \label{eq:diagGA} \\
\left( \begin{array}{c} \phi_1^\pm \\ \phi^\pm_2 \end{array} \right) &=&
R(\beta) \left( \begin{array}{c} G^\pm \\ H^\pm \end{array}
\right) \label{eq:diagGHpm}
\;, 
\eeq
leading to the physical Higgs states, a neutral light CP-even, $h$, a neutral heavy
CP-even, $H$, a neutral CP-odd, $A$, and two charged Higgs bosons, 
$H^\pm$. The massless pseudo-Nambu-Goldstone bosons $G^\pm$ and $G^0$
are absorbed by the longitudinal components of the massive gauge
bosons, the charged $W^\pm$ and the $Z$ boson, respectively. The
rotation matrices in terms of the mixing angles $\vartheta = \alpha$ and
$\beta$, respectively, read
\beq
R(\vartheta) = \left( \begin{array}{cc} \cos \vartheta & - \sin
    \vartheta \\ \sin \vartheta & \cos \vartheta \end{array} \right) \;.
\eeq
The mixing angle $\beta$ is related to the two VEVs as
\beq
\tan \beta = \frac{v_2}{v_1} \;, \label{eq:tanbetadef}
\eeq
with $v_1^2 + v_2^2 = v^2 \approx (246 \mbox{ GeV})^2$, while the mixing angle
$\alpha$ is expressed through
\beq
\tan 2\alpha = \frac{2 ({\cal M}_S)_{12}}{({\cal M}_S)_{11}-({\cal M}_S)_{22}} \;,
\eeq 
where $({\cal M}_S)_{ij}$ ($i,j=1,2$) denote the matrix elements of the
neutral CP-even scalar mass matrix ${\cal M}_S$. With 
\beq
M^2 \equiv \frac{m_{12}^2}{s_\beta c_\beta}
\eeq
we have \cite{Kanemura:2004mg}
\beq
\tan 2\alpha = \frac{s_{2\beta} (M^2- \lambda_{345} v^2)}{c_\beta^2
  (M^2-\lambda_1 v^2) -s_\beta^2 (M^2-\lambda_2 v^2)}
\;,  \label{eq:alphadef}
\eeq
where we have introduced the abbreviation
\beq
\lambda_{345} \equiv \lambda_3 + \lambda_4 + \lambda_5 
\eeq
and used short-hand notation $s_x \equiv \sin x$ etc.
\s

The minimization conditions of the Higgs potential require
the terms linear in the Higgs fields to vanish in the vacuum, {\it i.e.}
\beq
\left\langle\frac{\partial V}{\partial \Phi_1}\right\rangle = 
\left\langle\frac{\partial V}{\partial \Phi_2}\right\rangle = 0 \;, \label{eq:tadcond}
\eeq
where the brackets denote the vacuum. The corresponding coefficients,
the tadpole parameters $T_1$ and $T_2$, therefore have to be zero. The tadpole
conditions at lowest order are given by
\beq
\left\langle\frac{\partial V}{\partial \Phi_1}\right\rangle &\equiv&
\frac{T_1}{v_1} =  m_{11}^2 - m_{12}^2 \frac{v_2}{v_1} + \frac{\lambda_1 v_1^2}{2}
+ \frac{\lambda_{345} v_2^2}{2} \label{eq:tad1} \\
\left\langle\frac{\partial V}{\partial \Phi_2}\right\rangle &\equiv&
\frac{T_2}{v_2} =  m_{22}^2 - m_{12}^2 \frac{v_1}{v_2} + \frac{\lambda_2 v_2^2}{2}
+ \frac{\lambda_{345} v_1^2}{2} \;. \label{eq:tad2}
\eeq
There are various possibilities to choose the set of independent
parameters that parametrizes the Higgs potential $V$. Thus, 
Eqs.~(\ref{eq:tad1}) and (\ref{eq:tad2}) can be used to
replace $m_{11}^2$ and $m_{22}^2$ by the tadpole parameters $T_1$ and
$T_2$. The VEV $v$ can furthermore be expressed in terms of
the physical gauge boson masses $M_W$ and $M_Z$ and the electric
charge $e$. In the following, we will choose the set of independent parameters
such that the parameters can be related to as many physical quantities as
possible. Our set is given by the Higgs boson masses,
the tadpole parameters, the two mixing angles, the soft breaking
parameter, the massive gauge boson masses and the electric
charge. Additionally, we will need the fermion masses $m_f$ for the
Higgs decays into fermions which will be used for a process-dependent
definition of the angular counterterms. 
\beq
\mbox{\underline{Input parameters:}} \quad
m_h,\; m_H,\; m_A,\; m_{H^\pm},\; T_1,\; T_2,\; \alpha,\;
\tan\beta,\; m_{12}^2, \; M_W^2, \; M_Z^2,\; e , \; m_f \;.
\eeq

\section{Renormalization \label{sec:renorm}}
\setcounter{equation}{0}
In this section we will present the various renormalization schemes that
we will apply in the renormalization of the 2HDM and that will
be investigated with respect to their gauge parameter dependence and their
numerical stability. We will use these schemes in sample processes 
given by the EW one-loop corrected decays of the charged Higgs
boson into a $W^\pm$ and a CP-even Higgs boson, $H^\pm \to W^\pm h/H$,
and of the heavy $H$ into a $Z$ boson pair, $H\to ZZ$.  
The computation of the EW one-loop corrections leads to ultraviolet
(UV) divergences. In the charged Higgs decay we will furthermore
encounter infrared (IR) divergences because of massless photons
running in the loops. The UV divergences in the virtual corrections are 
canceled by the renormalization of the parameters involved in the EW
corrections of the process, while the IR ones are subtracted by
taking into account the real corrections. The renormalization of the
above decay processes requires the renormalization of the electroweak
sector and of the Higgs sector. We will also compute the EW
one-loop corrections to the decays of $H$ and $A$ into $\tau$
leptons, $H/A \to \tau \tau$. These processes will be exploited for a
process-dependent definition of the angular counterterms, which will
be presented as a possible renormalization scheme among others. The
corrections to the decays into $\tau$ leptons also require the
renormalization of the fermion sector.
Note, that the renormalization of the CKM matrix, which we will
  assume to be real, will not play a role in our renormalization procedure.
We start by replacing the relevant parameters by the renormalized ones and their corresponding
 counterterms: \s

\noindent
\underline{\it Gauge sector:}
The massive gauge boson masses and the electric charge are
replaced by
\beq
M_W^2 &\to& M_W^2 + \delta M_W^2 \\
M_Z^2 &\to& M_Z^2 + \delta M_Z^2 \\
e &\to& (1+\delta Z_e) \, e \;.
\eeq
Equally, the VEV $v$, which will be expressed in terms of these
parameters, is replaced by 
\beq
v &\to& v + \delta v \;.
\eeq
The gauge boson fields are renormalized by their field
renormalization constants $\delta Z$,
\beq
W^\pm &\to& \left( 1 + \frac{1}{2} \delta Z_{WW} \right) W^\pm \\
\left( \begin{array}{c} Z \\ \gamma \end{array} \right) &\to& 
\left( \begin{array}{cc} 1 + \frac{1}{2} \delta Z_{ZZ} & \frac{1}{2}
    \delta Z_{Z\gamma} \\ \frac{1}{2} \delta Z_{\gamma Z} & 1 +
    \frac{1}{2} \delta Z_{\gamma\gamma} \end{array} \right) 
\left( \begin{array}{c} Z \\ \gamma \end{array} \right) \;.
\eeq
\underline{\it Fermion sector:} The counterterms to the fermion masses
$m_f$ are defined through
\beq
m_f \to m_f + \delta m_f \;.
\eeq
The bare left- and right-handed fermion fields
\beq
f_{L/R} \equiv P_{L/R} f \quad \;, \quad \mbox{ with } \quad P_{L/R} =
(1\mp \gamma_5)/2 \;,
\eeq
are replaced by their corresponding
renormalized fields according to
\beq
f_{L/R} \to \left( 1 + \frac{1}{2} \delta Z^{L/R}_f \right) f_{L/R} \;.
\eeq
\underline{\it Higgs sector:} The renormalization is performed in the
mass basis and the mass counterterms are defined through
\beq
m_h^2 &\to& m_h^2 + \delta m_h^2 \\
m_H^2 &\to& m_H^2 + \delta m_H^2 \\
m_A^2 &\to& m_A^2 + \delta m_A^2 \\
m_{H^\pm}^2 &\to& m_{H^\pm}^2 + \delta m_{H^\pm}^2 \;.
\eeq
The fields are replaced by the renormalized ones and the field
renormalization constants as
\beq
\left( \begin{array}{c}  H \\ h \end{array} \right) &\to&
\left( \begin{array}{cc} 1 + \frac{1}{2} \delta Z_{HH} & \frac{1}{2}
    \delta Z_{Hh} \\ \frac{1}{2} \delta Z_{hH} & 1 + \frac{1}{2}
    \delta Z_{hh} \end{array} \right) \left( \begin{array}{c} H \\
    h \end{array} \right) \label{eq:renconst1} \\
\left( \begin{array}{c}  G^0 \\ A \end{array} \right) &\to&
\left( \begin{array}{cc} 1 + \frac{1}{2} \delta Z_{G^0 G^0} & \frac{1}{2}
    \delta Z_{G^0 A} \\ \frac{1}{2} \delta Z_{AG^0} & 1 + \frac{1}{2}
    \delta Z_{AA} \end{array} \right) \left( \begin{array}{c} G^0 \\
    A \end{array} \right) \label{eq:renconst2} \\
\left( \begin{array}{c}  G^\pm \\ H^\pm \end{array} \right) &\to&
\left( \begin{array}{cc} 1 + \frac{1}{2} \delta Z_{G^\pm G^\pm} & \frac{1}{2}
    \delta Z_{G^\pm H^\pm} \\ \frac{1}{2} \delta Z_{H^\pm G^\pm} & 1 + \frac{1}{2}
    \delta Z_{H^\pm H^\pm} \end{array} \right) \left( \begin{array}{c} G^\pm \\
    H^\pm \end{array} \right)  \label{eq:renconst3}
\eeq
and the mixing angles by
\beq
\alpha &\to& \alpha + \delta \alpha \\
\beta &\to& \beta + \delta \beta \;.
\eeq
While the tadpoles vanish at leading order, the terms linear in the
Higgs fields get loop contributions at higher orders. Therefore, also
the tadpole parameters $T_1$ and $T_2$ have to be renormalized in
order to fulfill the tadpole conditions Eqs.~(\ref{eq:tadcond}). The
tadpoles are hence replaced as 
\beq
T_1 \to T_1 + \delta T_1 \qquad \mbox{and} \qquad T_2 \to T_2 + \delta
T_2 \;. \label{eq:tadpshift}
\eeq

\subsection{Renormalization conditions}
The finite parts of the counterterms are fixed by the renormalization
conditions. Throughout we will fix the renormalization constants for
the masses and fields through on-shell (OS) conditions. The
renormalization schemes differ, however, in the treatment of the
tadpoles and of the mixing angles. We will describe two different
approaches for the treatment of the tadpoles. Both of them apply the same renormalization conditions for the tadpoles. They differ, however, in the way the
minimum conditions are applied when the mass counterterms are
generated. As a consequence, the tadpole counterterms can either
explicitly show up in the mass counterterms or not. The latter case, that we will call {\it
  'alternative tadpole'} or in short {\it 'tadpole' } scheme, has the
virtue that the mass counterterms are manifestly gauge independent,
while  in the former one, named {\it 'standard tadpole'} or simply
{\it 'standard'} scheme, this is not the case. The authors of
Ref.~\cite{Kanemura:2004mg} have combined the standard tadpole scheme
with the definition of the counterterms through off-diagonal wave
function renormalization constants. This {\it 'KOSY'} scheme, denoted
by the initials of the authors, leads to manifestly
gauge-dependent decay amplitudes, as we will show. In the alternative
tadpole scheme not only this problem does not occur, but in addition, the angular
counterterms are explicitly gauge independent. If the
angular counterterms are defined in a {\it 'process-dependent'} scheme
via a physical process, the decay
amplitude is gauge independent irrespective of the treatment of the
tadpoles. The only difference lies in the gauge independence of the
angular counterterms in case the alternative tadpole scheme is
adopted. In the following, the renormalization conditions of the
various schemes will be introduced.  

\subsubsection{Standard Tadpole Scheme \label{sec:standard}}
We start by presenting the usual, {\it i.e.}~'standard', approach in
the renormalization of the 2HDM as also applied in 
\cite{Kanemura:2004mg,Kanemura:2015mxa}. 
The gauge bosons are renormalized through OS conditions, which implies 
the following counterterms for the masses,
\beq
\delta M_W^2 = \mbox{Re} \, \Sigma^T_{WW} (M_W^2) \quad \mbox{and}
\quad
\delta M_Z^2 = \mbox{Re}  \, \Sigma^T_{ZZ} (M_Z^2) \;, \label{eq:mwmzct}
\eeq
where the superscript $T$ denotes the transverse part of the
respective self-energy $\Sigma$. 
In order to guarantee the correct OS properties the wave
function renormalization constants have to be introduced as 
\beq
\delta Z_{WW} &=& - \mbox{Re} \left.\frac{\partial \Sigma^T_{WW}
  (p^2)}{\partial p^2}\right|_{p^2=M_W^2} \label{eq:sigmaww} \\
\left( \begin{array}{cc} \delta Z_{ZZ} & \delta Z_{Z\gamma} \\ \delta Z_{\gamma Z} &
  \delta Z_{\gamma\gamma} \end{array} \right) &=&
\left( \begin{array}{cc} - \mbox{Re} \left.\frac{\partial \Sigma^T_{ZZ}
  (p^2)}{\partial p^2}\right|_{p^2=M_Z^2} & 
2 \frac{\Sigma^T_{Z\gamma} (0)}{M_Z^2} \\
 -2 \mbox{Re} \frac{\Sigma^T_{Z\gamma} (M_Z^2)}{M_Z^2} & 
- \left.\frac{\partial \Sigma^T_{\gamma\gamma} (p^2)}{\partial
    p^2}\right|_{p^2=0} \end{array}\right) \;. \label{eq:sigmagamz}
\eeq
The electric charge is defined to be the full electron-positron photon
coupling for OS external particles in the Thomson limit, implying that
all corrections to this vertex vanish OS and for zero momentum
transfer. The counterterm for the electric charge in terms of the
transverse photon-photon and photon-$Z$ self-energies reads
\cite{Denner:1991kt}
\beq
\delta Z_e^{\alpha(0)} &=& \frac{1}{2} \left.\frac{\partial
  \Sigma^T_{\gamma\gamma} (k^2)}{\partial k^2}\right|_{k^2=0} +
\frac{s_W}{c_W} \frac{\Sigma_{\gamma Z}^T (0)}{M_Z^2} \;,
\label{eq:deltaze0}
\eeq
where $s_W/c_W \equiv \sin \theta_W/\cos \theta_W$ and $\theta_W$
denotes the Weinberg angle. Note that the sign in the 
second term of Eq.~(\ref{eq:deltaze0}) differs from the one in \cite{Denner:1991kt} 
due to our sign conventions in the covariant derivative of Eq.~(\ref{eq:covdiv}).
In our computation, however, we will use the fine structure constant
at the $Z$ boson mass $\alpha (M_Z^2)$ as input, so that the results
are independent of large logarithms due to light fermions $f\ne
t$. The counterterm $\delta Z_e$ is therefore modified as
\cite{Denner:1991kt} 
\beq
\delta Z_e^{\alpha (M_Z^2)} &=& \delta Z_e^{\alpha(0)} - \frac{1}{2}
\Delta \alpha (M_Z^2) \\
\Delta \alpha (M_Z^2) &=& \left. \frac{\partial
    \Sigma^T_{\gamma\gamma} (k^2)}{\partial k^2}\right|_{k^2=0} -
\frac{\Sigma^T_{\gamma\gamma} (M_Z^2)}{M_Z^2} \;, \label{eq:delalphmz}
\eeq
where the transverse part of the photon self-energy
$\Sigma^T_{\gamma\gamma}$ in Eq.~(\ref{eq:delalphmz}) includes only
the light fermion contributions.
For the computation of the EW one-loop corrected Higgs decay widths we
also need to renormalize the coupling $g$, which can be related to $e$ and the
gauge boson masses as
\beq
g &=& \frac{eM_Z}{\sqrt{M_Z^2- M_W^2}} \;,
\eeq
so that its counterterm can be expressed in terms of the gauge
boson mass counterterms through 
\beq
\frac{\delta g}{g} = \delta Z_e - \frac{1}{2(1-M_Z^2/M_W^2)}
\left( \frac{\delta M_W^2}{M_W^2} - \frac{\delta M_Z^2}{M_Z^2} \right) \;.
\eeq
Defining the following structure for the fermion self-energies
\beq
\Sigma_f (p^2) = \slash{\!\!\! p} \Sigma^L_f (p^2) P_L + \slash{\!\!\!
  p} \Sigma^R_f
(p^2) P_R + m_f \Sigma^{Ls}_f (p^2) P_L + m_f \Sigma^{Rs}_f (p^2) P_R 
\eeq
the fermion mass counterterms applying OS conditions are given by
\beq
\frac{\delta m_f}{m_f} = \frac{1}{2} \mbox{Re} \left[ \Sigma^L_f
(m_f^2) + \Sigma^R_f (m_f^2) + \Sigma^{Ls}_f (m_f^2) + \Sigma^{Rs}_f
(m_f^2) \right] \;.
\eeq
The fermion wave function renormalization constants are determined from
\beq
\delta Z^{L/R}_f &=& -\mbox{Re} \Sigma^{L/R}_f (m_f^2) \nonumber
\\
&&- m_f^2
\frac{\partial}{\partial p^2} \mbox{Re} \left.\left( \Sigma^{L/R}_f (p^2) +
  \Sigma^{R/L}_f (p^2) + \Sigma^{L/Rs}_f (p^2) + \Sigma^{R/Ls}_f (p^2)
\right)\right|_{p^2=m_f^2} \,. 
\eeq
The OS conditions for the physical Higgs bosons yield the following
Higgs mass counterterms
\beq
\delta m_H^2 &=& \mbox{Re} [\Sigma_{HH} (m_H^2) - \delta T_{HH}] \;,
\qquad
\delta m_h^2 = \mbox{Re} [\Sigma_{hh} (m_h^2) - \delta T_{hh}] \;,
\label{eq:massctscalar}
\\
\delta m_A^2 &=& \mbox{Re} [\Sigma_{AA} (m_A^2) - \delta T_{AA}] \;,
\qquad
\delta m_{H^\pm}^2 = \mbox{Re} [\Sigma_{H^\pm H^\pm} (m_{H^\pm}^2) -
\delta T_{H^\pm H^\pm}] \;. \label{eq:massctpschar}
\eeq
The appearance of the tadpole counterterms in
Eqs.~(\ref{eq:massctscalar}) and (\ref{eq:massctpschar}) can be
understood by recalling that the parameters $m_{11}^2$ and $m_{22}^2$,
which enter the mass matrices, can be replaced by the tadpole
coefficients $T_1$ and $T_2$. Applying the shifts
Eq.~(\ref{eq:tadpshift}) and rotating into the mass eigenbasis yield
the above conditions in the OS scheme. The relations between the
tadpole counterterms in the mass basis and $\delta T_{1,2}$ are given by
\beq
\delta T_{HH}  &=& \frac{\delta T_1}{v_1} \cos^2 \vartheta +
\frac{\delta T_2}{v_2} \sin^2
\vartheta \;, \\
\delta T_{hh/AA/H^\pm H^\pm} &=& \frac{\delta T_1}{v_1} \sin^2 \vartheta +
\frac{\delta T_2}{v_2} \cos^2 \vartheta \;, \\[0.1cm]
\mbox{with } \vartheta &=& \left\{ \begin{array}{cll} \alpha & \mbox{ for
    } & \delta T_{HH,hh} \\
\beta & \mbox{ for } & \delta T_{AA,H^\pm H^\pm} \end{array} \right. \;.
\eeq
The tadpoles are renormalized
such that the correct vacuum is reproduced at one-loop order, leading
to the renormalization conditions
\beq
\delta T_1 = T_1 \qquad \mbox{and} \qquad 
\delta T_2 = T_2 \;. 
\eeq
The $T_{1,2}$ stand for the contributions coming from the corresponding
genuine Higgs boson tadpole graphs in the gauge basis. For the wave function renormalization constants the OS renormalization implies the following conditions 
\beq
\left( \begin{array}{cc} \delta Z_{HH} & \delta Z_{Hh} \\
  \delta Z_{hH} & \delta Z_{hh} \end{array} \right) \hspace*{-0.35cm}
&=& \hspace*{-0.35cm}
\left( \begin{array}{cc} - \mbox{Re} \left.\frac{\partial
        \Sigma_{HH} (k^2)}{\partial k^2}\right|_{k^2=m_H^2} &
    2 \frac{\mbox{Re} \left[\Sigma_{Hh} (m_h^2) - \delta
        T_{Hh}\right]}{m_H^2-m_h^2} \\[0.3cm] 
-2 \frac{\mbox{Re} \left[\Sigma_{Hh} (m_H^2) - \delta
    T_{Hh}\right]}{m_H^2-m_h^2} & 
- \mbox{Re} \left.\frac{\partial \Sigma_{hh} (k^2)}{\partial
    k^2}\right|_{k^2=m_h^2} 
 \end{array} \right) \label{eq:wavefunc1} \\[0.2cm]
\left( \begin{array}{cc} \delta Z_{G^0 G^0} & \delta Z_{G^0 A} \\
  \delta Z_{A G^0} & \delta Z_{AA} \end{array} \right)
\hspace*{-0.35cm} &=& \hspace*{-0.35cm}
\left( \begin{array}{cc} - \mbox{Re} \left.\frac{\partial
        \Sigma_{G^0 G^0} (k^2)}{\partial k^2}\right|_{k^2=0} &
    -2 \frac{\mbox{Re}\left[\Sigma_{G^0 A} (m_A^2) - \delta T_{G^0
          A}\right]}{m_A^2} \\[0.3cm]
2 \frac{\mbox{Re} \left[\Sigma_{G^0 A} (0) - \delta T_{G^0 A} \right]}{m_A^2} & 
- \mbox{Re} \left.\frac{\partial \Sigma_{AA} (k^2)}{\partial
    k^2}\right|_{k^2=m_A^2} \end{array} \right) \label{eq:wavefunc2} \\[0.2cm]
\left( \begin{array}{cc} \delta Z_{G^\pm G^\pm} & \delta Z_{G^\pm H^\pm} \\
  \delta Z_{H^\pm G^\pm} & \delta Z_{H^\pm H^\pm} \end{array} \right)
\hspace*{-0.35cm} &=& \hspace*{-0.35cm}
\left( \begin{array}{cc} - \mbox{Re} \left.\frac{\partial
        \Sigma_{G^\pm G^\pm} (k^2)}{\partial k^2}\right|_{k^2=0} &
    -2\frac{\mbox{Re} \left[\Sigma_{G^\pm H^\pm} (m_{H^\pm}^2) -
        \delta T_{G^\pm H^\pm}\right]}{m_{H^\pm}^2} \\
2 \frac{\mbox{Re} \left[\Sigma_{G^\pm H^\pm} (0) - \delta T_{G^\pm
      H^\pm}\right]}{m_{H^\pm}^2} &  
- \mbox{Re} \left.\frac{\partial \Sigma_{H^\pm H^\pm} (k^2)}{\partial
    k^2}\right|_{k^2=m_{H^\pm}^2} 
 \end{array} \right) 
\;. \label{eq:wavefunc3}
\eeq

\subsubsection{The KOSY Scheme}
We now turn to the renormalization conditions for the mixing
angles. The renormalization scheme chosen in \cite{Kanemura:2004mg},
the 'KOSY' scheme, uses the standard tadpole scheme. For the
renormalization of the mixing angles it is based on the idea of making
the counterterms $\delta \alpha$ and $\delta \beta$ appear in the
inverse propagator matrix and hence in the wave function renormalization
constants, in a way that is consistent with the internal relations of the 2HDM. This
can be achieved by renormalizing in the mass basis $(f_1, f_2)^T$, but temporarily
switching to the gauge basis $(\gamma_1, \gamma_2)^T$, and back again,
\beq
\left( \begin{array}{c} f_1 \\ f_2 \end{array} \right) = 
R(\vartheta)^T \left( \begin{array}{c} \gamma_1 \\
    \gamma_2 \end{array} \right) &\to& R(\vartheta + \delta
\vartheta)^T \sqrt{Z_\gamma} \left( \begin{array}{c} \gamma_1 \\
    \gamma_2 \end{array} \right) \nonumber 
\eeq
\beq
= \underbrace{R(\delta \vartheta)^T R(\vartheta)^T \sqrt{Z_\gamma}
  R(\vartheta)}_{\equiv \sqrt{Z_f}}
R(\vartheta)^T \left( \begin{array}{c} \gamma_1 \\
    \gamma_2 \end{array} \right) 
= \sqrt{Z_f} \left( \begin{array}{c} f_1 \\ f_2 \end{array} \right) \;.
\eeq
The fields $f_i$ and $\gamma_i$ ($i=1,2$) and the mixing
angle $\vartheta$ stand here for any of the
field pairs in the mass and gauge basis, respectively, defined in
Eqs.~(\ref{eq:diagHh})-(\ref{eq:diagGHpm}), together with their
corresponding mixing angle, {\it i.e.}~$(f_i; \gamma_i; \vartheta) =
(H,h; \rho_i; \alpha)$, $(G^0,A; \eta_i; \beta)$ and $(G^\pm, H^\pm;
\phi^\pm_i; \beta)$. With the field
renormalization matrix $\sqrt{Z_\gamma}$ in the gauge basis being a
real symmetric matrix the following parametrization of the field
renormalization matrices in the mass basis can be chosen
\cite{Kanemura:2004mg,Kanemura:2015mxa} 
\beq
\sqrt{Z_f} &=& R(\delta \vartheta)^T
\left( \begin{array}{cc} 1 + \frac{1}{2} \delta Z_{f_1 f_1} & \delta
    C_{f} \\ \delta C_{f} & 1+\frac{1}{2} \delta Z_{f_2
      f_2} \end{array} \right) \nonumber \\
&=& \left( \begin{array}{cc} 1 + \frac{1}{2} \delta Z_{f_1 f_1} &
    \delta C_{f} + \delta \vartheta \\ \delta C_{f} - \delta
    \vartheta & 1 + \frac{1}{2} \delta Z_{f_2 f_2} \end{array}\right)
+ {\cal O}(\delta^2) \;.
\label{eq:ansatz}
\eeq
The off-diagonal elements are identified with the off-diagonal wave
function renormalization constants in the mass basis. For the CP-even
scalar sector we obtain 
\beq
\frac{1}{2} \delta Z_{Hh}^{\text{OS}} &=& \delta C_h + \delta \alpha
\\
\frac{1}{2} \delta Z_{hH}^{\text{OS}} &=& \delta C_h - \delta \alpha 
\eeq
and hence
\beq
\delta \alpha &=& \frac{1}{4} (\delta Z_{Hh}^{\text{OS}} - \delta
  Z_{hH}^{\text{OS}}) \label{eq:alphact1} \\
\delta C_h &=& \frac{1}{4} (\delta Z_{Hh}^{\text{OS}} + \delta
  Z_{hH}^{\text{OS}}) \;.
\eeq
The superscript 'OS' indicates the OS renormalization scheme for the
wave function constants. The counterterm $\delta C_h$ will not be used
again. While the mixing angle $\beta$ diagonalizes both the charged and
the CP-odd mass matrices and we have altogether four off-diagonal wave
function constants in the charged and CP-odd Higgs sector,
Eq.~(\ref{eq:ansatz}) implies 
only three free parameters to be fixed, namely $\delta \beta$, $\delta
C_A$ and $\delta C_{H^\pm}$. Consequently, one has to choose three out
of four possible conditions and not all scalar fields can be OS at
the same time. If we choose {\it e.g.}~the OS renormalized $\delta Z_{G^0
  A}^{\text{OS}}$, $\delta Z_{G^\pm H^\pm}^{\text{OS}}$ and $\delta
Z_{H^\pm G^\pm}^{\text{OS}}$ to fix the counterterms, we ensure 
$H^\pm$ to be OS. This scheme can hence be used in the process $H^\pm
\to W^\pm h/H$, where we have an external charged Higgs
boson.\footnote{Note that, aiming at OS renormalized fields, this
  scheme cannot be used in processes where both $A$ and $H^\pm$ are
  external fields without applying an additional finite rotation to
  render both fields OS.} This yields the following possible first set of counterterms,
\beq
\delta \beta^{(1)} &=& \frac{1}{4} (\delta Z^{\text{OS}}_{G^\pm H^\pm} -  \delta
Z^{\text{OS}}_{H^\pm G^\pm}) \label{eq:betact1} \\
\delta C_{H^\pm}^{(1)} &=& \frac{1}{4} (\delta Z^{\text{OS}}_{H^\pm G^\pm} +  \delta
Z^{\text{OS}}_{G^\pm H^\pm}) \\
\delta C_{A}^{(1)} &=& \frac{1}{2} \delta Z^{\text{OS}}_{A G^0} +
\delta \beta^{(1)} \;.
\eeq
Choosing on the other hand the set $\delta Z_{G^0 A}^{\text{OS}}$,
$\delta Z_{A G^0}^{\text{OS}}$ and $\delta Z_{H^\pm
  G^\pm}^{\text{OS}}$ we get a second possible set
\beq
\delta \beta^{(2)} &=& \frac{1}{4} (\delta Z^{\text{OS}}_{G^0 A} -  \delta
Z^{\text{OS}}_{A G^0}) \label{eq:betact2} \\
\delta C_{H^\pm}^{(2)} &=& \frac{1}{2} \delta Z^{\text{OS}}_{H^\pm
  G^\pm} +  \delta \beta^{(2)} \\
\delta C_{A}^{(2)} &=& \frac{1}{4} (\delta Z^{\text{OS}}_{A G^0} +
\delta Z^{\text{OS}}_{G^0 A}) \;.
\eeq
There are two more sets that can be chosen. However, we are not going to use them and hence they will not be repeated here. 
Replacing the OS conditions given in 
Eqs.~(\ref{eq:wavefunc1}), (\ref{eq:wavefunc2}) and (\ref{eq:wavefunc3}) in
Eqs.~(\ref{eq:alphact1}), (\ref{eq:betact1}) and (\ref{eq:betact2}), respectively, yields
the following counterterms for the mixing angles $\alpha$ and $\beta$
\beq
\delta \alpha &=& \frac{\mbox{Re} [\Sigma_{Hh} (m_H^2) +\Sigma_{Hh}
  (m_h^2) - 2 \delta T_{Hh}]}{2(m_H^2-m_h^2)} \label{eq:standdelalp} \\
\delta \beta^{(1)} &=& -\frac{\mbox{Re} [\Sigma_{G^\pm H^\pm} (0) +
  \Sigma_{G^\pm H^\pm} (m_{H^\pm}^2) - 2 \delta T_{G^\pm H^\pm}]}{2
  m_{H^\pm}^2} \qquad 
\mbox{or} \label{eq:beta1stand} \\
\delta \beta^{(2)} &=& -\frac{\mbox{Re}[\Sigma_{G^0 A} (0) + \Sigma_{G^0
    A} (m_{A}^2) - 2 \delta T_{G^0 A}]}{2 m_A^2} \;. \label{eq:beta2stand}
\eeq 
As already mentioned before and as we will demonstrate later in detail
for the example of the charged Higgs boson decay, the application of
this renormalization scheme not only makes a gauge-independent definition of the
  counterterms impossible, but more seriously, leads to unphysical
  gauge-dependent decay amplitudes. The computation of the
loop-corrected amplitude in the 
general $R_\xi$ gauge shows that after including all counterterms but
the ones for the angles, there remains a residual gauge
dependence that is UV-divergent. The angular counterterms must
therefore reveal exactly the same UV-divergent gauge dependence but
with opposite sign. The counterterm $\delta \alpha$ is found to
  have exactly this 
UV-divergent $\xi$-dependent counterpart, needed to render the
amplitude gauge independent. However, in addition, $\delta
\alpha$ and $\delta \beta$ contain $\xi$-dependent finite terms, which
reintroduce a gauge dependence into the amplitude. 
To get rid of these finite gauge-dependent terms in $\delta \beta$,
the authors of Ref.~\cite{Kanemura:2015mxa} 
suggest to drop the assumption that $\sqrt{Z_f}$ is symmetric, thereby
yielding additional renormalization conditions. These are then
exploited to move the gauge dependence of 
$\delta \beta$ into $\delta C_{f}$\footnote{More specifically it is 
  moved into $\delta C_{AG^0}$ and $\delta C_{G^0 A}$, that due to the
  non-symmetric $\sqrt{Z_f}$ are now two independent counterterms. For details,
  we refer the reader to the original reference.}. While this scheme
would in principle allow to 
eliminate the gauge dependence of $\delta \beta$, it cannot be applied
in processes that involve the renormalization of $\alpha$. The
UV-divergent $\xi$-dependent counterterm $\delta \alpha$ is needed to
cancel the UV-divergent $\xi$-dependent counterpart in the
loop-corrected amplitude, that is encountered 
in the standard renormalization scheme. 
In practice, however, this procedure cannot be applied, as it lacks an
unambiguous prescription on how to extract the truly gauge-independent
parts from the loop-corrected amplitude and from the
counterterms. The extraction of the 
gauge-independent part is not straightforward as the loop functions
$A_0$ and $B_0$ \cite{'tHooft:1978xw,Passarino:1978jh} 
which appear in the angular counterterms, can be
rewritten in terms of higher $n$-point scalar integrals that contain
the gauge parameter $\xi$ besides additional gauge-independent
components. \s

\subsubsection{Alternative Tadpole Scheme \label{sec:tadpscheme}}
\begin{figure}[ht!]
\begin{center}
\includegraphics[width=14cm,trim = 0mm 15mm 1mm 3mm,]{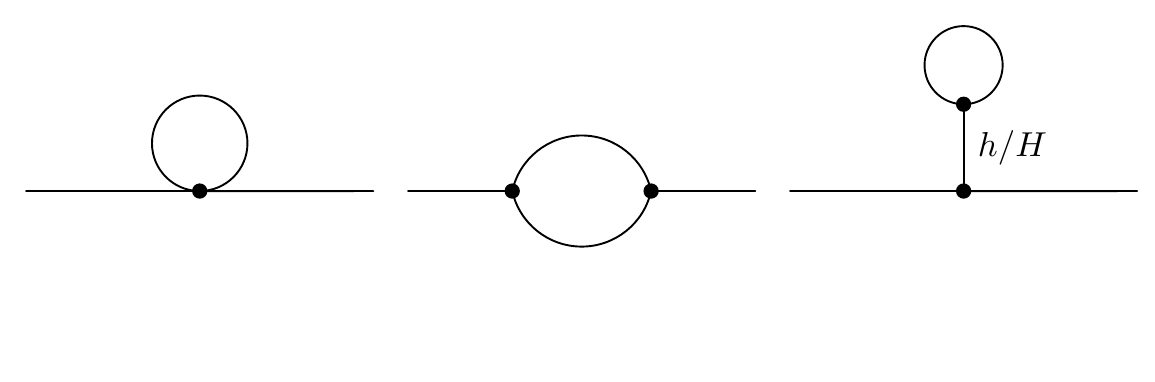}
\end{center}
\caption{Generic diagrams contributing to the self-energy $\Sigma^{\text{tad}}$.}
\label{fig:sigtad}
\end{figure}

We now present a renormalization scheme that fulfills the requirements
for a possible gauge-independent definition of the angular counterterms.
It relies on the application of the renormalization scheme worked out in
Ref.~\cite{Fleischer:1980ub}. In
Appendix~\ref{app:tadpole} we show in detail how this scheme works
and in particular we present its extension from the SM case
\cite{Fleischer:1980ub} to the 2HDM. The generic diagrams contributing
to the self-energies defined in this 'alternative tadpole' scheme, called 
$\Sigma^{\text{tad}}$ in the following, are shown in
Fig.~\ref{fig:sigtad}. Besides the generic one-particle irreducible
(1PI) diagrams depicted by the first two topologies in
Fig.~\ref{fig:sigtad}, they also contain the tadpole diagrams
connected to the self-energies through the CP-even Higgs bosons $h$
and $H$ that are represented by the third topology.
The application of the tadpole scheme alters the structure of the
mass counterterms and of the off-diagonal wave function renormalization 
constants\footnote{Note, that the application of the tadpole scheme also
  requires a change of all those vertices, where tadpole contributions
  now have to be taken into account, namely wherever it is possible to add a neutral
scalar. This will be discussed later in the 
computation of the loop-corrected decay widths.}
such that now the loop-corrected amplitude including all counterterms
but those for the angles does not encounter a UV-divergent $\xi$
dependence any more. Hence, also the angular counterterms can
and even have to be defined in a gauge-independent way by
applying appropriate renormalization conditions. 
\s 

Besides the angular counterterms, also the mass counterterms,
  defined via OS conditions become gauge independent in the tadpole
  scheme. This has been shown for the electroweak sector in
  \cite{Gambino:1999ai}. All counterterms of the electroweak sector
  have exactly the same structure as in the standard 
scheme, but the self-energies $\Sigma$ appearing in
Eqs.~(\ref{eq:mwmzct})-(\ref{eq:deltaze0}) 
have to be replaced by the self-energies $\Sigma^{\text{tad}}$
containing the tadpole contributions. Note however, that there are no
tadpole contributions for $\Sigma_{\gamma Z}^T$ so that
\beq
\Sigma_{\gamma Z}^{\text{tad},T} = \Sigma_{\gamma Z}^T \;.
\eeq
Furthermore, due to the fact that the tadpoles are independent of the external momentum the derivatives of the self-energies do not change,  
\beq
\frac{\partial \Sigma_{xy}^{\text{tad},T}}{\partial k^2} = \frac{\partial
  \Sigma_{xy}^T}{\partial k^2} \qquad \mbox{for} \qquad 
xy= WW, ZZ, \gamma\gamma,HH,hh,G^0G^0,G^\pm G^\pm,H^\pm H^\pm\;.
\label{eq:tadpderivatives}
\eeq
The Higgs mass counterterms become
\beq
\delta m_H^2 &=& \mbox{Re} [\Sigma^{\text{tad}}_{HH} (m_H^2)] \;,
\qquad
\delta m_h^2 = \mbox{Re} [\Sigma^{\text{tad}}_{hh} (m_h^2)] \;,
\label{eq:massctscalartadp}
\\
\delta m_A^2 &=& \mbox{Re} [\Sigma^{\text{tad}}_{AA} (m_A^2)] \;,
\qquad
\delta m_{H^\pm}^2 = \mbox{Re} [\Sigma^{\text{tad}}_{H^\pm H^\pm}
(m_{H^\pm}^2)]  \;. \label{eq:massctchartadp}
\eeq
And for the Higgs wave function renormalization constants we obtain 
\beq
\left( \begin{array}{cc} \delta Z_{HH} & \delta Z_{Hh} \\
  \delta Z_{hH} & \delta Z_{hh} \end{array} \right) \hspace*{-0.2cm}
&=& \hspace*{-0.2cm} 
\left( \begin{array}{cc} - \mbox{Re} \left.\frac{\partial
        \Sigma^{\text{tad}}_{HH} (k^2)}{\partial
        k^2}\right|_{k^2=m_H^2} & 2
    \frac{\mbox{Re} \left[\Sigma^{\text{tad}}_{Hh}
        (m_h^2)\right]}{m_H^2-m_h^2} \\[0.3cm] 
-2 \frac{\mbox{Re} \left[\Sigma^{\text{tad}}_{Hh}
    (m_H^2)\right]}{m_H^2-m_h^2}  & 
- \mbox{Re} \left.\frac{\partial \Sigma^{\text{tad}}_{hh}
    (k^2)}{\partial k^2}\right|_{k^2=m_h^2} 
 \end{array} \right) \label{eq:wavefunc1tadp} \\[0.2cm]
\left( \begin{array}{cc} \delta Z_{G^0 G^0} & \delta Z_{G^0 A} \\
  \delta Z_{A G^0} & \delta Z_{AA} \end{array} \right)
\hspace*{-0.2cm} &=& \hspace*{-0.2cm} 
\left( \begin{array}{cc} - \mbox{Re} \left.\frac{\partial
        \Sigma^{\text{tad}}_{G^0 G^0} (k^2)}{\partial
        k^2}\right|_{k^2=0} & -2 \frac{\mbox{Re}
      \left[\Sigma^{\text{tad}}_{G^0 A} (m_A^2)\right] }{m_A^2}
  \\[0.3cm] 
2 \frac{\mbox{Re} \left[ \Sigma^{\text{tad}}_{G^0 A} (0) \right]}{m_A^2} & 
- \mbox{Re} \left.\frac{\partial \Sigma^{\text{tad}}_{AA}
    (k^2)}{\partial k^2}\right|_{k^2=m_A^2} 
 \end{array} \right) \label{eq:wavefunc2tadp} \\[0.2cm]
\left( \begin{array}{cc} \delta Z_{G^\pm G^\pm} & \delta Z_{G^\pm H^\pm} \\
  \delta Z_{H^\pm G^\pm} & \delta Z_{H^\pm H^\pm} \end{array} \right) 
\hspace*{-0.2cm} &=& \hspace*{-0.2cm}
\left( \begin{array}{cc} - \mbox{Re} \left.\frac{\partial
        \Sigma^{\text{tad}}_{G^\pm G^\pm} (k^2)}{\partial
        k^2}\right|_{k^2=0} & -2 \frac{\mbox{Re}
      \left[\Sigma^{\text{tad}}_{G^\pm H^\pm} (m_{H^\pm}^2)
      \right]}{m_{H^\pm}^2} \\[0.3cm] 
2 \frac{\mbox{Re} \left[ \Sigma^{\text{tad}}_{G^\pm H^\pm} (0)
  \right]}{m_{H^\pm}^2} & 
- \mbox{Re} \left.\frac{\partial \Sigma^{\text{tad}}_{H^\pm H^\pm} (k^2)}{\partial
    k^2}\right|_{k^2=m_{H^\pm}^2} 
 \end{array} \right) 
\label{eq:wavefunc3tadp}
\eeq
keeping in mind that Eq.~(\ref{eq:tadpderivatives}) holds.
Applying the same procedure for the definition of the angular
counterterms as in the standard scheme, but with the different
treatment of the tadpoles, the angular counterterms in the tadpole
scheme read
\beq
\delta \alpha &=& \frac{\mbox{Re} \left[ \Sigma^{\text{tad}}_{Hh} (m_H^2)
  +\Sigma^{\text{tad}}_{Hh}
  (m_h^2) \right]}{2(m_H^2-m_h^2)} \label{eq:deltalphatad}\\ 
\delta \beta^{(1)} &=& -\frac{\mbox{Re}
  \left[\Sigma^{\text{tad}}_{G^\pm H^\pm} (0) + 
  \Sigma^{\text{tad}}_{G^\pm H^\pm} (m_{H^\pm}^2)\right]} {2 m_{H^\pm}^2} 
\label{eq:deltbetatad1} \\
\delta \beta^{(2)} &=& -\frac{\mbox{Re} \left[\Sigma^{\text{tad}}_{G^0 A} (0) +
  \Sigma^{\text{tad}}_{G^0 A} (m_{A}^2)\right]} {2 m_A^2} \;.
\label{eq:deltbetatad2}
\eeq 
Compared to the standard scheme, the self-energies are replaced by the
$\Sigma^{\text{tad}}$ and no tadpole counterterms appear any more. \s

The application of the tadpole scheme not only allows for a
gauge-independent definition of the angular counterterms but also requires it
in order to ensure a gauge-independent physical decay
amplitude. Note that the counterterms
    (\ref{eq:deltalphatad})-(\ref{eq:deltbetatad2}) still contain a
    $\xi$ dependence and hence, a $\xi$-independent definition has yet to be found. In the MSSM, several schemes for the
renormalization of $\tan\beta$ have been proposed and used, see {\it e.g.}
\cite{Freitas:2002um,Pierce:1992hg,Dabelstein:1994hb,Chankowski:1992er,Coarasa:1996qa,Shan:1998bv,Heinemeyer:2004gx,Fujimoto:2007bn,Baro:2009gn}. The
renormalization prescriptions have been discussed in
detail in \cite{Freitas:2002um} with respect to their gauge
dependence, process independence and numerical stability (see also
\cite{Baro:2008bg}). Renormalization prescriptions making use of
physical quantities like Higgs boson masses or physical processes 
clearly lead to a gauge-independent prescription. However, they were found to be
numerically unstable in the former case, while the latter
case may be viewed as unsatisfactory, as the definition via a specific
process makes $\tan\beta$ a non-universal, flavour-dependent
quantity \cite{Freitas:2002um}. Finally, $\overline{\mbox{DR}}$
prescriptions lead in the $R_\xi$ gauge to gauge-independence of
$\delta \tan\beta$ in the MSSM at one-loop level, but not at two-loop
level \cite{Freitas:2002um,Yamada:2001ck}.
We now present a renormalization scheme that
leads to $\xi$-independent $\delta \alpha$ and $\delta \beta$ and also
addresses the problem of extracting the gauge-independent part
unambiguously. \\

\noindent
{\bf On-shell tadpole-pinched scheme \label{subsubsec:tadpOSpin}} \\[0.2cm]
The scheme we propose here combines the virtues of the tadpole scheme with the 
unambiguous extraction of the truly gauge-independent parts of the
angular counterterms. It is based on the renormalization schemes
presented in \cite{Baro:2009gn} and in 
\cite{Yamada:2001px,Espinosa:2002cd}. The former defines the
angular  counterterms in a physical way as residues of poles appearing in
one-loop corrections, while in
\cite{Yamada:2001px,Espinosa:2002cd} the pinch 
technique\footnote{There has been some discussion on the pinch technique (PT). 
In
Refs.~\cite{Abbott:1980hw,Abbott:1981ke,KlubergStern:1974xv,KlubergStern:1975hc,Boulware:1980av,Hart:1984jy,Denner:1994xt,Denner:1994nn}
the background field method (BFM) was advocated in order to obtain
gauge invariant definitions of the counterterms, which, however, also
has drawbacks (see {\it
  e.g.}~\cite{Binosi:2004qe,Binosi:2009qm}). We apply the PT only in 
the definition of the angular counterterms at one-loop level and not
for the complete one-loop process, so that we do not run into possible
problems with 
regard to the PT. Also, note that for specific examples it
has been shown that the PT is connected to the BFM in case
the Feynman gauge is chosen for the background fields.}
\cite{Cornwall:1989gv,Papavassiliou:1989zd,Papavassiliou:1994pr,Degrassi:1992ue,Binosi:2009qm}
is used to extract the truly gauge-independent parts of the angular 
counterterms. Both methods lead to the same gauge-independent
definitions of the counterterms. With the help of the PT it is possible to define the pinched self-energies $\overline{\Sigma}$. The self-energies are
  related to the tadpole self-energies evaluated in the Feynman gauge as 
\beq
\overline{\Sigma} (p^2) = \left.\Sigma^{\text{tad}}
  (p^2)\right|_{\xi=1} + \Sigma^{\text{add}} (p^2) \;,
\label{eq:sigadddef}
\eeq
where $\xi$ stands for the gauge fixing parameters $\xi_Z$, $\xi_W$ and
$\xi_\gamma$ of the $R_\xi$ gauge. 
Note, that in order to apply the PT the tadpole scheme has to be
  used.\footnote{In Ref.~\cite{Bojarski:2015kra} the pinch
    technique was applied by evaluating the self-energies in the
    Feynman gauge, without taking into account that the self-energies
    have to be used in the tadpole scheme. We explicitly verified that
  the thus obtained quantities in \cite{Bojarski:2015kra} differ from
  the results that would be obtained by applying the pinch technique
  correctly. Furthermore, in our opinion the numerical verification of
  the gauge independence 
  performed in \cite{Bojarski:2015kra} is not valid, as it
is applied to the self-energy already fixed to be the one in the
Feynman gauge, so that a true check of gauge independence is
precluded.} For better readability we omitted the superscript 'tad'
in  $\overline{\Sigma}$. The self-energy $\Sigma^{\text{add}}$  in
Eq.~(\ref{eq:sigadddef})  is an additional contribution that is
explicitly independent of the gauge fixing parameter 
$\xi$. Applying \cite{Espinosa:2002cd} we arrive at the following
counterterms 
\beq
\delta \alpha &=& \frac{\mbox{Re}\left(\left[\Sigma_{Hh}^{\text{tad}} (m_H^2) +
  \Sigma^{\text{tad}}_{Hh} (m_h^2) \right]_{\xi=1} +
\Sigma^{\text{add}}_{Hh} (m_H^2) + \Sigma^{\text{add}}_{Hh} (m_h^2)
\right)}{2 (m_H^2 - m_h^2)}  
\label{eq:posalpha}
\\
\delta \beta^{(1)} &=& - \frac{\mbox{Re}\left(\left[\Sigma_{G^\pm
        H^\pm}^{\text{tad}} (0) + 
  \Sigma^{\text{tad}}_{G^\pm H^\pm} (m_{H^\pm}^2) \right]_{\xi=1} +
  \Sigma^{\text{add}}_{G^\pm H^\pm} (0)
  + \Sigma^{\text{add}}_{G^\pm H^\pm} (m_{H^\pm}^2)\right)}{2 m_{H^\pm}^2} 
\label{eq:posbeta1} \\
\delta \beta^{(2)} &=& - \frac{\mbox{Re}\left(\left[\Sigma_{G^0 A}^{\text{tad}} (0) +
  \Sigma^{\text{tad}}_{G^0 A} (m_A^2) \right]_{\xi=1} +
  \Sigma^{\text{add}}_{G^0 A} (0)
  + \Sigma^{\text{add}}_{G^0 A} (m_A^2)\right)}{2 m_A^2} 
\;. \label{eq:posbeta2}
\eeq
These angular counterterms are different from the ones obtained in the
KOSY scheme, so that the classification as an independent renormalization
scheme is justified. The additional contribution
$\Sigma^{\text{add}}_{Hh}$ has been given in
\cite{Espinosa:2002cd} for the MSSM. We have derived the remaining two
contributions  $\Sigma^{\text{add}}_{G^0 A}$ and
$\Sigma^{\text{add}}_{G^\pm H^\pm}$. Altogether we have 
\beq
\Sigma^{\text{add}}_{Hh} (p^2) &=& \frac{g^2 s_{\beta -\alpha}
  c_{\beta-\alpha}}{32 \pi^2 c_W^2} \left( p^2 - \frac{m_H^2 +
    m_h^2}{2} \right) \Big\{ B_0 (p^2; m_Z^2, m_A^2) -B_0 (p^2; m_Z^2,
m_Z^2) \nonumber
\\
&& + 2 c_W^2 \left[ B_0 (p^2; m_W^2, m_{H^\pm}^2) -B_0 (p^2; m_W^2,
  m_W^2) \right] \Big\}
\label{eq:sigaddhh} \\
\Sigma^{\text{add}}_{G^0 A} (p^2) &=& \frac{g^2 s_{\beta -\alpha}
  c_{\beta-\alpha}}{32 \pi^2 c_W^2} \left( p^2 - \frac{m_A^2}{2}
\right) \left[ B_0 (p^2; m_Z^2,m_H^2) - B_0 (p^2; m_Z^2, m_h^2)
\right] \label{eq:sigaddga} \\
\Sigma^{\text{add}}_{G^\pm H^\pm} (p^2) &=& \frac{g^2 s_{\beta -\alpha}
  c_{\beta-\alpha}}{16 \pi^2} \left( p^2 - \frac{m_{H^\pm}^2}{2}
\right) \left[ B_0 (p^2; m_W^2,m_H^2) - B_0 (p^2; m_W^2, m_h^2)
\right] \;, \label{eq:sigaddghpm} 
\eeq 
where $B_0$ is the scalar two-point function
\cite{'tHooft:1978xw,Passarino:1978jh}. \\[0cm] 

\noindent{\bf $p_\star$ tadpole-pinched scheme \label{subsubsec:tadppstarpin}} \\[0.1cm] 
As indicated by the name, this scheme differs from the OS
tadpole-pinched scheme solely in the scale at which the self-energies,
appearing in the definition of the angular counterterms, are evaluated. 
The self-energies are evaluated at the average
of the particle momenta squared \cite{Espinosa:2002cd},
\beq
p_\star^2 = \frac{m_{\phi_1}^2 + m_{\phi_2}^2}{2} \;,
\eeq
with $(\phi_1,\phi_2)=(H,h)$, $(G^\pm,H^\pm)$ and $(G^0,A)$,
respectively, and we will henceforth refer to this scheme as the
$p_\star$-scheme.
When the self-energies are evaluated at $p_\star^2$ the additional
self-energies $\Sigma^{\text{add}}$ vanish, as can easily be seen from
Eqs.~(\ref{eq:sigaddhh})-(\ref{eq:sigaddghpm}), and the 
  pinched self-energies are given by the  
tadpole self-energies $\Sigma^{\text{tad}}$ computed in the
Feynman gauge, {\it i.e.} 
\beq
\overline{\Sigma} (p_\star^2) = \left.\Sigma^{\text{tad}}
  (p_\star^2)\right|_{\xi=1} \;.
\eeq
The angular counterterms then read
\beq
\delta \alpha &=& \frac{\mbox{Re} \left[\overline{\Sigma}_{Hh} \left(
      \frac{m_h^2 + 
      m_H^2}{2} \right) \right]}{m_H^2-m_h^2} \label{eq:delalphstar} \\
\delta \beta^{(1)} &=& - \frac{\mbox{Re}
  \left[\overline{\Sigma}_{G^\pm H^\pm} \left( 
    \frac{m_{H^\pm}^2}{2} \right)\right]}{m_{H^\pm}^2} \label{eq:delbet1star}
\\
\delta \beta^{(2)} &=& - \frac{\mbox{Re} \left[\overline{\Sigma}_{G^0 A}
  \left( \frac{m_A^2}{2} \right) \right]}{m_A^2} \label{eq:delbet2star} \;.
\eeq

\subsubsection{Process-dependent Scheme \label{sec:procdepscheme}}
We will also investigate the renormalization of the mixing angles
through a physical process. Provided the alternative tadpole scheme is
applied, this leads to a manifestly gauge-independent 
definition of the mixing angle counterterms. In order to fix the
respective angular counterterm we will 
require the next-to-leading order (NLO) Higgs decay width, in
which the angle appears, to be equal to the leading order (LO) one, {\it i.e.} 
\beq
\Gamma_{\text{virt}} + \Gamma_{\text{c.t.}} = 0 \;,
\eeq
where $\Gamma_{\text{virt}}$ denotes the contribution of all virtual one-loop corrections to the decay width and $\Gamma_{\text{c.t.}}$ the counterterm
contributions. This implies (for NLO processes that do not encounter
real corrections, see below)
\beq
\Gamma^{\text{NLO}} = \Gamma^{\text{LO}}
\eeq
and allows to fix the angular counterterm via the decay process. This
scheme has some drawbacks, however, {\it
  cf.}~\cite{Freitas:2002um}. Conceptually, it is not satisfying 
as the definition of the mixing angles becomes non-universal and
flavour-dependent. From a calculational point of view, it is involved
as it requires the computation of loop-corrected three-particle
vertices. 
Another problem is related to the choice of the process that
  defines the counterterm. The definition through a
  process receiving QED corrections that cannot be separated from
  the rest of the EW corrections would entail real radiative
  corrections in the counterterm. This is precluded, however, as this
  counterterm would inevitably depend on some detector sensitivity $\Delta E$  via
  the photon phase space cut and thereby introduce a dependence
  on the experimental setting. This forbids {\it e.g.} the definition
  of the angular counterterms appearing in the loop corrected decay $H^\pm
  \to W^\pm h$ through the process $H^\pm \to W^\pm H$. 
Finally, care has to be taken to choose a process that is phenomenologically
accessible. This eliminates {\it e.g.}~the choice of $H \to ZZ$.
With the 125~GeV Higgs boson being very SM-like and hence coupling with full SM strength to the $Z$ bosons, sum rules lead to a tiny coupling
of the heavy Higgs boson to massive gauge bosons and hence a very small $H\to
ZZ$ decay width. In this paper we choose, as proposed in
\cite{Freitas:2002um}, the decays $H \to \tau \tau$ and $A \to \tau \tau$ in  
order to define $\delta \beta$ via the latter and $\delta \alpha$ via
the former. In both decays the QED corrections form a UV-finite
subset of the full EW one-loop corrections. 

\section{One-Loop EW Corrected Decay Widths \label{sec:oneloopdec}}
\setcounter{equation}{0}
In this section we present the EW one-loop 
corrections to the processes\footnote{The top quark loop corrections
  to $H^\pm \to W^\pm h$ have been calculated in \cite{Santos:1996hs}.}
\beq
H^\pm &\to& W^\pm h \quad\;\;\,\hspace*{0.03cm} \mbox{and} \quad W^\pm
H \;,  \label{eq:chardec} \\ 
H &\to& ZZ \;, \label{eq:zdec} \\ 
H &\to& \tau\tau \quad \mbox{and} \quad A \to \tau \tau\;. \label{eq:fermdec}
\eeq 
The charged Higgs decays (\ref{eq:chardec}) will serve us to discuss in detail the
renormalization of the mixing angles $\alpha$ and $\beta$ in view of a
gauge-independent definition. In this context, the fermionic decays
(\ref{eq:fermdec}) will be used for a 
process-dependent definition of the angular counterterms. 
Note that we could have equally well chosen $h\to \tau\tau$ instead of $H \to \tau
\tau$. The numerical implications of the different
renormalization schemes shall 
be investigated in the subsequent section. This will be done not only
for the charged Higgs decays, but also for another sample process, the
heavy Higgs decay into a $Z$ boson pair (\ref{eq:zdec}).

\subsection{Electroweak One-Loop Corrections to $H^\pm \to W^\pm h/H$} 
\begin{figure}[t]
\begin{center}
\includegraphics[width=15cm]{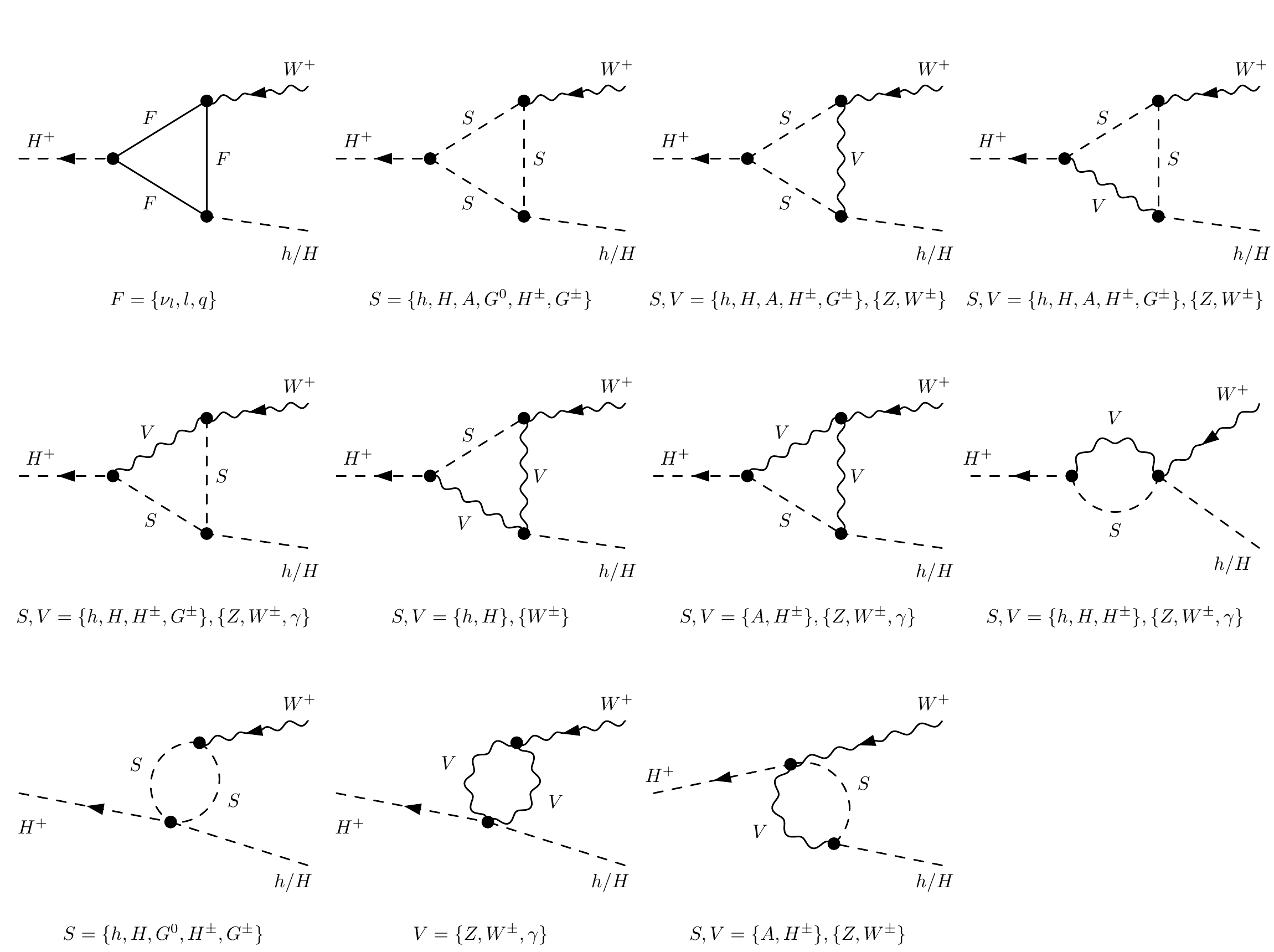}
\end{center}
\caption{Generic diagrams contributing to the vertex
  corrections in $H^\pm \to W^\pm h/H$.}
\label{fig:detailsvirt}
\end{figure}

\begin{figure}[b!]
\begin{center}
\includegraphics[width=15cm]{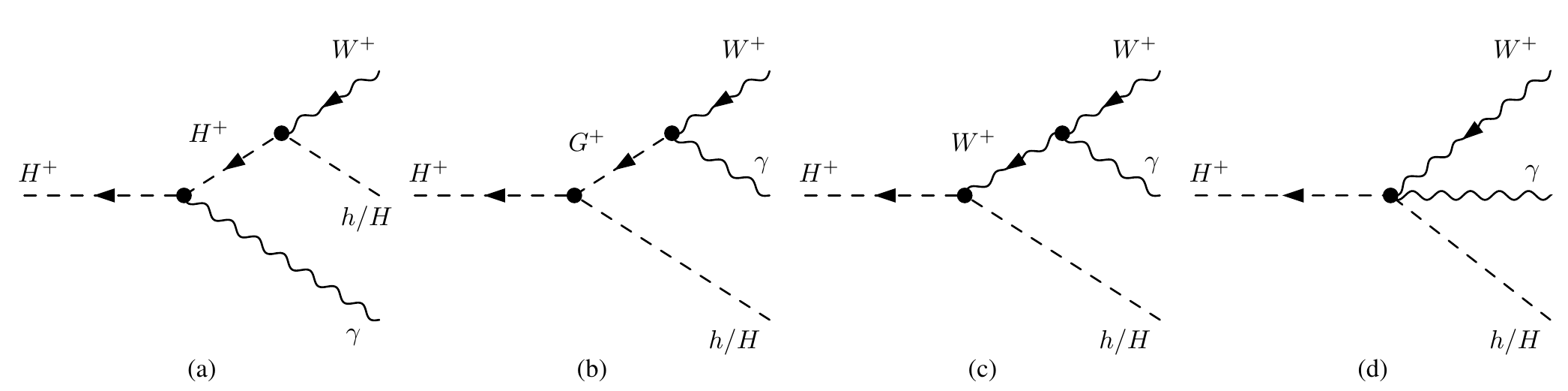}
\end{center}
\caption{Feynman diagrams contributing to the real
  corrections.}
\label{fig:realcontr}
\end{figure}

The decays of the charged Higgs boson into the charged $W^\pm$ boson and a
CP-even Higgs boson $\phi= h$ or $H$,
\beq
H^\pm \to W^\pm \phi \; ,
\eeq
depend through the couplings on the mixing angle combinations
\beq
g_{H^\pm W^\pm \phi} = \left\{ \begin{array}{cll}
- \cos (\beta - \alpha) & \mbox{for} & \phi = h \\
\sin (\beta-\alpha) & \mbox{for} & \phi = H 
\end{array} \right. \;,
\eeq
\begin{figure}[b!]
\begin{center}
\includegraphics[width=13cm]{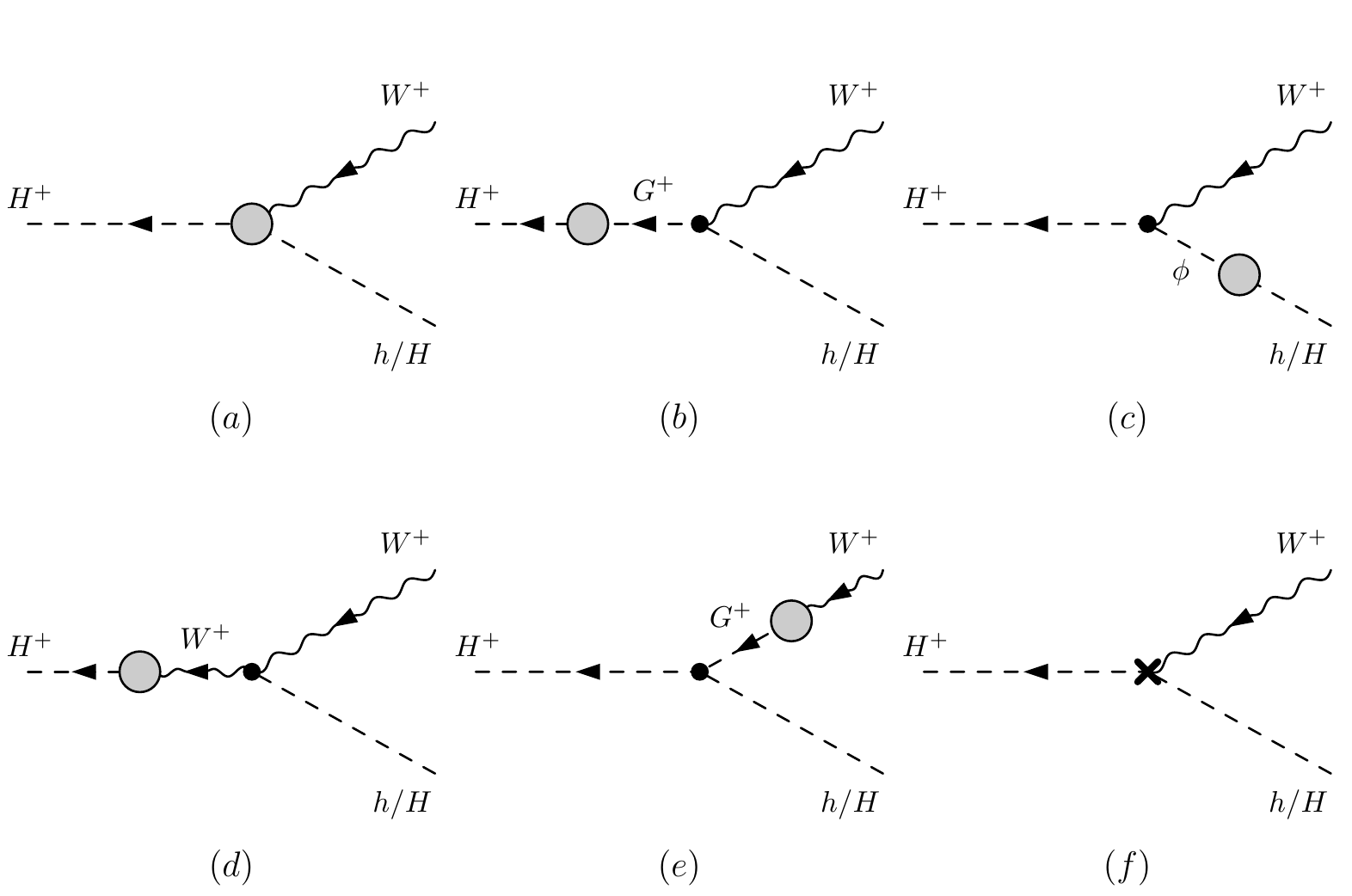}
\end{center}
\caption{Generic diagrams contributing to the virtual corrections of
  the decays $H^\pm \to W^\pm
  h/H$: vertex corrections (a) and corrections to the
  external legs (b)-(e). Diagram (f)
  displays the counterterm.}
\label{fig:virtualcorrs}
\end{figure}
and the LO decay width is given by
\beq
\Gamma^{\text{LO}} (H^\pm \to W^\pm \phi) = \frac{G_F g_{H^\pm W^\pm
    \phi}^2 }{8 \sqrt{2} \pi m_{H^\pm}^3} \,\lambda^3 (m_{H^\pm}^2,
M_W^2, m_\phi^2) \;,
\eeq
with 
\beq
\lambda (x,y,z) \equiv (x^2+y^2+z^2-2xy-2xz-2yz)^{\frac{1}{2}} \;.
\eeq
The NLO decay width can be written as
\beq
\Gamma^{\text{NLO}} = \Gamma^{\text{LO}} + \Gamma^{(1)} \;.
\eeq
The one-loop correction $\Gamma^{(1)}$ consists of the virtual
corrections, the counterterm contributions and the real
corrections. The counterterms cancel the UV divergences and the real
corrections the IR divergences encountered in the virtual
corrections. The diagrams contributing to the latter are depicted in
Fig.~\ref{fig:virtualcorrs} and show the pure vertex corrections
(a) and the corrections (b)-(e) to the external legs. The counterterm
diagram is shown in (f). 
The vertex corrections comprise the 1PI diagrams given by the
triangle diagrams with scalars, fermions and gauge bosons in the
loops, as shown in the first two rows of Fig.~\ref{fig:detailsvirt}, 
and the diagrams involving four-particle vertices (last
four diagrams of Fig.~\ref{fig:detailsvirt}). 
The corrections to the external legs in Fig.~\ref{fig:virtualcorrs}
(b) and (c) vanish due to the OS renormalization of the scalars, while
the vanishing of the mixing contribution (d) is ensured by a Slavnov-Taylor
identity \cite{Williams:2011bu}\footnote{This requires the formulation of the
  gauge fixing Lagrangian in terms of already renormalized fields when
  adding it to the bare 2HDM Lagrangian so that it need not be
  renormalized, {\it cf.}~Refs.~\cite{Ross:1973fp,Baulieu:1983tg}. See
also \cite{Santos:1996vt} for details. \label{foot:legcancel}} and the
one of (e) by the Ward identity for an OS $W^\pm$ boson. 
The vertex contributions with a
photon in the loop lead to IR divergences that need to be canceled
by the real corrections. These are computed from the diagrams
displayed in Fig.~\ref{fig:realcontr}. They consist of the proper
bremsstrahlung contributions (a)-(c), where a photon is radiated from the
charged initial and final state particles, and the diagram (d) involving a
four-particle vertex with a photon. Note, that this last diagram leads
to an IR-finite contribution. 
The NLO contributions
factorize from the LO amplitude, so that the one-loop corrected decay
width can be cast into the form
\beq
\Gamma^{\text{NLO}} (H^\pm \to W^\pm \phi) = \Gamma^{\text{LO}} \left[
  1 + \Delta^{\text{virt}} + 
  \Delta^{\text{ct}} + \Delta^{\text{real}} \right] \;.
\eeq
The counterterm contribution $\Delta^{\text{ct}}$ is given in terms of
the wave function renormalization constants, the coupling and angle
counterterms. For $\phi \equiv h$ it reads
\beq
\Delta^{\text{ct}} = \delta Z_{WW} + \delta Z_{H^\pm H^\pm} + \delta 
Z_{hh} + \frac{s_{\beta-\alpha}}{c_{\beta-\alpha}} \left(\delta 
Z_{G^\pm H^\pm} - \delta Z_{Hh} \right) + 2 \frac{\delta g}{g} - 2
t_{\beta-\alpha} \, (\delta \beta - \delta \alpha) \;, \label{eq:cthpwphl}
\eeq 
and for $\phi \equiv H$,
\beq
\Delta^{\text{ct}} = \delta Z_{WW} + \delta Z_{H^\pm H^\pm} + \delta
Z_{HH} - \frac{c_{\beta-\alpha}}{s_{\beta-\alpha}} \left(\delta
Z_{G^\pm H^\pm} + \delta Z_{hH} \right) + 2 \frac{\delta g}{g} + 
\frac{2(\delta \beta-\delta\alpha)}{t_{\beta-\alpha}} \;. \label{eq:cthpwphh}
\eeq
As the expressions for the counterterm $\Delta^{\text{ct}}$ and the virtual
and real contributions $\Delta^{\text{virt}}$ and
$\Delta^{\text{real}}$ in terms of scalar one-, two- and three-point
functions are rather lengthy, we do not display them explicitly here.

\subsection{Electroweak One-Loop Corrections to $H \to \tau\tau$ and 
  $A\to \tau\tau$}
The LO decay width for the process $H\to \tau\tau$ reads
\beq
\Gamma^{\text{LO}} (H\to \tau\tau) = \frac{G_F g^2_{H\tau\tau} m_H
  m_\tau^2}{4\sqrt{2}\pi} \left( 1-\frac{4 m_\tau^2}{m_H^2} \right)^{\frac{3}{2}} \;,
\eeq
with the coupling modification factor $g_{H\tau\tau}$ in the 2HDM,
which depends on the 2HDM type. We give in Table~\ref{tab:coupfac}
the coupling factors for all neutral Higgs bosons to
$\tau$ leptons in the different realizations of the 2HDM. 
\begin{table}[t!]
\begin{center}
\begin{tabular}{c||cccc}
type & I & II & lepton-specific & flipped \\ \hline\hline
$g_{h\tau\tau}$ & $c_\alpha/s_\beta$ & $- s_\alpha/c_\beta$
& $-s_\alpha/c_\beta$ & $c_\alpha/s_\beta$\\
$g_{H\tau\tau}$ & $s_\alpha/s_\beta$ & $c_\alpha/c_\beta$
& $c_\alpha/c_\beta$ & $s_\alpha/s_\beta$ \\
$g_{A\tau\tau}$ & $-1/t_\beta$ & $t_\beta$ & $t_\beta$ & $- 1/t_\beta$
\end{tabular}
\caption{Neutral Higgs boson couplings to $\tau$ leptons in different
  realizations of the 2HDM.\label{tab:coupfac}}
\end{center}
\end{table}
For the decay $A\to \tau\tau$ the LO decay width is 
\beq
\Gamma^{\text{LO}} (A\to \tau\tau) = \frac{G_F g^2_{A\tau\tau} m_A
  m_\tau^2}{4\sqrt{2}\pi} \sqrt{1-\frac{4 m_\tau^2}{m_A^2}}  \;,
\eeq
with $g_{A\tau\tau}$ given in Table~\ref{tab:coupfac}.
These two processes can hence be used to define the
counterterms for $\alpha$ and $\beta$. \s

\begin{figure}[t!]
\begin{center}
\includegraphics[width=15cm]{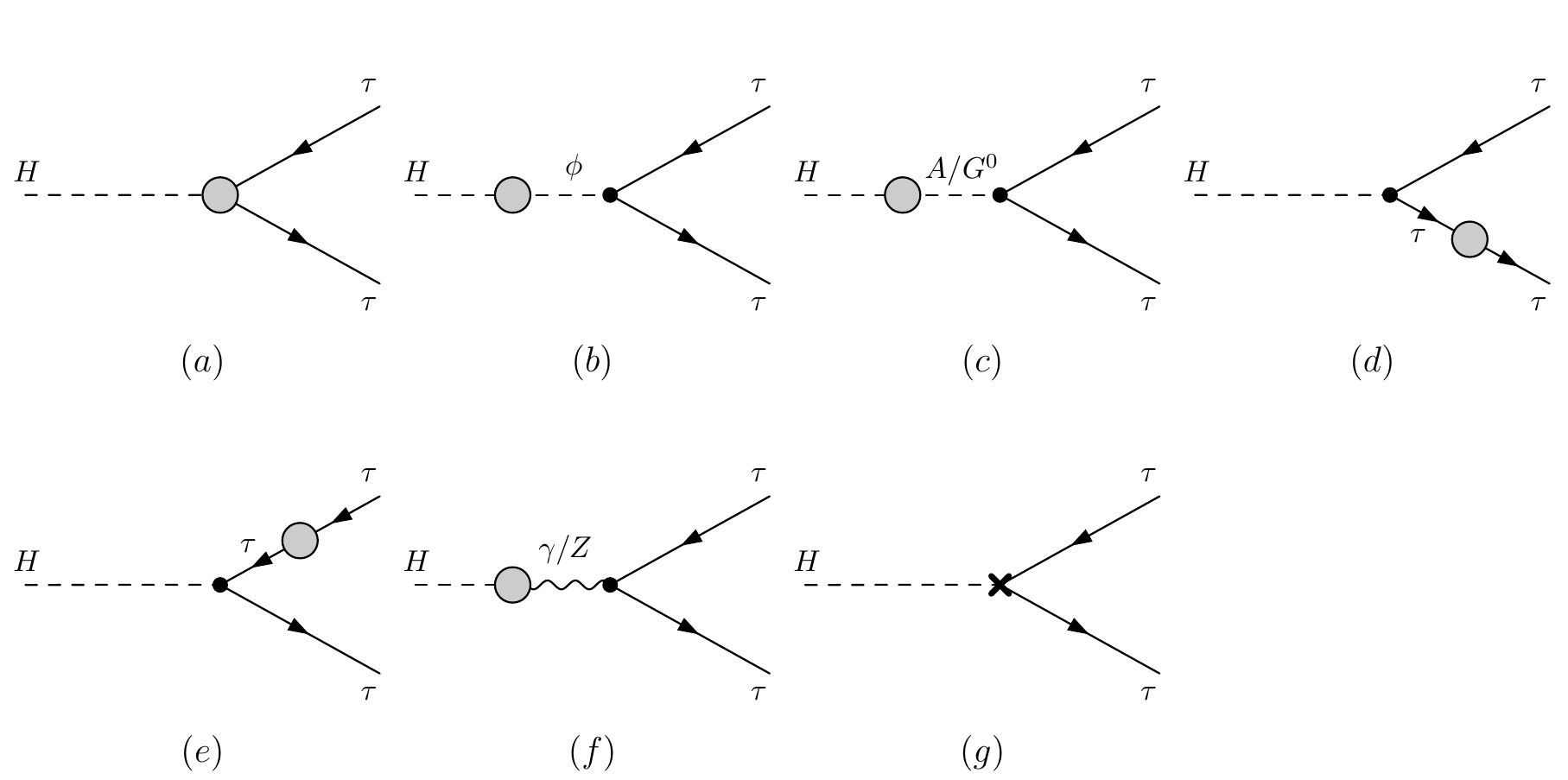}
\end{center}
\caption{Generic diagrams contributing to the virtual corrections of
  $H \to \tau \tau$: vertex corrections (a) and corrections to the
  external legs (b)-(f) where $\phi \equiv h/H$. Diagram (g) displays
  the counterterm.} 
\label{fig:htotautaugenericnlo}
\end{figure}

\begin{figure}[t!]
\begin{center}
\includegraphics[width=14cm]{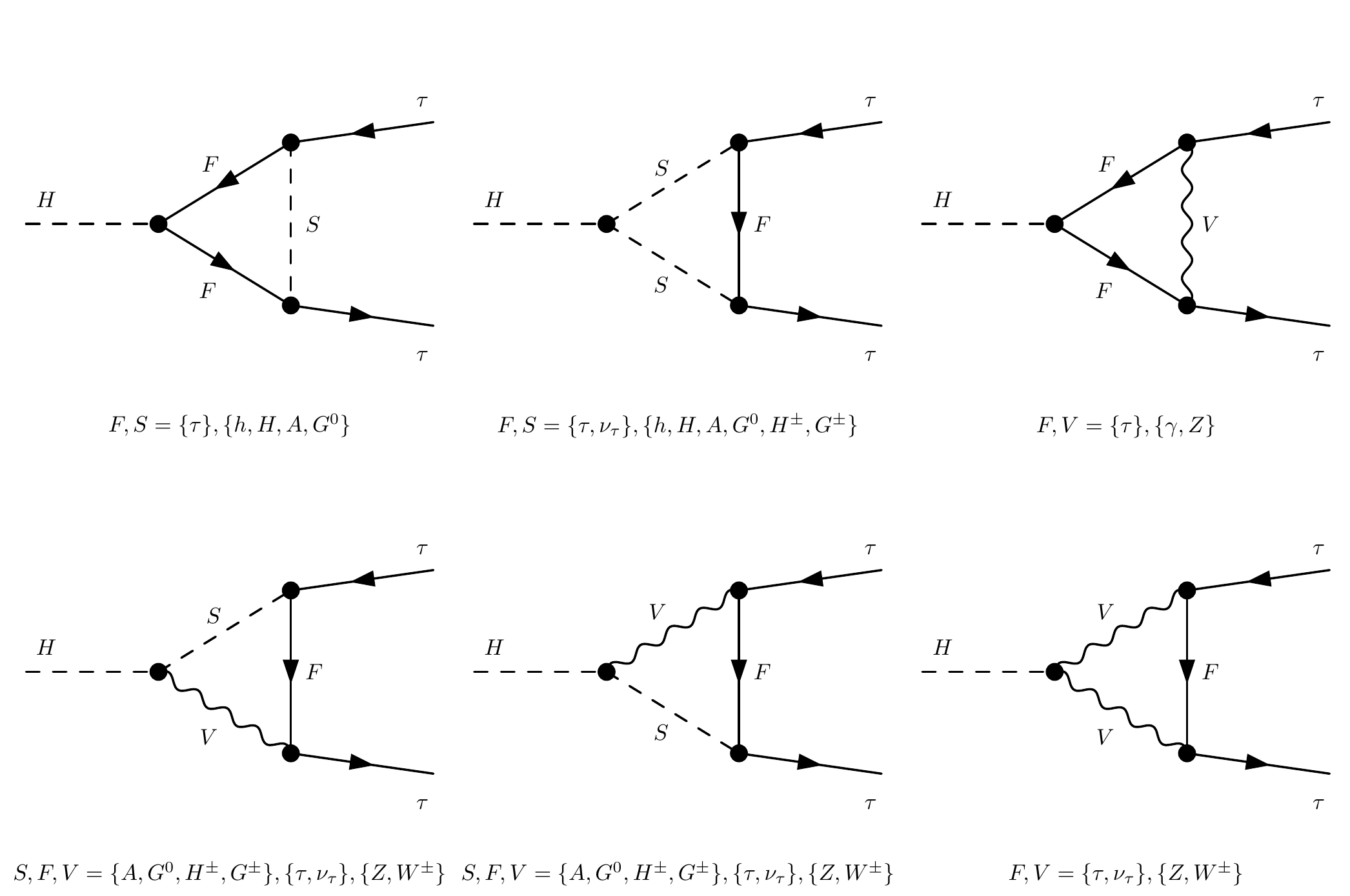}
\end{center}
\caption{Generic diagrams contributing to the vertex
  corrections in $H\to \tau\tau$.}
\label{fig:htotautauvertex}
\end{figure}

The EW NLO corrections to $H\to \tau \tau$ consist of the virtual
corrections, the counterterms and the real corrections. 
The generic contributions to the virtual corrections are depicted in
Fig.~\ref{fig:htotautaugenericnlo}. 
The 1PI diagrams of the vertex corrections are shown in
Fig.~\ref{fig:htotautauvertex} and consist of the triangle
diagrams with scalars, fermions, massive gauge bosons and photons in
the loop. 
The corrections to the external legs in
Fig.~\ref{fig:htotautaugenericnlo} (b), (d) and (e) vanish because of the OS
renormalized $H$ and $\tau$, respectively. Diagram (c) is zero because
of CP conservation. Diagram (f) finally vanishes because of a
Slavnov-Taylor identity. The real corrections consist of the diagrams
where a photon is radiated off either of the final state $\tau$ leptons. 
We explicitly checked that all NLO corrections factorize from the LO
width so that the NLO decay width can be cast into the form
\beq
\Gamma^{\text{NLO}} (H\to \tau\tau) = \Gamma^{\text{LO}} \left[ 1 +
  \Delta^{\text{virt}} + \Delta^{\text{ct}} + \Delta^{\text{real}} \right] \;.
\eeq
For $\Delta^{\text{ct}}$ we have
\beq
\Delta^{\text{ct}} &=&  \delta Z_{HH} + \frac{g_{h\tau\tau}}{g_{H\tau\tau}}
\delta Z_{hH} + \delta Z_{\tau\tau}^L + \delta Z_{\tau\tau}^R 
+ 2 \frac{\delta g}{g} + 2 \frac{\delta m_\tau}{m_\tau} - \frac{\delta
M_W^2}{M_W^2} \nonumber \\
&& + \frac{2 g_{h\tau\tau}}{g_{H\tau\tau}} \, \delta \alpha  
+ 2 g_{A\tau\tau} \, \delta \beta \;. 
\eeq
Note, that the pure QED contributions in $\Delta^{{\text{virt}}}$ and $\Delta^{\text{ct}}$
can be separated from the weak
contributions in a gauge-invariant way and form a UV-finite subset by
themselves. This is important as it allows 
to define the angular counterterm via this process through the purely
weak NLO contributions, see also the discussion in section
\ref{sec:procdepscheme}. Requiring the following renormalization
condition for the process-dependent definition of $\delta\alpha$, 
\beq
\Gamma^{\text{LO}} (H\to \tau\tau) \stackrel{!}{=} \Gamma_{\text{weak}}^{\text{NLO}}
(H\to \tau\tau) \;, \label{eq:prochcond}
\eeq
and imposing this condition only on the weak part of the decay width
we arrive at the process-dependent counterterm definition
\beq
\delta \alpha^{H\to \tau\tau} &=& 
-\frac{g_{H\tau\tau}}{2 g_{h\tau\tau}} \left[ 
\delta Z_{HH} 
+ \frac{g_{h\tau\tau}}{g_{H\tau\tau}} \delta Z_{hH} + \delta
Z_{\tau\tau}^{L,\text{weak}} + 
\delta Z_{\tau\tau}^{R,\text{weak}} + 2 \frac{\delta g}{g} 
+ 2 \frac{\delta m_\tau^{\text{weak}}}{m_\tau} - \frac{\delta M_W^2}{M_W^2} 
\right. \nonumber \\
&& \left. + \, 2 g_{A\tau\tau} \, \delta \beta + 
\Delta^{\text{virt,weak}}_{H\to \tau\tau} \right] \;. \label{eq:alphact}
\eeq
The superscript 'weak' indicates that in the respective counterterms and
in the virtual correction only the purely weak contributions are taken
into account. For example for $\Delta^{\text{virt,weak}}_{H\to
  \tau\tau}$ this means that corrections stemming from diagrams in
Fig.~\ref{fig:htotautauvertex} that involve photons are dropped. \s
 
The counterterm $\delta \tan\beta$ or $\delta \beta$, respectively, 
which is necessary in (\ref{eq:alphact}), can be defined in a
process-dependent scheme via the NLO decay $A \to \tau\tau$ as outlined in
the following. Again the NLO contributions consist of virtual,
counterterm and real diagrams. 
\begin{figure}[b!]
\begin{center}
\includegraphics[width=15cm]{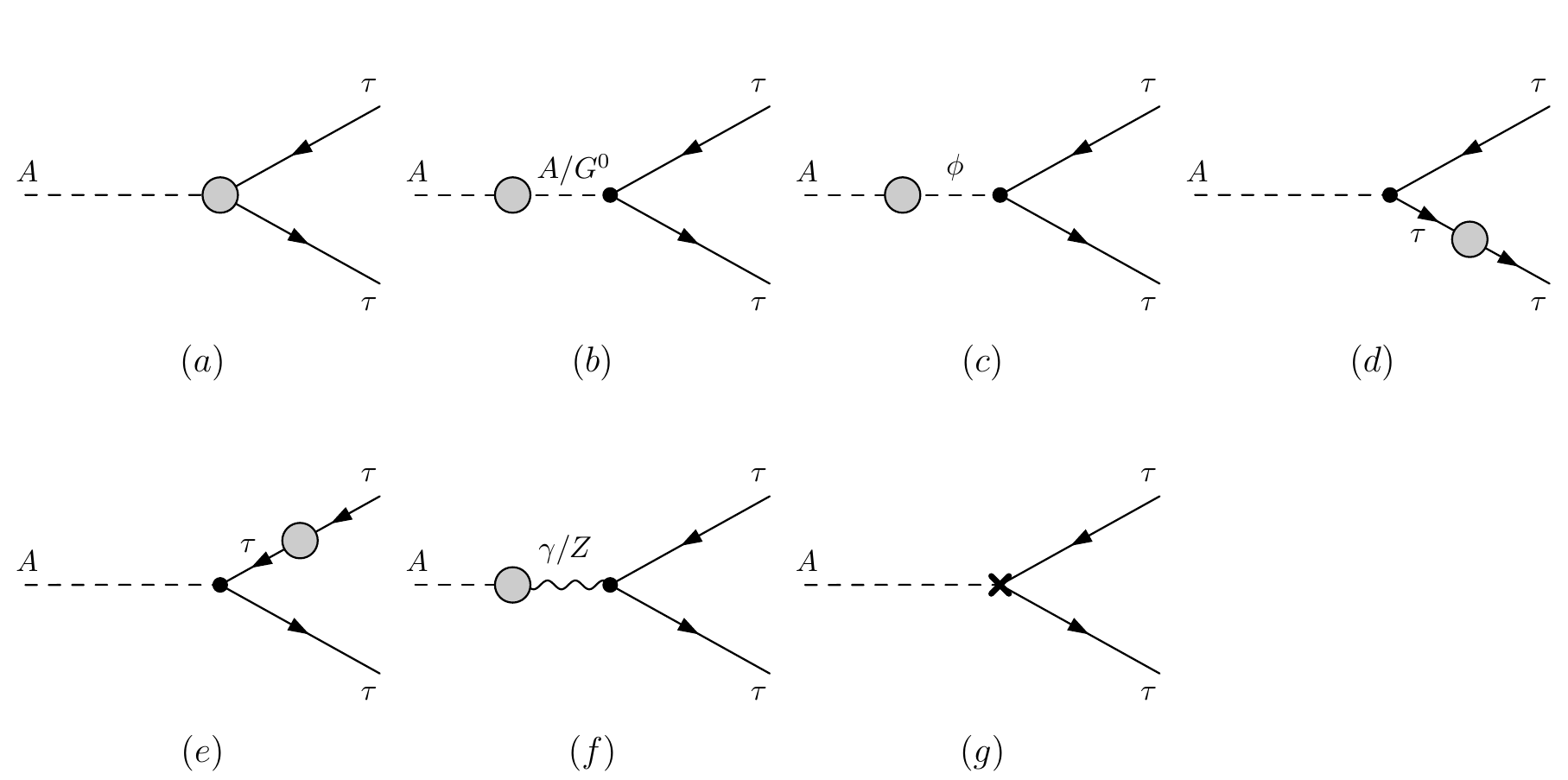}
\end{center}
\caption{Generic diagrams contributing to the virtual corrections of
  $A \to \tau \tau$: vertex corrections (a) and corrections to the
  external legs (b)-(f), where $\phi \equiv h/H$. Diagram (g) displays
  the counterterm.} 
\label{fig:atotautaugenericnlo}
\end{figure}
\begin{figure}[t!]
\begin{center}
\includegraphics[width=13cm]{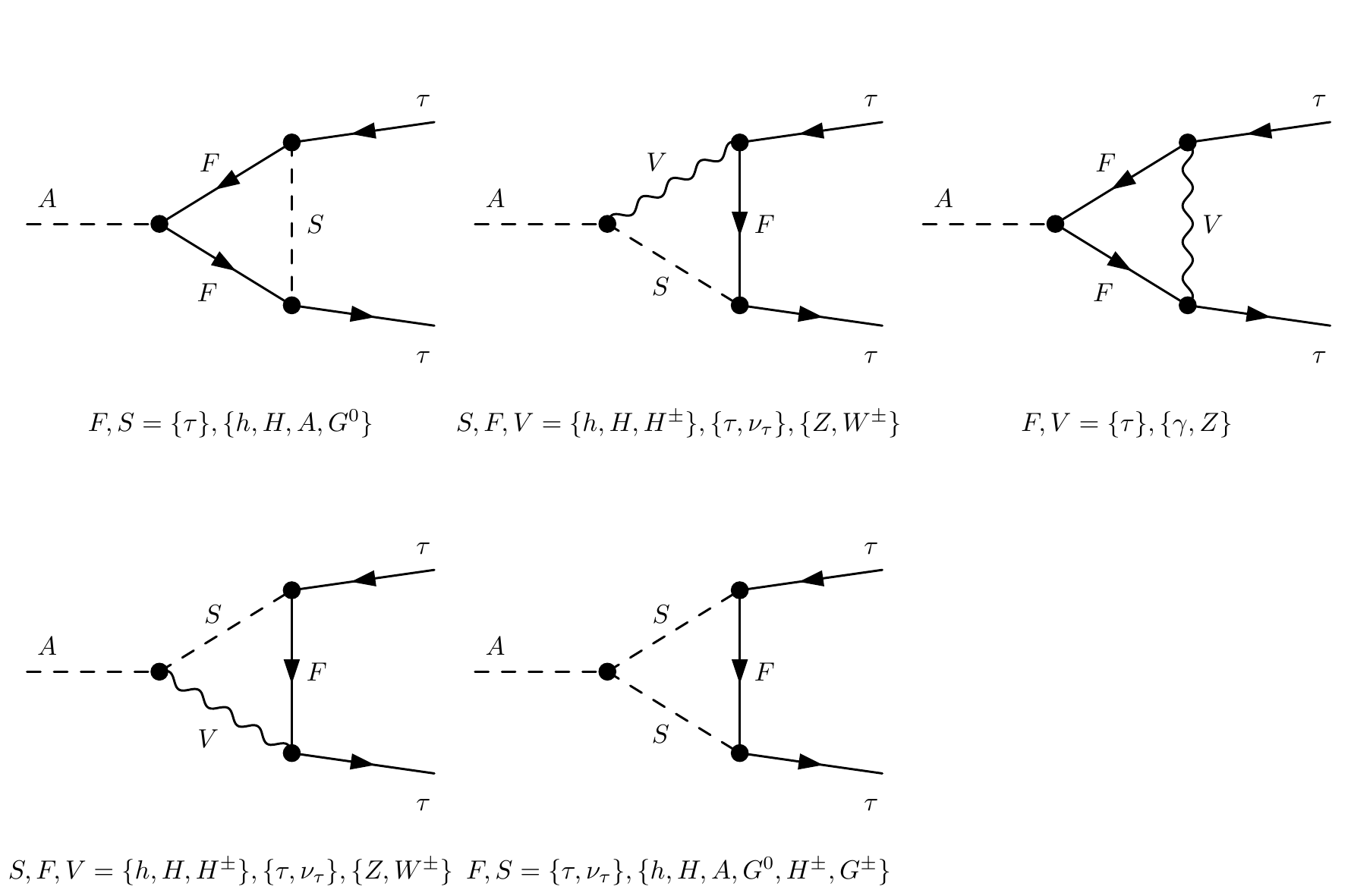}
\end{center}
\caption{Generic diagrams contributing to the vertex
  corrections in $A\to \tau\tau$.}
\label{fig:atotautauvertex}
\end{figure}
The generic ones for the former two are
shown in Fig.~\ref{fig:atotautaugenericnlo} and the 1PI diagrams of
the vertex corrections are summarized in
Fig.~\ref{fig:atotautauvertex}. The loops contain scalars, fermions,
massive gauge bosons and photons. The loops with photons induce IR divergences
that are canceled by the real corrections.
The corrections to the external legs in Fig.~\ref{fig:atotautaugenericnlo}
(b), (d) and (e) vanish due to OS renormalization conditions, those in
(c) because of CP invariance and those in (f) because of a
Slavnov-Taylor identity. Also in this process the pure QED
corrections can be separated from the remainder in a gauge-invariant way and
form a UV-finite subset so that the NLO decay width can be used for
the process-dependent definition of the counterterm $\delta \beta$ through
the requirement
\beq
\Gamma^{\text{LO}} (A\to \tau \tau) \stackrel{!}{=}
\Gamma^{\text{NLO}}_{\text{weak}} (A \to \tau \tau) \;. \label{eq:procacond}
\eeq
With the factorization
\beq
\Gamma^{\text{NLO}} (A\to \tau\tau) = \Gamma^{\text{LO}} \left[ 1 +
  \Delta^{\text{virt}} + \Delta^{\text{ct}} + \Delta^{\text{real}} \right] 
\eeq
and the counterterm
\beq
\Delta^{\text{ct}} &=&  \delta Z_{AA} - \frac{1}{g_{A\tau\tau}}
\delta Z_{G^0 A} + \delta Z_{\tau\tau}^L + \delta Z_{\tau\tau}^R 
+ 2 \frac{\delta g}{g} + 2 \frac{\delta m_\tau}{m_\tau} - \frac{\delta
M_W^2}{M_W^2} \nonumber \\
&& 
+ \frac{2(1+g^2_{A\tau\tau})}{g_{A\tau\tau}} \, \delta \beta 
\eeq
we arrive by imposing the condition (\ref{eq:procacond}) at 
\beq
\delta \beta^{A\to \tau\tau} =
&&\frac{-g_{A\tau\tau}}{2(1+g_{A\tau\tau}^2)} \left[  
\delta Z_{AA} - \frac{1}{g_{A\tau\tau}}
\delta Z_{G^0 A} + \delta Z_{\tau\tau}^{L,\text{weak}} + \delta
Z_{\tau\tau}^{R,\text{weak}}  
+ 2 \frac{\delta g}{g} + 2 \frac{\delta
      m_\tau^{\text{weak}}}{m_\tau} \right. \nonumber \\
&& \left. - \frac{\delta M_W^2}{M_W^2} +
    \Delta^{\text{virt,weak}}_{A\to \tau\tau}\right] \;. 
\eeq
Again the superscript 'weak' denotes the purely weak contributions to
the respective counterterms and to the virtual corrections. 
Thus, $\Delta^{\text{virt,weak}}_{A\to \tau\tau}$ is given by the
purely weak virtual corrections to $A \to \tau \tau$ at NLO which are
computed from the diagrams in Fig.~\ref{fig:atotautauvertex}
discarding those with photons in the loop. 

\subsection{The gauge (in)dependence of the angular counterterms
\label{sec:gaugeindep}}
\underline{\it The question of gauge dependence in the standard scheme:}
In order to investigate the question whether the angular counterterms can be
defined in a gauge-independent way, we have calculated the one-loop
corrected decay width for the charged Higgs decays in the general $R_\xi$ gauge. 
When we apply the standard scheme, the computation of the NLO amplitude ${\cal
  M}_{H^\pm \to W^\pm h}$ including all counterterms but the one for
the angles - {\it i.e.}~$\delta c_{\beta-\alpha}$ is set to zero -
yields an amplitude that depends on the gauge parameters as follows,
\beq
\left. {\cal M}_{H^\pm \to W^\pm
    h}\right|^{\text{standard}}_{\text{NLO, } \xi, \, \delta 
    c_{\beta-\alpha} = 0} &=& 
-\frac{g \Lambda _5 c_{\beta - \alpha}
  s^2_{\beta - \alpha} \, p_1 \cdot \epsilon^*(p_3)}{32\pi ^2 (m_{H}^2 -
  m_{h}^2)} \left[ 
2 M_W^2 (1-\xi_W) \alpha _W \right.  \nonumber \\ 
&& \left. + M_Z^2 (1-\xi_Z) \alpha_Z \right]  
\;, \label{eq:nlogd} 
\eeq
where we have introduced the abbreviation
($V\equiv W,Z$)
\begin{equation}
\begin{split}
\alpha _V &= \frac{1}{(1-\xi _V) m_V^2} \left[ A_0 (m_V^2) - A_0 (\xi
  _V m_V^2) \right] 			
\end{split}
\end{equation}
in terms of the scalar one-point function $A_0$
\cite{'tHooft:1978xw,Passarino:1978jh}. With $p_1$ we denote the 
incoming four-momentum of $H^\pm$ and with 
$\epsilon^* (p_3)$ the polarization vector of the outgoing $W^\pm$ boson with
four-momentum $p_3$ and 
\beq
\Lambda_5 \equiv \frac{2 m_{12}^2}{v^2 s_\beta c_\beta} \;.
\label{eq:lambda5def}
\eeq
Note that $\alpha_V$ is UV-divergent. 
This result shows explicitly what we have already stated before: In
the standard renormalization scheme, the NLO
decay amplitude without the angular counterterms has a residual
UV-divergent gauge dependence. This can only be 
canceled by the angular counterterms. Therefore, the
counterterms cannot be defined in a gauge-independent way. This 
gauge dependence is independent of the renormalization scheme chosen
for the angular counterterms. It is purely due to the treatment
  of the tadpoles. Let us investigate what happens if we apply the
  KOSY scheme, which yields the renormalization conditions 
Eq.~(\ref{eq:standdelalp}) and Eq.~(\ref{eq:beta1stand}) or
Eq.~(\ref{eq:beta2stand}), respectively. 
Introducing the UV-finite integral
\beq 
\beta _{Vj} (p^2) = \frac{1}{(1-\xi _V) m_V^2} \left[ B_0 (p^2 ;
  m_V^2 , m_j^2) - B_0 (p^2 ; \xi _V m_V^2 , m_j^2) \right]  
\eeq 
in terms of the scalar two-point function $B_0$, we find the following
gauge-dependent results for the angular counterterms,
\beq
\delta \alpha &=&\left. \delta \alpha \right| _{\xi = 1} \nonumber \\[0.1cm]
&&- \frac{\Lambda _5 c_{\beta - \alpha}
  s_{\beta - \alpha}}{32\pi ^2 (m_{H}^2 - m_{h}^2)} \left[ 2
  M_W^2 (1-\xi_W) \alpha _W + M_Z^2  (1-\xi _Z) \alpha _Z \right]
\nonumber \\ 
&&+ (1-\xi _Z) \frac{g^2 c_{ \beta - \alpha } s_{\beta - \alpha }}{256
  \pi ^2 c_W^2 } \bigg\{ 2m_{A}^2 
\Big[ \beta _{ZA} (m_{H}^2) - \beta _{ZA} (m_{h}^2) \Big] \nonumber\\ 
&&\hspace*{0.2cm} + m_{H}^2 \Big[ \beta _{Z\xi Z} (m_{H}^2) -
2\beta _{ZA} (m_{H}^2) \Big] - m_{h}^2 \Big[ \beta _{Z\xi Z}
(m_{h}^2) - 2\beta _{ZA} (m_{h}^2) \Big] \bigg\} \nonumber \\
&&+ (1-\xi _W) \frac{g^2 c_{\beta - \alpha } s _{\beta
    - \alpha }}{128 \pi ^2 } \bigg\{ 2m_{H^\pm }^2 \Big[ \beta
_{WH^\pm } (m_{H}^2) - \beta _{WH^\pm } (m_{h}^2) \Big] \nonumber
\eeq
\beq
&&\hspace*{0.2cm} + m_{H}^2 \Big[ \beta _{W\xi W} (m_{H}^2) -
2\beta _{WH^\pm } (m_{H}^2) \Big] - m_{h}^2 \Big[ \beta _{W\xi W}
(m_{h}^2) - 2\beta _{WH^\pm } (m_{h}^2) \Big] \bigg\} ,
\raisetag{5.0\baselineskip} 
\label{eq:delalpstandxi}
\eeq
and
\begin{equation}
\begin{split}
\delta \beta ^{(1)} = &\left. \delta \beta ^{(1)} \right| _{\xi = 1} \\
&+ \left( 1- \xi _W \right) \frac{g^2 c_{\beta - \alpha }
  s_{\beta - \alpha } }{128\pi ^2} \bigg\{ m_{h}^2 \Big[
\beta _{Wh} (m_{H^\pm }^2) - \beta _{W h} (0) \Big]  \\ 
&+ m_{H^+}^2 \Big[ \beta _{WH} (m_{H^\pm }^2) - \beta _{W h}
(m_{H^\pm }^2) \Big] + m_{H}^2 \Big[ \beta _{WH} (0) - \beta _{W
  H} (m_{H^\pm }^2) \Big] \bigg\} ~. 
\end{split}
\label{eq:delbetstandxi}
\end{equation}
Here the symbol $\big|_{\xi=1}$ represents the counterterm result obtained for
$\xi=\xi_W= \xi_Z=1$. The result for $\delta \beta^{(2)}$ looks
similar with the appropriate mass replacements and $\xi_W \to \xi_Z$. 
The second line in Eq.~(\ref{eq:delalpstandxi}) has the appropriate
structure to cancel the remaining UV-divergent gauge dependence in the
amplitude (\ref{eq:nlogd}). However, the additional finite terms in
(\ref{eq:delalpstandxi}) and (\ref{eq:delbetstandxi}) proportional to
the $\beta$-integrals defined above, reintroduce a gauge dependence
into the amplitude. In \cite{Kanemura:2015mxa} it was argued that the
gauge dependence of $\delta \beta$ can be moved into the unphysical
counterterm $\delta C_f$, see Eq.~(\ref{eq:ansatz}). Yet, lacking a
method to define uniquely the gauge-dependent parts in the standard
scheme, where the PT cannot be applied, it remains unclear, how this
could be accomplished. The situation is even worse for
$\delta\alpha$, where we necessarily have to  retain  the
gauge-dependent part proportional to the UV-divergent $A_0$ functions,
but must move the rest into $\delta C_f$. 
To summarize, this result shows that not only is it impossible to
arrive at a gauge-independent definition of $\delta\alpha$ in the
standard scheme, but it also explicitly demonstrates that the KOSY
scheme leads to an unphysical gauge dependence of the decay amplitude,
which one cannot be disposed of in a straightforward way. This is not only true
for the charged Higgs bosons decays we are discussing. In fact, the
investigation of the origin of this gauge dependence shows, that the
standard tadpole scheme inevitably leads to gauge-dependent decay
widths in case the KOSY scheme is applied for the mixing angles. \s

If we define the angular counterterms via a physical 
process, however, namely through the decay widths $H\to \tau \tau$
and $A \to \tau \tau$, compute the contribution of the counterterm
$\delta c_{\beta-\alpha}$, and extract the $\xi$-dependent parts we
obtain the following, 
\beq
\left. {\cal M}_{H^\pm \to W^\pm
    h}\right|^{\text{standard}}_{\text{ct, } \xi, \, \delta
    c_{\beta-\alpha} \text{ only}}  &=& 
 \frac{g \Lambda _5 c_{\beta - \alpha}
  s^2_{\beta - \alpha} \, p_1 \cdot \epsilon^* (p_3)}{32\pi ^2
  (m_{H}^2 - m_{h}^2)} \left[ 2 
  M_W^2 (1-\xi_W) \alpha_W \right. 
\nonumber \\
&& \left. + M_Z^2 (1-\xi _Z) \alpha _Z \right]
\;. \label{eq:delalbetproc}
\eeq
It is exactly the same as Eq.~(\ref{eq:nlogd}) but with opposite sign, so
that altogether the EW one-loop corrected decay width is gauge
independent and UV-finite as required. The standard treatment of the
tadpoles combined with a process-dependent definition hence leads to a
gauge-independent physical result, as it should be. The counterterms,
however, necessarily contain a gauge dependence. \s

\noindent
\underline{\it Gauge-independent angular counterterms:}
For the angular counterterms to be gauge-independent the loop-corrected
amplitude including all counterterms but the angular ones must be
independent of $\xi$. This can be achieved by treating the
tadpoles according to Ref.~\cite{Fleischer:1980ub}, {\it cf.}~the
discussion in section \ref{sec:tadpscheme}. It means that in
the coun\-ter\-terms Eq.~(\ref{eq:cthpwphl}) and 
Eq.~(\ref{eq:cthpwphh}), respectively, the self-energies
$\Sigma$ and the tadpole counterterms $\delta T$, contained in the wave
function constants, the scalar mass counterterms and the angular
counterterms, have to be replaced 
by $\Sigma^{\text{tad}}$ and $\delta T = 0$. 
Note, that the change to this tadpole scheme in principle implies new
vertices arising from constant tadpole contributions to the respective
original vertices, {\it cf.}~App.~\ref{app:tadpole}. In the  2HDM,
however, there is no quartic vertex 
between two scalars, a charged Higgs and a charged gauge boson,
$h/H-h/H-H^\pm-W^\mp$, where one of the external $h/H$ legs would
carry the additional tadpole contribution. Therefore, the process
$H^\pm \to W^\pm h/H$ does not receive additional tadpole diagrams. 
The counterterms $\delta\alpha$, $\delta \beta$, $\delta Z_{hH}$,
$\delta Z_{Hh}$ $\delta Z_{G^0 A}$ and $\delta Z_{G^\pm H^\pm}$ change however.
With these modifications the
gauge-dependent part of the amplitude with the angular counterterms set
to zero, becomes
\beq
\left.{\cal M}_{H^\pm \to W^\pm h}\right|_{\text{NLO, } \xi, \, \delta
  c_{\beta - \alpha} = 0}^{\text{tad}} = 0 \;.
\eeq
The amplitude without the mixing angle counterterm is
itself gauge-independent, so that it is possible to provide a
gauge-independent renormalization of the angular counterterms. 
\s

\noindent 
\underline{\it a) Gauge-independent tadpole-pinched scheme:}
The pinch technique allows to extract from the Green functions the
truly gauge-independent part. Combined with the tadpole scheme this leads
to manifestly gauge-independent angular counterterms. Choosing the OS
scale, 
they are given by Eqs.~(\ref{eq:posalpha})-(\ref{eq:sigaddghpm}). In the
$p_\star$ scheme the formulae simplify to
(\ref{eq:delalphstar})-(\ref{eq:delbet2star}). In the numerical
analysis we will apply both choices. 
\s

\noindent 
\underline{\it b) Gauge-independent process-dependent
  definition of the angular counterterms:} 
Another possibility to arrive at a truly gauge-independent definition
of the angular counterterms is the definition via the physical
processes $H/A\to \tau\tau$, provided of course that the framework of
the tadpole scheme is applied. \s

In the processes $H/A \to \tau \tau$ no new diagrams are introduced
when switching to the tadpole scheme, while the counterterms do change. 
In the tadpole scheme the process-dependent definition of
$\delta \alpha$ and $\delta\beta$ through the 
requirement Eq.~(\ref{eq:prochcond}) and Eq.~(\ref{eq:procacond}),
respectively, then indeed leads to gauge independence of both counterterms
and hence also of $\delta c_{\beta-\alpha}$, {\it i.e.}
\beq
(\delta c_{\beta-\alpha})^{\text{tad, proc-dep}}_{\xi} = 0 \;.
\eeq
We have seen in Eq.~(\ref{eq:delalbetproc}) that the treatment of the
tadpoles in the standard scheme cannot lead to gauge-independent
angular counterterms, although they are defined through a physical
process. In detail, this gauge parameter dependence stems from $\delta
\alpha$, whereas $\delta \beta$ is gauge-independent in the
process-dependent definition also without applying the tadpole
scheme. Thus we have
\beq
\delta \beta^{\text{proc-dep}}_{\xi} &=& \delta
\beta^{\text{tad, proc-dep}}_{\xi} = 0 \label{eq:delbetxi} \\
 \delta
\alpha^{\text{tad, proc-dep}}_{\xi} &=& 0 \label{eq:delalptadxi} \\
\delta \alpha^{\text{proc-dep}}_{\xi} &=& 
- \frac{\Lambda_5 c_{\beta-\alpha} s_{\beta-\alpha}}{32 \pi^2
  (m_H^2-m_h^2)} \left[2M_W^2(1-\xi_W) \alpha_W + M_Z^2 (1-\xi_Z)
  \alpha_Z \right]\,. \label{eq:delalpxi}
\eeq
This result shows two important things: First, the
process-dependent definition of the angular counterterms leads to
gauge-independent counterterms only if the tadpole scheme is
applied. Second, Eqs.~(\ref{eq:delbetxi})-(\ref{eq:delalpxi})
demonstrate, that in a process-dependent definition of the counterterms
the difference between the application of the tadpole and the standard
scheme is a gauge-dependent 
expression that solely depends on $A_0$ functions, which are
UV-divergent. As the 2HDM is renormalizable this implies that also in
the amplitude the difference in the application of the two
schemes must be UV-divergent and must have the same structure, since
the divergences have to cancel. In conclusion, this means: The
definition of the angular counterterms via any  
physical process leads for any NLO decay process to a gauge-independent
result, independently of the treatment
of the tadpoles.\s

In the following numerical analysis in section~\ref{sec:numerics} we will
apply all three types of renormalization schemes, the standard, the
tadpole-pinched and the process-dependent scheme, and compare them to each
other. We will do this for the sample processes $H^\pm \to W^\pm h/H$
and $H \to ZZ$. In order to describe also for this latter process the
implications of the tadpole scheme, required for a gauge-independent
definition of the angular counterterms, we briefly repeat the
ingredients of the EW one-loop corrections to $H \to ZZ$. 

\begin{figure}[b!]
\begin{center}
\includegraphics[width=13cm]{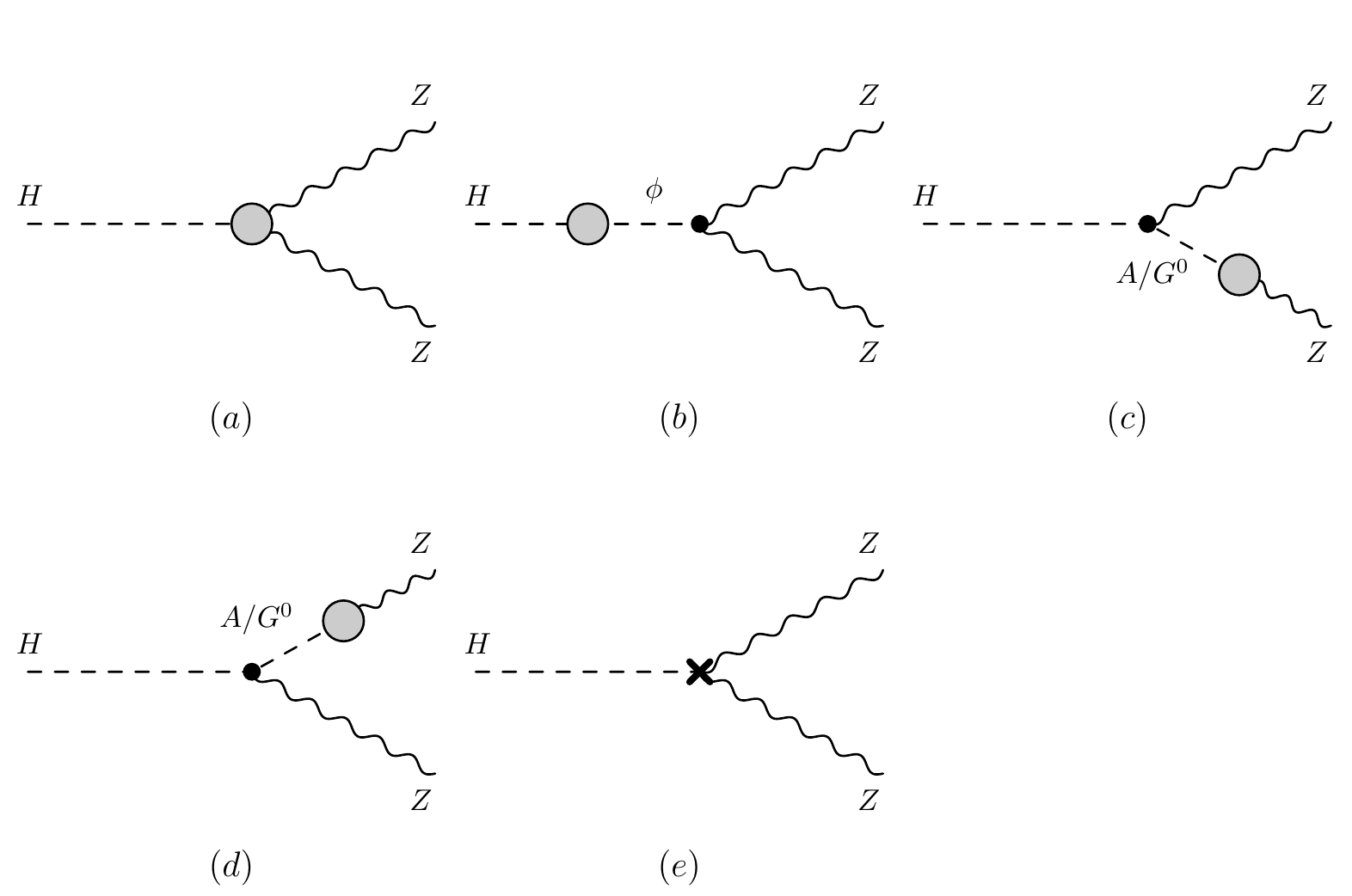}
\end{center}
\caption{Generic diagrams contributing to the virtual corrections of
 the decay $H \to ZZ$: vertex corrections (a) and corrections to the
 external legs (b)-(d) where $\phi \equiv h,H$. Diagram (e)
 displays the counterterm.}
\label{fig:genericnlohzz}
\end{figure}
\subsection{Electroweak One-Loop Corrections to $H \to ZZ$}
\begin{figure}[t!]
\begin{center}
\includegraphics[width=14cm]{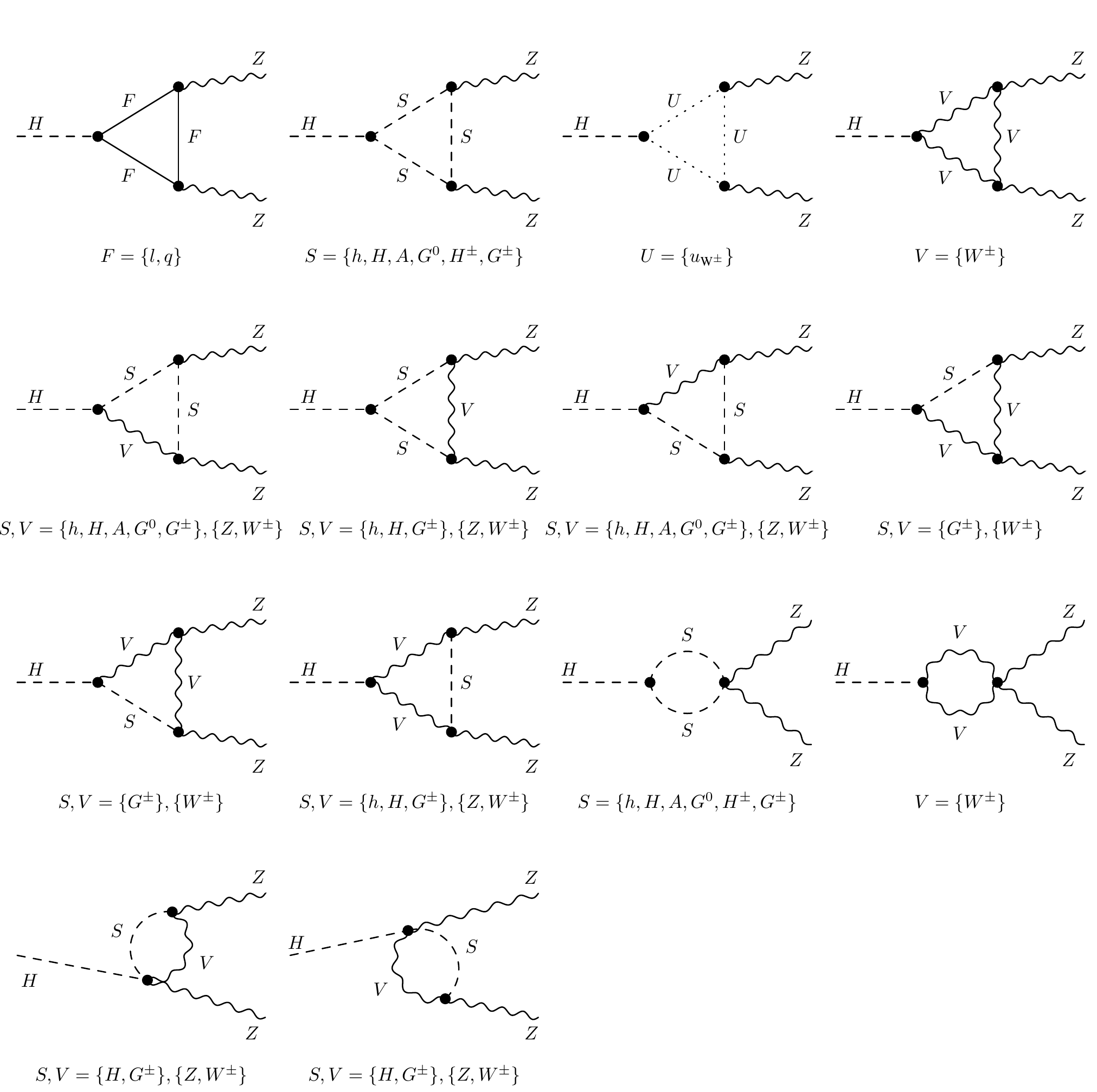}
\end{center}
\caption{Generic diagrams contributing to the vertex
  corrections in $H\to ZZ$. The ghost particles are denoted by $U$.}
\label{fig:detailsvirthtozz}
\vspace*{-0.2cm}
\end{figure}

\begin{figure}[t!]
\begin{center}
\includegraphics[width=16cm]{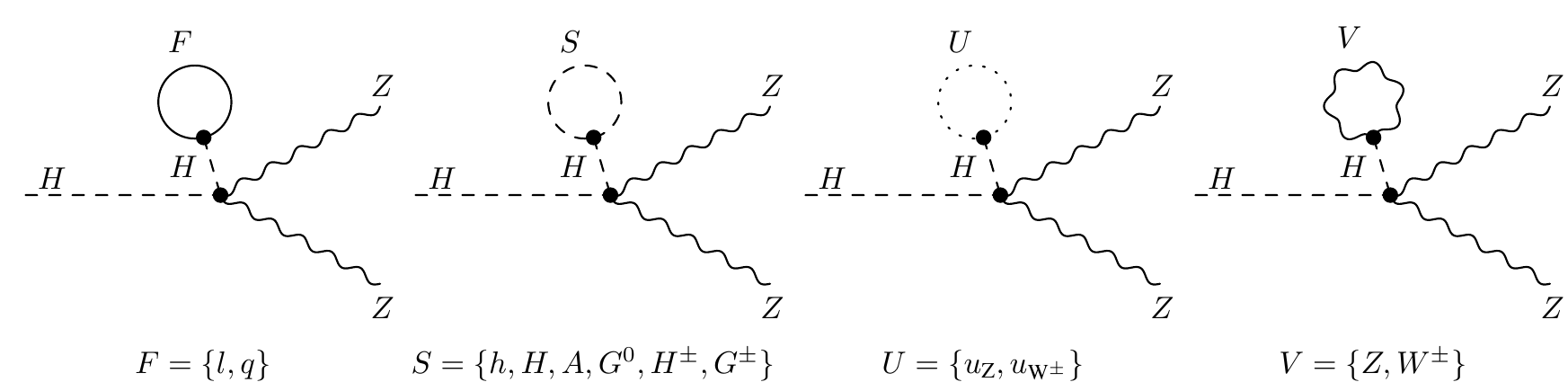}
\end{center}
\caption{Additional vertex diagrams in the tadpole scheme contributing
  to the decay $H\to ZZ$.}
\label{fig:hzztadpdiags}
\end{figure}

The LO decay width for the process
\beq
H \to ZZ
\eeq
is given by
\beq
\Gamma^{\text{LO}} (H\to ZZ) = \frac{G_F g_{HZZ}^2}{32
  \sqrt{2} \pi m_H} (m_H^4-4 m_H^2 m_Z^2 + 12 m_Z^4) \sqrt{1-\frac{4
    M_Z^2}{m_H^2}} 
\eeq
and depends on the mixing angles through the coupling factor
\beq
g_{HZZ} = c_{\beta -\alpha} \;.
\eeq
The NLO decay width consists of virtual corrections and the counterterm
contributions to cancel the UV divergences. There are neither IR
divergences nor real corrections. The generic diagrams for the virtual
corrections and the counterterm are depicted in Fig.~\ref{fig:genericnlohzz}. 
The 1PI diagrams contributing to the vertex corrections are given by the
triangle diagrams with scalars, fermions, massive gauge bosons and
ghost particles in the loops, as shown in the first three rows of
Fig.~\ref{fig:detailsvirthtozz}, 
and by the diagrams involving four-particle vertices (last
four diagrams of Fig.~\ref{fig:detailsvirthtozz}). 
The corrections to the external leg in Fig.~\ref{fig:genericnlohzz}
(b) vanish due to the OS renormalization of the $H$. The mixing
contributions (c) and (d) vanish because of the Ward identity for the
OS $Z$ boson. The counterterm amplitude is given by
\beq
{\cal M}_{H\to ZZ}^{\text{ct}} &=& \frac{e
    c_{\beta-\alpha} 
  M_W}{c_W^2 s_W} \epsilon^{*} (p_3) \cdot \epsilon^{*} (p_4) 
\nonumber \\
&\times& \left[ 
\frac{\delta g}{g} + \frac{\delta c_{\beta-\alpha}}{c_{\beta-\alpha}}
+ \frac{\delta M_Z^2}{M_Z^2} - \frac{\delta M_W^2}{2 M_W^2} +
\frac{s_{\beta-\alpha}}{c_{\beta-\alpha}} \, \frac{\delta Z_{hH}}{2} +
\frac{\delta Z_{HH}}{2} + \delta Z_{ZZ}
\right] \,,
\eeq
where the $\epsilon^{\mu *}$ denote the polarization vectors of the
outgoing $Z$ bosons with four-momentum $p_3$ and $p_4$, respectively. 
If the tadpole scheme is applied, the $HZZ$ vertex is modified by
additional tadpole contributions, which lead to further diagrams, that
have to be taken into account in the computation of the decay
width. They are shown in Fig.~\ref{fig:hzztadpdiags}. 
As the formula for the vertex corrections and counterterms in terms of
the scalar one-, two- and three-point functions are quite lengthy, we
do not display them explicitly here. 

\section{Numerical Analysis \label{sec:numerics}}
\setcounter{equation}{0}
For the computation of the NLO EW corrections to the Higgs decay
widths described in the previous section we have performed two
independent calculations. Both of them employed the {\tt Mathematica} package
{\tt FeynArts 3.9} \cite{Kublbeck:1990xc,Hahn:2000kx} to generate the
amplitudes at LO and NLO in the
general $R_\xi$ gauge. To this end, the model file for a CP-conserving
2HDM was used, which is already implemented in the
package. Additionally, all tadpole and self-energy amplitudes, needed
for the definition of the counterterms and wave function
renormalization constants, have been generated in the general $R_\xi$
gauge. The contraction of the Dirac matrices and formulation of the
results in terms of Passarino-Veltman functions has been done with
{\tt FeynCalc 8.2.0} \cite{Mertig:1990an} in one calculation and with
{\tt FormCalc} \cite{Hahn:1998yk} in the other. The dimensionally
regularized \cite{'tHooft:1972fi,Bollini:1972ui} integrals have been
evaluated numerically with the help of the {\tt C++} 
library {\tt LoopTools 2.12} \cite{Hahn:1998yk}. \s

For one of the two calculations the Python progam {\tt
  2HDMCalc} was developed that links {\tt FeynArts}, generates the
needed counterterms dynamically from the 2HDM Lagrangian by calling a
{\tt Mathematica} script and combines the LO, NLO and counterterms
calculated by {\tt FeynCalc} into the full
partial decay widths. These are then evaluated numerically by linking
{\tt LoopTools}. Finally, the LO and NLO partial decay widths are
written out for all renormalization schemes of the mixing angles
introduced above. 
The outcome of this program was compared to the
results of the second independent computation. All results agree within
numerical errors. \s

In the following we specify the input parameters that we used for the
numerical evaluation. As explained in section~\ref{sec:renorm} we use
the fine structure constant $\alpha$ at the $Z$ boson mass scale,
given by \cite{Agashe:2014kda} 
\beq
\alpha (M_Z^2) = \frac{1}{128.962} \;.
\eeq
The massive gauge boson masses are set to \cite{Agashe:2014kda,Denner:2047636}
\beq
M_W = 80.385 \mbox{ GeV} \qquad \mbox{and} \qquad M_Z = 91.1876 \mbox{
GeV} \;.
\eeq
For the lepton masses we choose \cite{Agashe:2014kda,Denner:2047636}
\beq
m_e = 0.510998928 \mbox{ MeV} \;, \quad
m_\mu = 105.6583715 \mbox{ MeV} \;, \quad
m_\tau = 1.77682 \mbox{ GeV} \;.
\eeq
These and the light quark masses, which we set \cite{LHCHXSWG}
\beq
m_u = 100 \mbox{ MeV} \;, \quad m_d = 100 \mbox{ MeV} \;, \quad
m_s = 100 \mbox{ MeV} \;,
\eeq
have only a small influence on our results. 
In order to be consistent with the ATLAS and CMS analyses, we follow
the recommendation of the LHC Higgs Cross Section Working Group
(HXSWG) \cite{Denner:2047636,Dittmaier:2011ti} and use the following OS value for the
top quark mass
\beq
m_t = 172.5 \mbox{ GeV} \;. 
\eeq
The charm and bottom quark OS masses are set to
\beq
m_c = 1.51 \mbox{ GeV} \qquad \mbox{and} \qquad
m_b = 4.92 \mbox{ GeV} \;,
\eeq
as recommended by \cite{Denner:2047636}.
Omitting CP violation we consider the CKM matrix to be real, with the
CKM matrix elements given by \cite{Agashe:2014kda}
\beq
V_{\text{CKM}} = \left( \begin{array}{ccc} V_{ud} & V_{us} & V_{ub} \\
V_{cd} & V_{cs} & V_{cb} \\ V_{td} & V_{ts} & V_{tb} \end{array}
\right) = \left( \begin{array}{ccc} 0.97427 & 0.22536 & 0.00355 \\
    -0.22522 & 0.97343 & 0.0414 \\ 0.00886 & -0.0405 &
    0.99914 \end{array} \right) \;.
\eeq
The SM-like Higgs mass value, denoted by $m_{H_\text{SM}}$, has been
set to \cite{Aad:2015zhl}
\beq
m_{H^\text{SM}} = 125.09 \mbox{ GeV} \;. 
\eeq
Note, that in the 2HDM, depending on the chosen parameter set, it is
possible that either the lighter or the heavier of the two CP-even
neutral Higgs bosons can be the SM-like Higgs boson.  \s

The IR divergences in the computation of the NLO corrections to the
process $H^\pm \to W^\pm H/h$ require the inclusion of the real
corrections to regularize the decay width. This introduces a
dependence on the detector sensitivity $\Delta E$ for the resolution of
the soft photons from the real corrections. We showed that this
dependence is small \cite{masterlorenz}. For our analysis we fixed the value to
\beq
\Delta E = 10 \mbox{ GeV} \;.
\eeq

In the subsequently presented plots we only used 2HDM parameter sets
that are not yet excluded by experiment and that fulfill certain
theoretical constraints. These data sets have been generated with the
tool {\tt ScannerS} \cite{Coimbra:2013qq}.\footnote{We thank Marco
  Sampaio, one of the authors of {\tt ScannerS}, who kindly provided
  us with the necessary data sets.} The applied theoretical
constraints require that the chosen CP-conserving vacuum is the global minimum
\cite{Barroso:2013awa}, that the 2HDM potential is bounded from below
\cite{Deshpande:1977rw} and that tree-level unitarity holds
\cite{Kanemura:1993hm,Akeroyd:2000wc}. For consistency with
experimental data the following conditions have been imposed. The
electroweak precision constraints
\cite{Peskin:1991sw,PhysRevD.45.2471,Grimus:2008nb,Haber:2010bw,ALEPH:2010aa,Baak:2011ze,Baak:2012kk} 
have to be satisfied, {\it i.e.}~the $S,T,U$ variables
\cite{Peskin:1991sw} predicted by 
the model are within the 95\% ellipsoid centered on the best fit point
to the EW data. Indirect experimental constraints are due to loop
processes involving charged Higgs bosons, that depend on $\tan\beta$
via the charged Higgs coupling to the fermions. They are mainly due to
$B$ physics observables
\cite{Mahmoudi:2009zx,Deschamps:2009rh,Hermann:2012fc} and the
measurement of $R_b$
\cite{Denner:1991ie,Grant:1994ak,Haber:1999zh,Freitas:2012sy}. We have
included the most recent bound of $m_{H^\pm} \gsim 480$~GeV for the 
type II and flipped 2HDM \cite{Misiak:2015xwa}. The results
from LEP \cite{Abbiendi:2013hk}
and the recent ones from the LHC
\cite{Aad:2014kga,Khachatryan:2015qxa}\footnote{The results reported
  in the recent ATLAS paper \cite{Aad:2015typ}  have not been translated into
  bounds so far.}
constrain the charged Higgs mass to
be above ${\cal O} (100 \mbox{ GeV})$ depending on the model type. In
order to check the compatibility with the LHC Higgs data {\tt ScannerS}
is interfaced with {\tt SusHi} \cite{Harlander:2012pb} which computes 
the Higgs production cross sections through gluon fusion and
$b$-quark fusion at NNLO QCD. All other 
production cross sections are taken at NLO QCD \cite{LHCHXSWG}. The
2HDM decays were obtained from {\tt HDECAY}
\cite{Djouadi:1997yw,Harlander:2013qxa}.  Note that in the computation
of these processes all EW corrections were consistently neglected, as
they are not available for the 2HDM. The exclusion limits were checked
by using {\tt HiggsBounds} \cite{Bechtle:2008jh,Bechtle:2011sb,Bechtle:2013wla} 
and the compatibility with the observed signal for the 125~GeV Higgs
boson was tested with {\tt HiggsSignals} \cite{Bechtle:2013xfa}. For
further details we refer to \cite{Ferreira:2014dya}. \s

In our numerical analysis we investigate the applicability
of the various proposed renormalization schemes. The goal is to find 
a renormalization scheme for the 2HDM, that is process independent, gauge
independent and numerically stable. All results that we show are for
the 2HDM type II.

\subsection{Gauge dependence of the KOSY scheme}
\begin{figure}[t!]
\centering
\includegraphics[width=0.7\linewidth, clip]{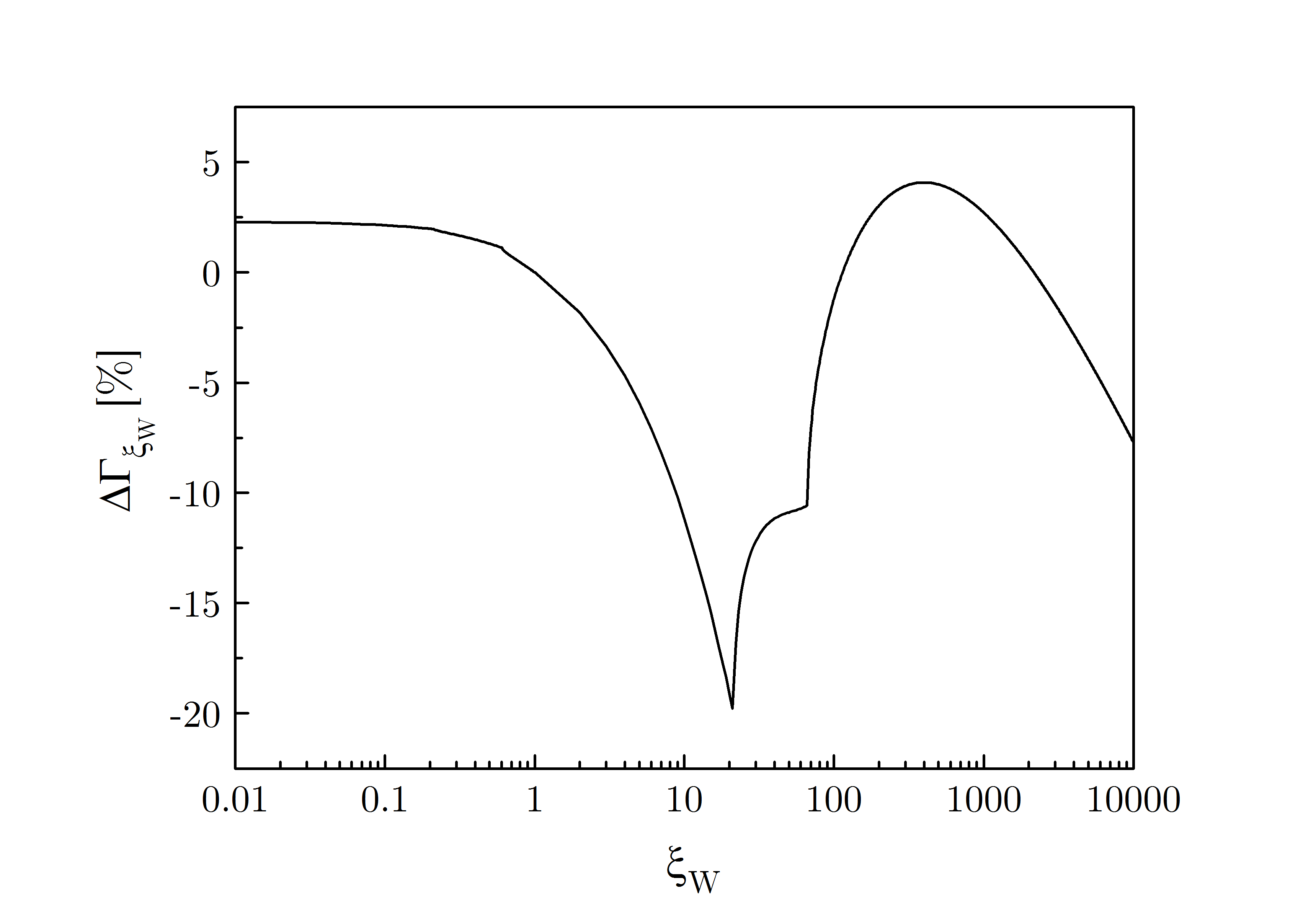}
    \caption{Gauge dependence of the decay $H^\pm \to W^\pm h$ within
      the KOSY$^c$ scheme. The 2HDM parameters are given by {\it Scen1}
      defined in Eq.~(\ref{eq:scen1}).}
\label{fig:xidep}
\end{figure}

We start by analyzing the gauge dependence of the partial decay width, 
introduced through the renormalization of the mixing angles $\alpha$
and $\beta$ in the KOSY scheme. As an example we choose the charged
Higgs boson decay into the $W$ boson and the light CP-even
scalar $h$ corresponding to $H^{\text{SM}}$, $H^\pm \to W^\pm h$. For
the renormalization of $\beta$ we use the charged sector and call the
renormalization scheme accordingly KOSY$^c$. The corresponding
angular counterterm $\delta \beta^{(1)}$ is defined in
Eqs.~(\ref{eq:beta1stand}), while $\delta \alpha$ is given by  
Eq.~(\ref{eq:standdelalp}).
The size of the gauge dependence will be quantified by 
\beq
\Delta \Gamma_{\xi} \equiv \frac{\Gamma^{\text{NLO}}_\xi -
  \Gamma^{\text{NLO}}_{\xi=1}}{\Gamma^{\text{NLO}}_{\xi=1}} \;.
\eeq
It parametrizes the deviation of the NLO partial decay width for an
arbitrarily chosen gauge parameter $\xi$ in the $R_\xi$ gauge from the
reference decay width chosen to be the NLO width in the Feynman gauge,
normalized to the reference value. For simplicity we only vary the
gauge parameter $\xi_W$ and set $\xi_Z=1$.  
The 2HDM scenario {\it Scen1} that we investigate is defined by the input
parameters
\beq
\begin{array}{llll}
\hspace*{-0.2cm} \mbox{\it \underline{Scen1:}} & \; m_{H^\pm} =
780 \mbox{ GeV}\,, & \; m_H = 742.84 \mbox{ GeV}\,, & \;
m_A = 700.13 \mbox{ GeV}\,, \\ 
\hspace*{-0.2cm} & \; \tan\beta = 1.46 \,, & \; \alpha = -0.57 \,, & \;
m_{12}^2 = 2.076\cdot 10^5 \mbox{ GeV}^2 \,.
\label{eq:scen1}
\end{array}
\eeq
Figure \ref{fig:xidep} shows the $\xi_W$ dependence of our process,
$\Delta_{\xi_W}^{H^\pm W^\pm h}$, as a function of $\xi_W$. The kinks in
the figure are due to threshold effects in the $B_0$ functions entering
the counterterms. In detail, the kinks are given by the
following parameter configurations and counterterms
\begin{center}
\begin{tabular}{ccccc}  
\toprule
Kink & $\xi _W$ & Kinematic point & Origin \\
\midrule
1 & ~ 0.2137 ~ & ~ $m_{H^\pm} \approx m_H + \sqrt{\xi _W} m_W$ ~ & $\delta \beta^{(1)}$ \\ 
2 & ~ 0.60539 ~ & ~ $m_{h} \approx \sqrt{\xi _W} m_W + \sqrt{\xi _W}
m_W$ ~ & $\delta \alpha$ \\
3 & ~ 21.3491 ~ & ~ $m_{H} \approx \sqrt{\xi _W} m_W + \sqrt{\xi _W}
m_W$ ~ & $\delta \alpha$ \\
4 & ~ 66.3763 ~ & ~ $m_{H^\pm} \approx m_h + \sqrt{\xi _W} m_W$ ~ & $\delta \beta^{(1)}$ \\
\bottomrule
\end{tabular}
\end{center} 

\vspace*{0.2cm}
With a relative variation of the NLO width of up to 20\% due to the
change of the gauge parameter, the figure clearly
demonstrates the gauge dependence of the NLO decay width in the KOSY
scheme. The explicit calculation shows that for large values of
$\xi_W$ the partial decay width drops as 
$ - (m_H^2 - m_h^2) \ln(\xi _W)$. This explicit gauge dependence makes a
practical use of the KOSY scheme impossible as it leads to non-physical
gauge dependences in the decay widths.

\subsection{The processes $\Gamma (H^\pm \to  W^\pm h/H)$ at NLO}
\begin{figure}[t!]
\centering
\includegraphics[width=0.48\linewidth, clip]{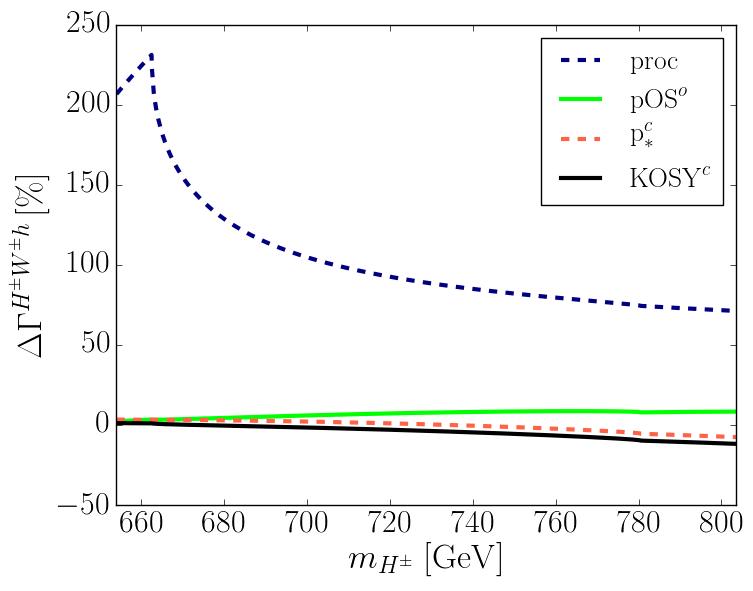}
\hspace*{0.2cm}
\includegraphics[width=0.48\linewidth, clip]{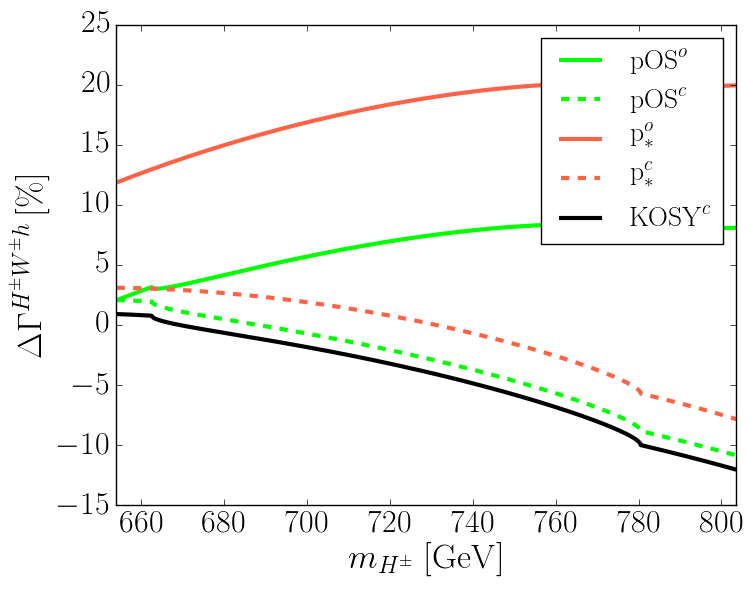}
  \caption{Relative NLO corrections to $H^\pm \to W^\pm h$ for
      various renormalization schemes as defined in
      Eq.~(\ref{eq:schemenotation}), with the
      2HDM parameters given by {\it Scen2}, Eq.~(\ref{eq:scen2}); left:
      with, right: without the process-dependent renormalization.}
\label{fig:delhpmwpmhl}
\end{figure}

We move on to the investigation of the size of the NLO corrections to the
processes $H^\pm \to W^\pm h/H$ and their dependence on the
renormalization scheme. In our scenarios $h$ corresponds to the
SM-like Higgs bosons. We define the quantity
\beq
\Delta \Gamma \equiv
\frac{\Gamma^{\text{NLO}}-\Gamma^{\text{LO}}}{\Gamma^{\text{LO}}} \;,
\label{eq:deltagamdef}
\eeq
which measures the relative size of the NLO corrections compared to the LO
decay width. For the discussion of the $H^\pm \to W^\pm h$ decay we
chose among the generated valid scenarios again the one given by
{\it Scen1}, but this time vary the charged Higgs boson mass. For
distinction, we call it {\it Scen2} and it is given by
\beq
\begin{array}{llll}
\hspace*{-0.2cm} \mbox{\it \underline{Scen2:}} & \; m_{H^\pm} =
(654...804) \mbox{ GeV}\,, & \; m_H = 742.84 \mbox{ GeV}\,, & \;
m_A = 700.13 \mbox{ GeV}\,, \\ 
\hspace*{-0.2cm} & \; \tan\beta = 1.46 \,, & \; \alpha = -0.57 \,, & \;
m_{12}^2 = 2.076\cdot 10^5 \mbox{ GeV}^2 \,.
\label{eq:scen2}
\end{array}
\eeq
For $H^\pm \to W^\pm H$ we chose {\it Scen3} where the mass $m_A$ is varied,
\beq
\begin{array}{llll}
\hspace*{-0.2cm} \mbox{\it \underline{Scen3:}} & \; m_{H^\pm} = 745.54
\mbox{ GeV}\,, & \; m_H = 594.55 \mbox{ GeV}\,, & \; m_A = (704...735)
\mbox{ GeV}\,, \\ 
\hspace*{-0.2cm} & \; \tan\beta = 1.944 \,, & \; \alpha =-0.458 \,, & \;
m_{12}^2 = 1.941\cdot 10^5 \mbox{ GeV}^2 \,.
\label{eq:scen3}
\end{array}
\eeq
In Fig.~\ref{fig:delhpmwpmhl} we show the relative NLO corrections for
$H^\pm \to W^\pm h$, $\Delta \Gamma^{H^\pm W^\pm h}$, as a function of the
charged Higgs boson mass for various renormalization schemes. We
denote them as
\beq
\begin{array}{lll}
\mbox{proc} &:& \quad \mbox{process-dependent}
\\
p_\star^{c,o} &:& \quad p_\star \mbox{ tadpole-pinched},\; \delta
\beta^{(1)} \mbox{ ('$c$') or } \delta \beta^{(2)} \mbox{ ('$o$')} \\
\mbox{pOS}^{c,o} &:& \quad \mbox{on-shell tadpole-pinched},\;
\delta\beta^{(1)} \mbox{ or } \delta \beta^{(2)} \\
\mbox{KOSY}^{c,o} &:& \quad \mbox{gauge-dependent scheme},\;
\delta\beta^{(1)}  \mbox{ or } \delta \beta^{(2)} \;.
\end{array}
\label{eq:schemenotation}
\eeq
The process-dependent renormalization refers to the renormalization of
$\alpha$ via the process $H \to \tau \tau$ and of $\beta$ via $A \to
\tau \tau$. The process-dependent renormalization can be performed by
applying either the standard or the alternative tadpole scheme. The
investigation of the decay widths shows, however, that all decays
discussed in this analysis, {\it i.e.}~$H^\pm \to W^\pm h/H$ and $H\to
ZZ$, are invariant with respect to a change of the tadpole scheme.\footnote{
For details on the cancellation of the contributions when changing
from the standard to the alternative tadpole scheme between the various building
blocks of the NLO decay widths, we refer the reader to \cite{MKrause2016}.
}
In the process-independent schemes we can choose to renormalize
$\beta$ either through the charged sector, with the counterterm given
by $\delta \beta^{(1)}$, or through the CP-odd sector, with the
counterterm given by $\delta \beta^{(2)}$. 
For the shown $m_{H^\pm}$ range the LO decay
width varies from $\Gamma^{\text{LO}}= 0.0750$~GeV at 
$m_{H^\pm} = 654$~GeV to $\Gamma^{\text{LO}}= 0.1474$~GeV at
$m_{H^\pm} = 804$~GeV. \s

In Fig.~\ref{fig:delhpmwpmhl} (left) we show results for the
process-dependent renormalization and for some 
representatives of the process-independent schemes, the pOS$^o$, the
$p_\star^c$ and for comparison also the KOSY$^c$ scheme.
As can be inferred from the left plot, the process-dependent
renormalization leads to much larger NLO corrections than the other
schemes. The NLO corrections can increase the LO width by more than a
factor of three. For the process-independent renormalization schemes on the
other hand, the NLO corrections are much milder and vary between about
$-11$ to $20$\% depending on the renormalization scheme and the charged
Higgs mass value (and discarding the unphysical KOSY scheme). 
This can be inferred from Fig.~\ref{fig:delhpmwpmhl} (right) which
displays the results for the process-independent 
schemes, where the $\beta$ renormalization is performed both through the charged
and through the CP-odd sector.\footnote{In all plots we show the
  gauge-dependent results of 
  the KOSY scheme, however, only for $\beta$ renormalized via $\delta
  \beta^{(1)}$ in order to keep a clear presentation of the plots.}
Provided that the same choice for the $\beta$ renormalization is made, the OS
tadpole-pinched scheme, pOS, leads to results closer to the KOSY
scheme than the $p_\star$ tadpole-pinched scheme. 
This is due to the fact that the KOSY and the pOS scheme use the scale
of the OS masses for the evaluation of the self-energies.
Also note that the schemes which rely on the CP-odd sector for the
renormalization of $\beta$, show a slightly weaker dependence on the
mass of the charged Higgs boson, as the latter enters the counterterm $\delta
\beta^{(2)}$ only through a few diagrams (namely the tadpole
contributions). An important conclusion, which can be drawn from the plots, is that 
the process-dependent renormalization scheme is less advisable due to the
induced unnaturally large NLO corrections compared to the results in
the other renormalization schemes. \s

Discarding the numerically unstable process-dependent scheme and the
unphysical KOSY scheme, 
we can use the comparison of the results for $p_\star^c$ and
$p_\star^o$ and the comparison of those for pOS$^c$ and pOS$^o$ to estimate the
remaining theoretical uncertainty due to missing higher order
corrections, based on a change of the renormalization scheme for
$\beta$. In the same way we can estimate the uncertainty based on a
variation of the renormalization scale by comparing the results for
pOS$^o$ and $p_\star^o$ or the results for pOS$^c$ and p$_\star^c$. 
In the investigated $m_{H^\pm}$ range from the lower to the upper end, 
the remaining uncertainty varies between 1\% and 11\%, when 
estimated from the scale change, and from close to 0  
to 18\%,  when estimated from the change of the
$\beta$ renormalization scheme.  
Note also that the results in the tadpole-pinched scheme, when evaluated at the OS
scale, are less affected by a change of the renormalization scheme for
$\delta \beta$ than in the $p_\star$ scheme. The 
renormalization of $\beta$ through the charged sector 
is less sensitive to the scale choice than $\delta \beta^{(2)}$, which
uses the CP-odd sector, as can be inferred by comparing $p_\star^c$
with pOS$^c$ on the one hand, and $p_\star^o$ and pOS$^o$ on the other
hand. Taking these as indicators for theoretical uncertainties, one might
draw the conclusion that the pOS$^c$ scheme would be the best choice
here. 
Finally, we note that the kinks, which are independent of the renormalization
scheme, are due to the thresholds in the
following counterterms and parameter configurations
\begin{center}
\begin{tabular}{ccccc}  
\toprule
Kink & Kinematic point & Origin \\
\midrule
1 & ~ $m_{H^\pm} (662.46 \mbox{ GeV}) = M_{H} (742.84 \mbox{ GeV}) - M_W$~ & $\delta Z_{H^\pm H^\mp}, \delta Z_{G^\pm H^\mp}$ \\[0.1cm]
2 & ~ $m_{H^\pm} (780.51 \mbox{ GeV}) = M_A (700.13 \mbox{ GeV}) + M_W$~ &
$\delta Z_{H^\pm H^\mp}, \delta Z_{G^\pm H^\mp}$ \\ 
\bottomrule
\end{tabular}
\end{center} 

\begin{figure}[t!]
\centering
\includegraphics[width=0.49\linewidth , clip]{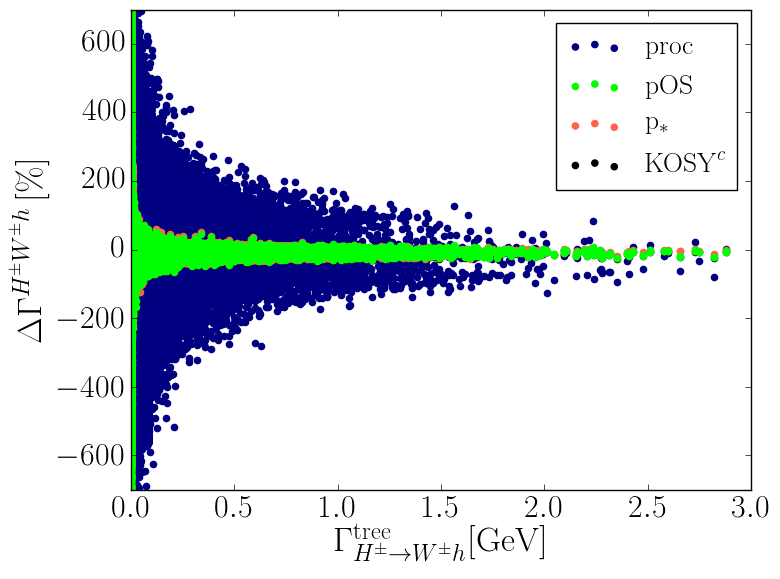}
\hspace*{0.2cm}
\includegraphics[width=0.48\linewidth , clip]{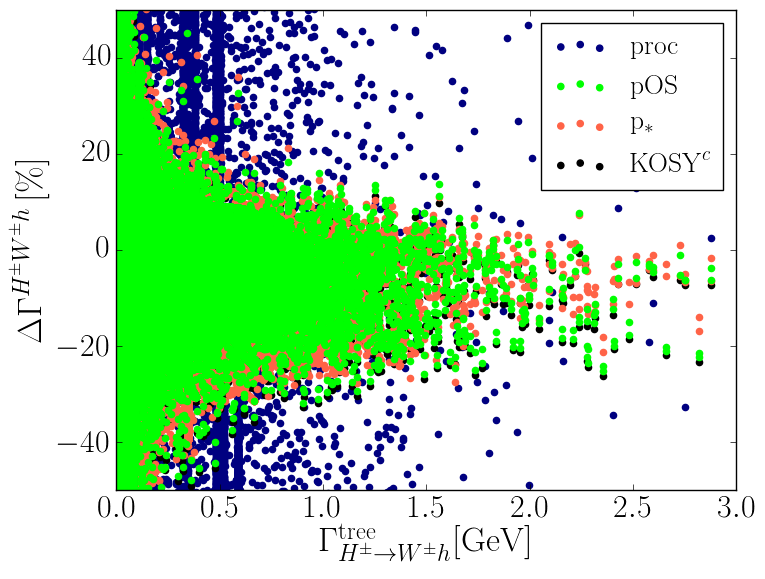}
    \caption{Scatter plots for the relative NLO corrections to $H^\pm
      \to W^\pm h$ for all parameter points passing the theoretical and
      experimental constraints as a function of the LO width; shown
      for various renormalization 
      schemes: process-dependent (blue), pOS tadpole-pinched (green),
      $p_\star$ tadpole-pinched (red), KOSY$^c$ (black). The right plot
    zooms into the central region.}
\label{fig:scatterhpmwpmhl}
\end{figure} 
\begin{figure}[b!]
\centering
\includegraphics[width=0.48\linewidth, clip]{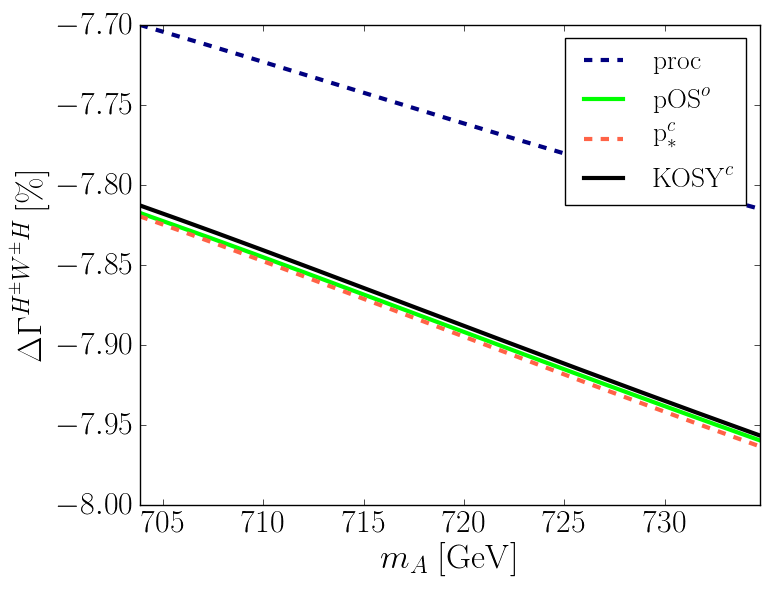}
\hspace*{0.2cm}
\includegraphics[width=0.48\linewidth,clip]{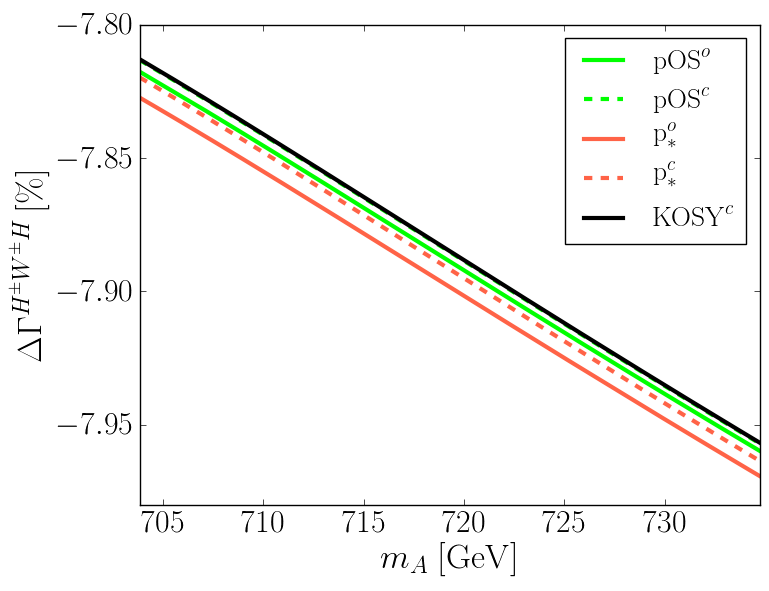} 
\caption{Relative NLO corrections to $H^\pm \to W^\pm H$ for
      various renormalization schemes, with the
      2HDM parameters given by {\it Scen3}, Eq.~(\ref{eq:scen3}); left:
      with, right: without the process-dependent renormalization. In
      the right plot the lines for KOSY$^c $ and pOS$^c$ lie on top of
    each other.}
\label{fig:nlohpmwpmhh}
\end{figure} 
\vspace*{0.2cm}
In Fig.~\ref{fig:scatterhpmwpmhl} we show the relative NLO corrections
for $H^\pm \to W^\pm h$ as a function of the LO width for all generated
scenarios compatible with the applied theoretical and experimental
constraints. The colours indicate the results for the
process-dependent scheme, the $p_\star$
tadpole-pinched schemes, the OS 
tadpole-pinched schemes and the KOSY$^c$ scheme. The plots demonstrate
that the 
process-dependent renormalization in general leads to relative NLO corrections
that are one to two orders of magnitude above those obtained in the
other schemes, which yield corrections of typically\footnote{We
  discard the region for very small LO widths, where the relative NLO
  corrections of course become very large, {\it cf.}~the definition of
  $\Delta \Gamma$, Eq.~(\ref{eq:deltagamdef}).} a few percent up 
to 40\%, as can be inferred from the right plot. \s

In Fig.~\ref{fig:nlohpmwpmhh} we show the relative NLO corrections for the
process $H^\pm \to W^\pm H$ with the parameters given by {\it Scen3},
Eq.~(\ref{eq:scen3}). In the plotted $m_{A}$ range the LO decay width,
which does not depend on $m_A$, is given by $\Gamma^{\text{LO}}= 4.0568$~GeV.
In the left plot we have included
the results for the process-dependent renormalization, for pOS$^o$,
$p_\star^c$ and KOSY$^c$. The right plot 
includes all renormalization schemes but the process-dependent
one. The relative corrections lie between about $-7.70$ to $-7.97$\% in
the investigated mass range.\footnote{The small $m_A$ mass range is
  due to the fact that all other parameter points for this scenario
  are excluded.} Altogether the results for all schemes lie very close to
each other, with the process-dependent scheme deviating the most from
the remaining schemes, although the difference in $\Delta \Gamma$ is
of maximally 0.16\% only. This behaviour can be understood
by looking at the counterterm for the NLO process, Eq.~(\ref{eq:cthpwphh}). The
contributions from the angular counterterms $\delta\alpha$ and
$\delta\beta$ come with the factor $1/t_{\beta-\alpha}$, which is numerically
very small in the SM-like limit $h \equiv H^{\text{SM}}$. Therefore
any difference in the renormalization schemes for the angles will
barely manifest itself in the total NLO corrections. 
 The zoomed in region in Fig.~\ref{fig:nlohpmwpmhh}  (right) again
 shows 
 that the KOSY scheme is closer to pOS than to the other schemes and
that the usage of the OS scale in $\delta\beta$ is less sensitive to a change of
 the renormalization scheme, while the  
renormalization of $\beta$ via the charged sector is less sensitive to
a scale change than the one through the CP-odd sector. 

\subsection{The process $\Gamma (H  \to  ZZ)$ at NLO}
We now turn to the discussion of the NLO corrections to the heavy Higgs
boson decay into a pair of $Z$ bosons, $H\to ZZ$. The scenario we have chosen is
given by
\beq
\begin{array}{llll}
\hspace*{-0.2cm} \mbox{\it \underline{Scen4:}} & \; m_{H^\pm} =
659.16 \mbox{ GeV}\,, & \; m_H = (690...809) \mbox{ GeV}\,, & \;
m_A = 705.44 \mbox{ GeV}\,, 
\\ 
\hspace*{-0.2cm} & \; \tan\beta = 1.24 \,, & \; \alpha = -0.61 \,, & \;
m_{12}^2 = 2.045\cdot 10^5 \mbox{ GeV}^2 \,.
\label{eq:scen4}
\end{array}
\eeq
\begin{figure}[b!]
\centering
\includegraphics[width=0.49\linewidth, clip]{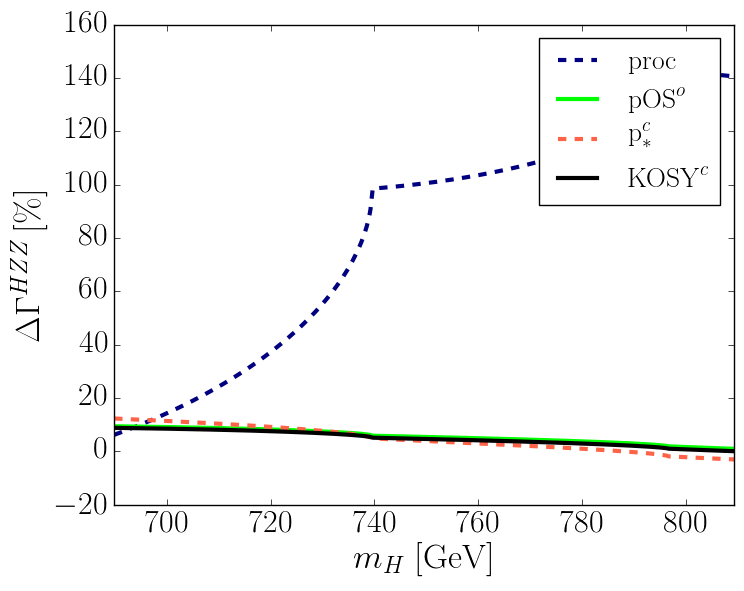}
\hspace*{0.2cm}
\includegraphics[width=0.48\linewidth, clip]{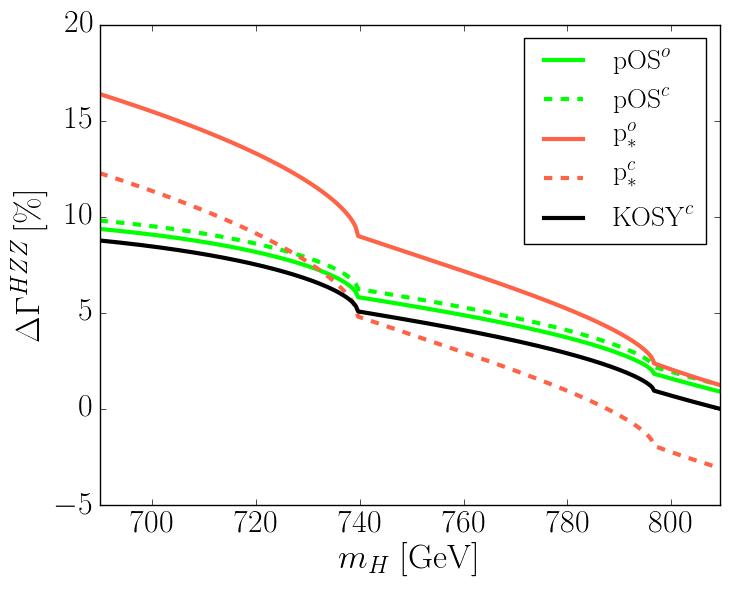}
    \caption{Relative NLO corrections to $H \to ZZ$ for
      various renormalization schemes, with the
      2HDM parameters given by {\it Scen4}, Eq.~(\ref{eq:scen4}); left:
      with, right: without the process-dependent renormalization.}
\label{fig:htozzscen4}
\end{figure} 
In Fig.~\ref{fig:htozzscen4} we show the relative NLO corrections
$\Delta \Gamma^{H\to ZZ}$ for the decay $H \to ZZ$ as a function of the
heavier CP-even Higgs mass $m_H$ for different renormalization
schemes. The LO width ranges from 0.2314 GeV to 0.3845 GeV in the plotted
$m_H$ range. The kinks are due to
\begin{center}
\begin{tabular}{ccccc}  
\toprule
Kink & Kinematic point & Origin \\
\midrule
1 & ~ $m_{H} (739.55 \mbox{ GeV}) = m_{H^\pm} (659.16 \mbox{ GeV}) + M_W$ ~
& $\delta Z_{HH}, \delta Z_{hH}$ 
\\[0.1cm]
2 & ~ $m_{H} (796.63 \mbox{ GeV}) = m_{A} (705.44 \mbox{ GeV})+ M_Z$ ~ &
$\delta Z_{HH}, \delta Z_{hH}$ \\
\bottomrule
\end{tabular}
\end{center} 

\vspace*{0.2cm}
In the left plot the process-dependent renormalization is
included. Additionally we show representatives for 
process-independent schemes, the pOS$^o$, the $p_\star^c$ and the
KOSY$^c$ scheme. Again the counterterm definition via tauonic heavy
Higgs decays leads to much larger corrections than the other
schemes. In the investigated mass range it can increase the LO decay
width by more than a factor of two. The observed coincidence of the
results for the process-independent and process-dependent
renormalization schemes at $m_H=690$~GeV is accidental. The
relative corrections in the process-dependent renormalization start to
increase quickly again for different $m_H$ values. The NLO 
increase in the process-independent schemes, on the 
other hand, ranges from about -3 to 17\% in the investigated parameter
range. The right plot shows the same behaviour we have seen
previously. The results in the KOSY and in the pOS scheme are closer
to each other than to the $p_\star$ scheme. Furthermore, the change of
the $\beta$ renormalization scheme affects the pOS scheme less than
the $p_\star$ scheme and the $\beta$ renormalization through the
charged sector is less sensitive to a change in the renormalization
scale than the one through the CP-odd sector. 
Overall, in the investigated mass range, the theoretical uncertainty
due to missing higher order corrections can be estimated to be of less
than a percent to around 6\% based on a scale change, and it ranges from
the permille level to about 4\% when estimated from the change of the
$\beta$ renormalization scheme, discarding the
numerically unstable process-dependent scheme. \s

\begin{figure}[t!]
\centering
\includegraphics[width=0.498\linewidth, clip]{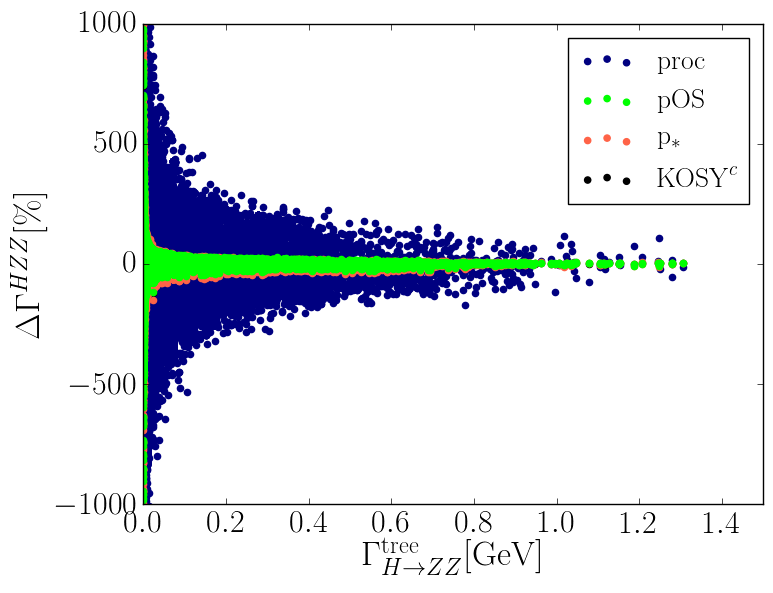}
\hspace*{0cm}
\includegraphics[width=0.48\linewidth, clip]{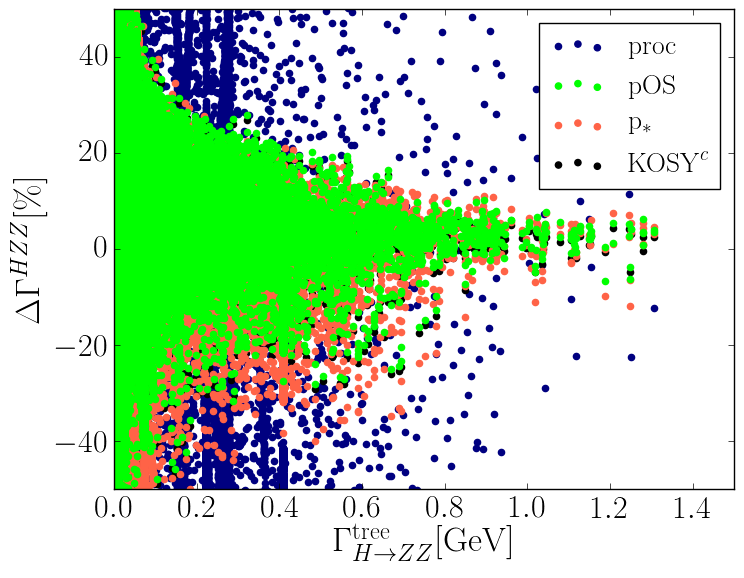}
\caption{Scatter plots for the relative NLO corrections to $H
      \to ZZ$ for all parameter points passing the theoretical and
      experimental constraints as a function of the LO width; shown
      for various renormalization 
      schemes: process-dependent (blue), pOS tadpole-pinched (green),
      $p_\star$ tadpole-pinched (red), KOSY (black). The right plot
    zooms into the central region.}
\label{fig:htozzscatter}
\end{figure} 
Figure~\ref{fig:htozzscatter} shows the relative NLO corrections $\Delta \Gamma^{H
  \to ZZ}$ for $H \to ZZ$ as a function of the LO width for all generated
scenarios compatible with the applied theoretical and experimental
constraints. The colours indicate the results for the various
renormalization schemes. The plots clearly demonstrate the
numerical instability of the process-dependent renormalization, which
exceeds the relative corrections in the other schemes by one to two
orders of magnitudes. For the process-independent schemes the relative
corrections are typically of the order of a few percent to 40\%,
discarding the region with small LO widths. \s

Altogether we conclude, that the choice of the KOSY scheme for the
renormalization of the angular counterterms is precluded due to its
manifest gauge dependence. The choice of the process-dependent scheme
is not advisable, as it leads to very large relative NLO corrections\footnote{
This statement of course only holds for scenarios where the
contributions from the angular counterterms are not parametrically
suppressed, in which case the NLO corrections obviously hardly depend
on the angular renormalization scheme.}. The process-independent
tadpole-pinched schemes lead to results that are manifestly
gauge-independent and numerically stable. Among these 
schemes the OS tadpole-pinched scheme turns out to be
more stable when changing the $\beta$ renormalization scheme than the
$p_\star$ scheme for our investigated scenarios. 

\section{Conclusions and Outlook\label{sec:concl}}
We have investigated the renormalization of the 2HDM with special
focus on the mixing angles $\alpha$ and $\beta$ which diagonalize the
Higgs mass matrices. These angles are highly relevant for the phenomenology
of the Higgs bosons as they enter the Higgs boson couplings and
therefore all Higgs observables. We have shown that if the tadpoles
are treated in the more usual approach, which we called 'standard tadpole', a
process-independent definition of 
the angular counterterms leads to gauge-dependent decay
amplitudes and thus to gauge-dependent physical observables. Therefore,
the counterterms $\delta 
\alpha$ and $\delta \beta$ either have to be defined through
a physical process, or the treatment of the tadpoles has to be
changed. Following the 'alternative tadpole' scheme as proposed in
\cite{Fleischer:1980ub} allows for a manifestly gauge-independent
definition of the masses and in particular of the mixing angles. \s

In this work we presented several distinct renormalization schemes and
investigated their implications by applying them to the NLO
EW corrections in the decays $H^\pm \to W^\pm h$, $H^\pm \to W^\pm H$
and $H \to ZZ$. It was explicitly shown that the scheme presented in
\cite{Kanemura:2004mg} leads to gauge-dependent decay widths. This
scheme applies the standard tadpole scheme and relates the angular
counterterms to the off-diagonal wave function renormalization
constants. By using the alternative tadpole scheme together with the
modified Higgs self-energies obtained from the application of the pinch technique
we introduced the 'tadpole-pinched' scheme as a manifestly
gauge-independent scheme for the angular counterterms. We
furthermore investigated the process-dependent definition of $\delta
\alpha$ and $\delta \beta$ through the decays $H \to \tau\tau$ and
$A\to \tau\tau$, respectively. In this scheme the angular counterterms
are gauge dependent when the standard tadpole scheme is applied, they
are gauge independent in case the alternative tadpole scheme is used. 
For the investigated decay processes and scenarios, the
process-dependent scheme turned out to lead to 
unnaturally large relative NLO corrections. Based on the
investigated parameter sets and decay widths this leads us
to the conclusion to propose the tadpole-pinched scheme as the
renormalization scheme for the mixing angles that is at the same time
process independent, gauge independent and numerically
stable. \s

In order to complete the renormalization of the 2HDM, also the
renormalization of the soft-breaking parameter $m_{12}^2$ has to be
investigated. This parameter appears in the couplings of the Higgs
self-interactions and hence impacts the Higgs-to-Higgs decay
widths. The renormalization of $m_{12}^2$ and the phenomenological
investigation of the implications of the higher order corrections for
Higgs phenomenology will be the subject of a follow-up paper. 

\subsubsection*{Acknowledgments} 
The authors acknowledge financial support from the DAAD
project ``PPP Portugal 2015'' (ID: 57128671).  
Hanna Ziesche acknowledges financial support from the Graduiertenkolleg ``GRK
1694: Elementarteilchenphysik bei h\"ochster Energie und h\"ochster
Pr\"azision''. 
We want to thank Marco Sampaio for kindly providing us with 2HDM data
sets. We are grateful to Fawzi Boudjema, Thi Nhung Dao, Ayres Freitas,
David Lopez-Val, Dominik St\"ockinger and Georg Weiglein for helpful
discussions. We want to thank Michael Spira and Augusto Barroso for useful
comments.   

\section*{Appendix}
\setcounter{equation}{0}
\begin{appendix}
\section{The Tadpole Scheme in the
  2HDM  \label{app:tadpole}}
In this section we will explain in detail the tadpole scheme, by
applying it to the 2HDM, and show
how to derive the relations for the mass counterterms and the wave
function renormalization constants. We will furthermore
derive which additional vertices have to be considered when performing
explicit calculations in this scheme. At the end of this appendix, in A.2, we
will give the complete list of rules for the application of the
tadpole scheme. 

\subsection{Derivation of the Tadpole Scheme \label{app:tadpole1}}
We start by setting the notation and by presenting the standard
scheme before we move on to the derivation of the tadpole scheme in
the 2HDM. \s

\begin{figure}[tb]
\centering
  	\includegraphics[width=0.9\linewidth , trim = 2mm 0mm 25mm 9mm, clip]{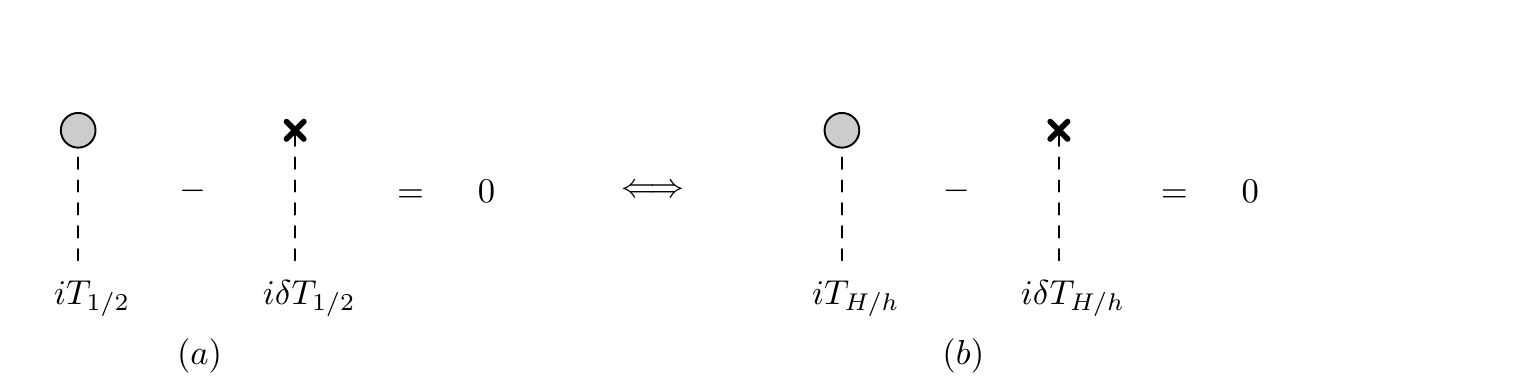}
    \caption{Renormalization condition for the tadpoles: (a) in the
      gauge basis, (b) in the mass basis.}
\label{fig:tadcond}
\end{figure}

\noindent
{\subsubsection{Setting of the notation and tadpole renormalization}} 
The expansion of the two Higgs doublets $\Phi_1$ and $\Phi_2$ about
the VEVs, {\it cf.}~Eq.~(\ref{eq:vevexpansion}), leads to the mass matrices that are
obtained from the terms bilinear in the Higgs fields in the 2HDM
potential. Due to CP- and 
charge conservation they decompose into $2\times 2$ matrices for the
neutral CP-even, neutral CP-odd and charged Higgs sector,
respectively. As we have seen in Sec.~\ref{sec:model} the minimum
conditions of the potential require the tree-level tadpole parameters $T_1$ and
$T_2$ to vanish. At lowest order they are given by Eqs.~(\ref{eq:tad1}) and
(\ref{eq:tad2}). These tadpole conditions can be exploited to eliminate $m_{11}$
and $m_{22}$. Higher order corrections, however, lead to non-vanishing
tadpole contributions that have to be taken into account. Applying 
Eqs.~(\ref{eq:tad1}) and (\ref{eq:tad2}) we arrive at the following
mass matrices
\begin{align}
	M_\rho ^2 &= \begin{pmatrix} m_{12}^2 \frac{v_2}{v_1} +
          \lambda _1 v_1^2 && -m_{12}^2 + \lambda _{345}  v_1 v_2\\
          -m_{12}^2 + \lambda _{345} v_1 v_2 && m_{12}^2
          \frac{v_1}{v_2} + \lambda _2 v_2^2 \end{pmatrix}
        + \begin{pmatrix} \frac{T_1}{v_1} && 0 \\ 0 &&
          \frac{T_2}{v_2} \end{pmatrix} 
\label{eq:rhomatrix} \\
	M_\eta ^2 &= \left( \frac{m_{12}^2}{v_1v_2} - \lambda _5
        \right) \begin{pmatrix} v_2^2 && -v_1v_2 \\ -v_1v_2 &&
          v_1^2 \end{pmatrix} + \begin{pmatrix} \frac{T_1}{v_1} && 0
          \\ 0 && \frac{T_2}{v_2} \end{pmatrix} 
\label{eq:etamatrix} \\
	M_{\phi^\pm} ^2 &= \left( \frac{m_{12}^2}{v_1v_2} - \frac{\lambda
            _4 + \lambda _5}{2} \right) \begin{pmatrix} v_2^2 &&
          -v_1v_2 \\ -v_1v_2 && v_1^2 \end{pmatrix} + \begin{pmatrix}
          \frac{T_1}{v_1} && 0 \\ 0 && \frac{T_2}{v_2} \end{pmatrix}
        ~. 
\label{eq:charmatrix}
\end{align}
Here we have explicitly kept the tadpole parameters although they
vanish at tree level. This helps us to keep track of their
non-vanishing contributions at higher orders when performing the
renormalization program. The mass matrices are diagonalized by the
rotation matrices $R$ rotating the scalar fields from the gauge basis into the 
mass basis, {\it
  cf.}~Eqs.~(\ref{eq:diagHh})-(\ref{eq:diagGHpm}), 
\beq 
D_\rho^2 &=& R(\alpha)^T M_\rho^2 R(\alpha) \label{eq:drho} \\
D_\eta^2 &=& R(\beta)^T M_\eta^2 R(\beta) \label{eq:deta} \\
D_{\phi^\pm}^2 &=& R(\beta)^T M_{\phi^\pm}^2 R(\beta) \label{eq:dphi} \;.
\eeq
The scalar mass eigenstates with same quantum
numbers, grouped into the doublets $(H,h)$, $(G^0,A)$ and $(G^\pm,
H^\pm)$, mix at higher orders. The wave function renormalization constants,
introduced in Eqs.~(\ref{eq:renconst1})-(\ref{eq:renconst3}) for the
three doublets, also develop non-vanishing mixing contributions and
form $2\times 2$ matrices with off-diagonal elements. In the following we
will use a generic notation and denote with $\phi_1$ and $\phi_2$ the
two scalars of the same doublet. With this notation we then have for
Eqs.~(\ref{eq:renconst1})-(\ref{eq:renconst3})
\beq
\left( \begin{array}{c} \phi_1 \\ \phi_2 \end{array} \right) \to
\sqrt{Z_\phi} \left( \begin{array}{c} \phi_1 \\
    \phi_2 \end{array} \right)
\approx \left( \mathbbm{1}_{2\times 2} + \frac{\delta Z_\phi}{2} \right)
\left( \begin{array}{c} \phi_1 \\ \phi_2 \end{array} \right) \;,
\eeq
with
\beq
\frac{\delta Z_\phi}{2} \equiv \left( \begin{array}{cc} \frac{\delta
      Z_{\phi_1 \phi_1}}{2} &  \frac{\delta
      Z_{\phi_1 \phi_2}}{2} \\ \frac{\delta
      Z_{\phi_2 \phi_1}}{2} & \frac{\delta
      Z_{\phi_2 \phi_2}}{2} \end{array} \right) \;.
\eeq
For the diagonal mass matrices, denoted from now on generically by
$D_\phi^2$, we introduce the counterterm matrix $\delta D_\phi^2$,
which is a symmetric $2\times 2$ matrix whose specific form will be
determined below. With these definitions the renormalized self-energy
$\hat \Sigma_\phi$ becomes
\begin{equation}
\begin{split}
	\hat{\Sigma } _\phi (p^2) &\equiv  \begin{pmatrix}
          \hat{\Sigma } _{\phi _1 \phi _1} (p^2) &&
          \hat{\Sigma } _{\phi _1 \phi _2} (p^2) \\
          \hat{\Sigma } _{\phi _2 \phi _1} (p^2) &&
          \hat{\Sigma } _{\phi _2 \phi _2} (p^2) \end{pmatrix} \\ 
	&= \Sigma _\phi (p^2) - \delta D^2_\phi + \frac{\delta Z
          _\phi^\dagger }{2} \left( p^2 \mathbbm{1} _{2\times 2} -
          D_\phi^2 \right) + \left( p^2 \mathbbm{1} _{2\times 2} - D_\phi ^2
        \right) \frac{\delta Z _\phi }{2} ~~ , 
\end{split}
\label{eq:renselfenergy}
\end{equation}
The self-energy $\Sigma_{\phi}$ is a symmetric $2\times 2$ matrix
containing the 1PI self-energies of the scalar doublet $(\phi_1,
\phi_2)$. We require OS renormalization conditions for the scalar
Higgs fields yielding the following conditions for the counterterm
$\delta D_\phi^2$ and the wave function renormalization constants
$\delta Z_\phi$, ($i=1,2$)
\beq
\mbox{Re} \left[ \delta D_{\phi_i \phi_i}^2 \right] &=& \mbox{Re} \left[ 
\Sigma_{\phi_i \phi_i} (m_{\phi_i}^2 )\right] 
\label{eq:renconda}
\\
\delta Z_{\phi_i \phi_i} &=& - \mbox{Re} \left[ \frac{\partial
    Z_{\phi_i \phi_i (p^2)}}{\partial p^2} \right]_{p^2 =
  m_{\phi_i}^2} 
\label{eq:rencondb} \\
\delta Z_{\phi_i \phi_j} &=& \frac{2}{m_{\phi_i}^2-m_{\phi_j}^2}
\mbox{Re} \left[ \Sigma_{\phi_i \phi_j} (m_{\phi_j}^2) - \delta
  D^2_{\phi_i \phi_j} \right] \;, \qquad i\ne j \;.
\label{eq:rencondc}
\eeq 
So far we have not specified $\delta D_\phi^2$. Its exact form depends
on the treatment of the tadpoles in the renormalization procedure and
will be elaborated below. In order to guarantee the correct
minimization conditions for the Higgs potential also at one-loop order,
the tadpoles are renormalized as
\beq
\hat{T}_i = T_i - \delta T_i = 0 \;, \qquad i = 1,2 \;, 
\eeq
where $T_1$ and $T_2$ are the sum of all one-loop tadpole
contributions to the fields $\rho_1$ and $\rho_2$, respectively, in
the gauge basis. Applying the renormalization conditions we have for
the tadpole counterterms the conditions
\beq
\delta T_i = T_i \;, \qquad i=1,2 \;. \label{eq:tadpolerenorm}
\eeq
In the mass basis we have 
\beq
\left( \begin{array}{c} \delta T_1 \\ \delta T_2 \end{array}
\right) = R(\alpha) \left( \begin{array}{c} \delta T_H \\ \delta T_h \end{array}
\right) = \left( \begin{array}{c} c_\alpha \delta T_H - s_\alpha T_h
    \\ s_\alpha \delta T_H + c_\alpha T_h \end{array}
\right) \;, \label{eq:reldelt12deltHh}
\eeq
and 
\beq
\delta T_H = T_H \qquad \mbox{and} \qquad \delta T_h = T_h \;. 
\label{eq:tadpolerenormh}
\eeq
The renormalization conditions for the tadpoles are shown pictorially in
Fig.~\ref{fig:tadcond}. \s 

\noindent
{\subsubsection{Mass counterterms and wave function renormalization constants in the standard scheme}} 
Regarding the renormalization of the masses, the
bare mass of each particle in the 2HDM is split into a physical mass
and a counterterm as specified in section \ref{sec:renorm}. The VEVs
$v_1$ and $v_2$, respectively $v$, are fixed at one-loop level such
that their values in the tree-level mass relations for the scalars,
derived by calculating explicitly
Eqs.~(\ref{eq:drho})-(\ref{eq:dphi}), lead to the OS physical masses
at one-loop level. The shift from the bare parameter to the physical
one-loop value is hence fully contained in the mass counterterms. In
generic notation the diagonalized bare mass matrices read
\begin{equation}
\begin{split}
D ^2_{\phi ,0}  &= \begin{pmatrix} m_{\phi _1 ,0} ^2 && 0 \\ 0 &&
  m_{\phi _2 ,0} ^2 \end{pmatrix} + R_\varphi ^T \begin{pmatrix}
  \frac{T_{1,0}}{v_1} && 0 \\ 0 && \frac{T_{2,0}}{v_2} \end{pmatrix}
R_\varphi  \;
\end{split}
\end{equation}
where the subscript $0$ denotes the bare quantities and $\varphi=
\alpha$ for the CP-even, $\varphi=\beta$ for the CP-odd and charged
doublets, respectively. We have explicitly kept the bare tadpole
parameters to keep track of their renormalization. Taking into account
the renormalization of the tadpole parameters given in
Eq.~(\ref{eq:tadpolerenorm}) we arrive at the NLO counterterm for the
mass matrix
\begin{equation}
\begin{split}
\delta D_\phi ^2 &\approx \begin{pmatrix} \delta m_{\phi _1 } ^2 && 0
  \\ 0 && \delta m_{\phi _2 } ^2 \end{pmatrix} + R_\varphi
^T \begin{pmatrix} \frac{\delta T_{1}}{v_1} && 0 \\ 0 && \frac{\delta
    T_{2}}{v_2} \end{pmatrix}  R_\varphi \equiv \begin{pmatrix} \delta
  m_{\phi _1 } ^2 && 0 \\ 0 && \delta m_{\phi _2 } ^2 \end{pmatrix}
+ \begin{pmatrix} \delta T_{\phi _1 \phi _1} && \delta T_{\phi _1 \phi
    _2} \\ \delta T_{\phi _1 \phi _2} && \delta T_{\phi _2 \phi
    _2} \end{pmatrix} \,, 
\end{split}
\label{eq:massmatrixct}
\end{equation}
where we have consistently neglected all terms beyond NLO. 
The explicit form of the $T_{\phi_i \phi_j}$ is found by using
Eq.~(\ref{eq:reldelt12deltHh}) and applying the rotation to the 
mass basis,
\begin{align}
\delta T_{H H} &= \frac{c _\alpha ^3 s _\beta +
  s _\alpha ^3 c _\beta }{v s _\beta
  c _\beta } \delta T_{H} - \frac{s_{2\alpha}
  s_{\beta - \alpha} }{v s_{2\beta} } \delta T_{h}
~,  \label{eq:tadpoleH2H2} \\
\delta T_{Hh} &= -\frac{s _{2\alpha} s _{\beta -
    \alpha} }{v s_{2\beta}} \delta T_{H} + \frac{s_{2\alpha} c _{\beta - \alpha} }{v s_{2\beta}}
\delta T_{h} ~, \\
\delta T_{hh} &= \frac{s_{2\alpha} c_{\beta -
    \alpha} }{v s_{2\beta}} \delta T_{H} -
\frac{s_\alpha^3 s_\beta - c_\alpha ^3
  c_\beta }{v s_\beta c_\beta } \delta
T_{h} ~, \\
\delta T_{G^0G^0} &= \frac{c_{\beta -\alpha}}{v} \delta
T_{H} + \frac{s_{\beta - \alpha} }{v} \delta T_{h} ~, \label{eq:tadpoleg0g0}
\\
\delta T_{G^0A} &= -\frac{s_{\beta - \alpha} }{v} \delta
T_{H} + \frac{c_{\beta - \alpha} }{v} \delta T_{h}
~,  \label{eq:tadpoleg0A} \\
\delta T_{AA} &= \frac{c_\alpha s_\beta ^3 +
  s_\alpha c_\beta ^3 }{v s_\beta
  c_\beta } \delta T_{H} - \frac{s_\alpha
  s_\beta ^3 - c_\alpha c_\beta ^3
}{v s_\beta c_\beta } \delta T_{h} ~,  \label{eq:tadpoleAA} \\
\delta T_{G^+G^+} &= \frac{c_{\beta -\alpha} }{v} \delta
T_{H} + \frac{s_{\beta - \alpha} }{v} \delta T_{h} ~, \\
\delta T_{G^+H^+} &= -\frac{s_{\beta - \alpha} }{v} \delta
T_{H} + \frac{c_{\beta - \alpha} }{v} \delta T_{h} ~, \\
\delta T_{H^+H^+} &= \frac{c_\alpha s_\beta ^3 +
  s_\alpha c_\beta ^3 }{v s_\beta
  c_\beta } \delta T_{H} - \frac{ s_\alpha
  s_\beta ^3 - c_\alpha c_\beta ^3
}{v s_\beta c_\beta } \delta T_{h}  ~.  \label{eq:tadpoleHpHm}
\end{align}
Insertion of Eq.~(\ref{eq:massmatrixct}) in the renormalization conditions
(\ref{eq:renconda})-(\ref{eq:rencondc}) we get the field strength
renormalization constants and mass counterterms in the standard scheme
\beq
\delta m_{\phi_i}^2 &=& \mbox{Re} \left[ \Sigma_{\phi_i \phi_i}
  (m_{\phi_i}^2) - \delta T_{\phi_i \phi_i} \right] 
\label{eq:standct1} \\
\delta Z_{\phi_i \phi_i} &=& - \mbox{Re} \left[ \frac{\partial
    \Sigma_{\phi_i \phi_i} (p^2)}{\partial p^2}
\right]_{p^2=m_{\phi_i}^2} 
\label{eq:standct2} \\
\delta Z_{\phi_i \phi_j} &=& \frac{2}{m_{\phi_i}^2 - m_{\phi_j}^2}
\mbox{Re} \left[ \Sigma_{\phi_i \phi_j} (m_{\phi_j}^2) - \delta
  T_{\phi_i \phi_j} \right]  \;, \qquad i\ne j \;. \label{eq:standct3}
\eeq
These formulae can easily be generalized to the fermion and gauge
boson sector. There, however, no tadpole counterterms will be
involved, as they are not part of the tree-level mass relations. The
counterterms introduced in
Eqs.~(\ref{eq:standct1})-(\ref{eq:standct3}) are in general
gauge dependent, which is not a problem, as long as all gauge
dependencies cancel in physical observables. Since the renormalized
masses must be gauge independent, the bare masses must be gauge
dependent as well. \s 

\noindent
{\subsubsection{Mass counterterms and wave function renormalization constants in the tadpole scheme}} 
We have seen that in the standard tadpole scheme
  the correct vacuum is 
reproduced by renormalizing the VEVs at higher orders
accordingly. Derived from the gauge-dependent loop-corrected
potential, the VEVs themselves are gauge dependent. The counterterms
and the bare masses, that are given in terms of the VEVs therefore
become gauge dependent, as the physical OS masses are gauge independent.  
In the tadpole scheme \cite{Fleischer:1980ub} the same renormalization
conditions as given in Eq.~(\ref{eq:tadpolerenorm}), respectively in
Eq.~(\ref{eq:tadpolerenormh}), are used. The crucial point, however,
is the inclusion of the minimization conditions of the
potential such that the mass and coupling counterterms can be defined
in a gauge-independent way. This is achieved in the following way: In
the alternative tadpole 
scheme the bare masses are expressed in terms of the
tree-level VEVs. As the tree-level VEVs are gauge independent, the
bare masses do not depend on the gauge choice either. In order to
still reproduce the correct minimum at higher orders, the VEVs acquire
a shift. This shift now affects the counterterms and not the bare
masses, as the latter are expressed in terms of the tree-level
VEVs. The gauge dependences related to the VEV shifts cancels those of
the counterterms, so that the counterterms become gauge independent
themselves. Together with the gauge-independent bare masses the OS
renormalized masses are gauge independent as they should be. The VEVs
are hence shifted when going from LO to NLO as
\beq
v_1 \to v_1 + \delta v_1 \qquad \mbox{and} \qquad v_2 \to v_2 + \delta
v_2 \;.
\label{eq:vevshift}
\eeq
We emphasize that $v_{1,2}$ represent the
  tree-level values of the 
VEVs. The shifts $\delta v_{1,2}$ are fixed by the minimization,
that is, by the tadpole conditions. The tadpole parameters
are given in terms of the VEVs, {\it cf.}~Eqs.~(\ref{eq:tad1}) and
(\ref{eq:tad2}), so that a shift in the VEVs corresponds to a shift in
the tadpole parameters. Note that we apply the term 'shift' here 
in order to describe the changes of the parameters due to the VEV
shifts, and to differ these from the counterterms for the 
chosen set of independent parameters. \s

The shifts in the VEVs are propagated into all parameters that depend
on the VEVs. These shifts are determined as follows:
$(i)$ Express the parameters in terms of $v_1$ and
$v_2$. $(ii)$ Perform the shifts Eq.~(\ref{eq:vevshift}) of the
VEVs. $(iii)$ Apply the tree-level relations between the VEVs and the
various parameters to remove the redundant parameters $m_{11}^2$,
$m_{22}^2$ and/or to simplify the expressions as convenient. \s

\begin{figure}[tb]
\centering
\includegraphics[width=\linewidth , trim = 19mm 13mm 1mm 7mm, clip]{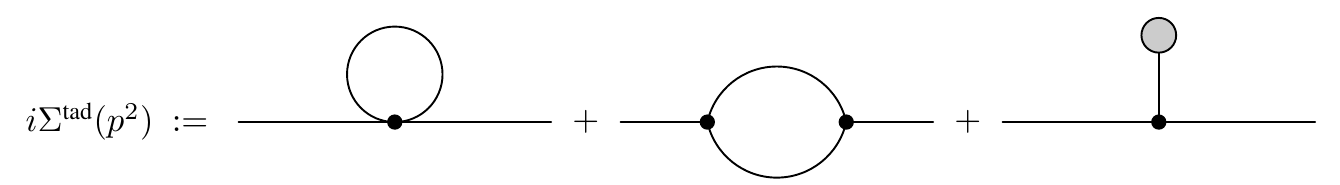}
\caption{Modified self-energy $i\Sigma ^\textrm{tad} (p^2)$ in the
  tadpole scheme, consisting of all 1PI self-energy diagrams together
  with the one-loop tadpole diagrams, indicated by a gray blob.
} 
\label{fig:tadpoleselfen}
\end{figure}

Thus by shifting and subsequently applying the tadpole conditions
Eqs.~(\ref{eq:tad1}) and (\ref{eq:tad2}) we obtain
\begin{align}
T_1 ~ &\rightarrow ~ T_1 + \bigg( m_{12}^2 \frac{v_2}{v_1} + \lambda
_1 v_1^2 \bigg) \delta v_1 + \bigg( - m_{12}^2 + \lambda _{345} v_{1}
v_{2} \bigg) \delta v_2 \equiv T_1 + \delta T_1  \label{eq:delt1shift}
\\ 
T_2 ~ &\rightarrow ~ T_2 + \bigg( - m_{12}^2 + \lambda _{345} v_{1}
v_{2} \bigg) \delta v_1 + \bigg( m_{12}^2 \frac{v_{1}}{v_{2}} +
\lambda _2 v_{2}^2 \bigg) \delta v_2 \equiv T_2 + \delta T_2 ~. \label{eq:delt2shift}
\end{align}
Since the VEVs are determined order by order by applying the
VEV shifts such that the tadpole conditions (\ref{eq:tad1}) and
(\ref{eq:tad2}) hold we identify on the right-hand side of both
equations the shift of the tadpole parameters induced by
the shift of the VEVs with the counterterms $\delta T_1$ and $\delta
T_2$.
By comparing the coefficients of $\delta v_{1,2}$ in Eqs.~(\ref{eq:delt1shift})
and (\ref{eq:delt2shift}) with the elements of the CP-even mass matrix
given in Eq.~(\ref{eq:rhomatrix}) the following relation between the
VEV shifts and the tadpole counterterms, that determine $\delta
v_{1,2}$ can be derived
\begin{equation}
\begin{pmatrix} \delta T_1 \\ \delta T_2 \end{pmatrix} =
\left. M^2_\rho \right| _{T_{i} = 0} ~ \begin{pmatrix} \delta v_1 \\
  \delta v_2 \end{pmatrix} ~.
\end{equation}
Rotation to the mass basis yields
\begin{equation}
\begin{pmatrix} \delta v_{H} \\ \delta v_{h} \end{pmatrix} =
\renewcommand*{\arraystretch}{2.3} \begin{pmatrix} \frac{\delta
    T_{H} }{m_{H}^2 } \\ \frac{\delta T_{h} }{m_{h}^2
  } \end{pmatrix} ~.
\end{equation}
By applying the renormalization condition depicted diagrammatically in
Fig.~\ref{fig:tadcond}, the shift can be interpreted as a connected
tadpole diagram, containing the Higgs tadpole and its propagator at
zero momentum transfer,
\begin{equation}
\delta v_{h_i} = \frac{-i}{m_{h_i}^2} i \delta T_{h_i} =
\frac{-i}{m_{h_i}^2} ~ \mathord{
  \left(\rule{0cm}{30px}\right. \vcenter{
    \hbox{ \includegraphics[height=57px , trim = 18mm 12mm 17mm 10mm,
      clip]{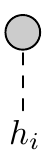} }
  } \left.\rule{0cm}{30px}\right) =
  \left(\rule{0cm}{30px}\right. \vcenter{
    \hbox{ \includegraphics[height=57px , trim = 17.5mm 12mm 15.4mm
      10mm,
      clip]{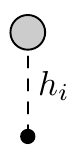}
    } } \left.\rule{0cm}{30px}\right) } ~,
\label{eq:vevshiftinterpret}
\end{equation}
where $h_i \in \{ H,h \}$ stands for the physical Higgs particles. For
the consistent application of the tadpole scheme the VEV shifts have
to be applied wherever the VEVs appear explicitly. As the calculation of
the tadpole diagrams is usually performed in the mass basis, but the
VEV shifts are introduced most conveniently in the gauge basis, we
give the relation between the two bases,
\begin{equation}
\renewcommand*{\arraystretch}{2.3} \begin{pmatrix} \delta v_1 \\
  \delta v_2 \end{pmatrix} = \begin{pmatrix} \frac{\delta
    T_{H}}{m_{H}^2} c_\alpha - \frac{\delta
    T_{h}}{m_{h}^2} s_\alpha \\ \frac{\delta
    T_{H}}{m_{H}^2} s_\alpha + \frac{\delta
    T_{h}}{m_{h}^2} c_\alpha \end{pmatrix} ~.
\label{eq:shiftbasesrelation} 
\end{equation}
For the illustration of the implications of the tadpole scheme we
consider a specific example, namely the NLO effects of the VEV shifts
on the CP-odd mass matrix given in Eq.~(\ref{eq:etamatrix}). The
application of the shifts requires the replacement of the tadpoles by
$T_i + \delta T_i$, with the $\delta T_i$ given in
Eqs.~(\ref{eq:delt1shift}) and (\ref{eq:delt2shift}), and the
replacement of all occurring VEVs by $v_i + \delta v_i$ so that we have
\begin{equation}
\begin{split}
M_\eta ^2 ~ \rightarrow ~ &M_\eta ^2 + \begin{pmatrix} \frac{\delta
    T_1}{v_1} && 0 \\ 0 && \frac{\delta T_2}{v_2} \end{pmatrix} 
  + \left( \frac{m_{12}^2}{v_1v_2}
  - \lambda _5 \right) \begin{pmatrix} 2v_2 \delta v_2 && -v_1 \delta
  v_2 - v_2 \delta v_1 \\ -v_1 \delta v_2 - v_2 \delta v_1 && 2v_1
  \delta v_1 \end{pmatrix} \\ 
\renewcommand*{\arraystretch}{2.3} \renewcommand*{\arraycolsep}{2.2pt}
&- \frac{m_{12}^2}{v_1v_2} \left( \frac{\delta v_1}{v_1} +
  \frac{\delta v_2}{v_2} \right) \begin{pmatrix} v_2^2 && -v_1v_2 \\
  -v_1v_2 && v_1^2 \end{pmatrix} + \begin{pmatrix} \frac{- T_1 \delta
    v_1}{v_1^2} && 0 \\ 0 && \frac{-T_2 \delta v_2}{v_2^2} \end{pmatrix} ~. 
\end{split}
\label{eq:metashift}
\end{equation}
Having applied the shifts, we can now use the tree-level relations
again to eliminate the last matrix in Eq.~(\ref{eq:metashift}), as the
tadpole parameters vanish at tree-level. The rotation to the mass
basis is performed by applying the rotation matrix $R(\beta)$ which
is defined as the matrix diagonalizing the {\it tree-level} mass matrix
$M_\eta^2$. We get
\begin{equation}
\begin{split}
D_\eta ~ \rightarrow & ~ D_\eta + \begin{pmatrix} \delta T_{G^0G^0} &&
  \delta T_{G^0A} \\ \delta T_{G^0A} && \delta
  T_{A A} \end{pmatrix} - \frac{\Lambda _5 v}{s_{2\beta}}
\left( s_\beta \delta v_1 + c _\beta \delta v_2
\right) \renewcommand*{\arraycolsep}{4.6pt} \begin{pmatrix} 0 && 0 \\
  0 && 1 \end{pmatrix} \renewcommand*{\arraycolsep}{3pt} \\ 
&\hspace*{0.1cm} + \frac{m_{A}^2}{v} \begin{pmatrix} 0 &&
  s_\beta \delta v_1 - c _\beta \delta v_2 \\
  s _\beta \delta v_1 - c _\beta \delta v_2 &&
  2\left( c _\beta \delta v_1 + s _\beta \delta v_2
  \right) \end{pmatrix} \\  
&\equiv D_\eta + \begin{pmatrix} \Delta D_{G^0G^0} && \Delta
  D_{G^0A} \\ \Delta D_{G^0A} && \Delta D_{A A} \end{pmatrix}
~,
\end{split}
\end{equation}
where we applied the definition of $\Lambda_5$
Eq.~(\ref{eq:lambda5def}) and the tree-level 
relation for the mass of the pseudoscalar \cite{Kanemura:2004mg,Branco:2011iw}
\beq
m_A^2 = v^2 \left( \frac{m_{12}^2}{v_1 v_2} -\lambda_5 \right) \;.
\eeq
We furthermore applied the definition of the tadpole matrix in the mass
basis, Eq.~(\ref{eq:massmatrixct}). In the last line we defined the
terms $\Delta_{G^0 G^0}$, $\Delta_{G^0 A}$ and $\Delta_{AA}$ that
contain all effects of the VEV shifts on the physical mass matrix
$D_\eta$. These shifts can be further evaluated. In order to do so, we
introduce the coupling constants for the trilinear Higgs couplings
\cite{Kanemura:2004mg}
\beq
g_{H G^0 G^0} &=& \frac{- c_{\beta-\alpha} m_H^2}{v} \label{eq:gH2GG} \\
g_{h G^0 G^0} &=& \frac{-s_{\beta-\alpha} m_h^2}{v} \label{eq:gH1GG} \\
g_{H A A} &=& \frac{-1}{v} \left( c_{\beta-\alpha} (2 m_A^2 - m_H^2) +
 \frac{s_{\alpha+\beta}}{s_{2\beta}} (2 m_H^2 - v^2 \Lambda_5)
\right) \label{eq:gH2AA} \\
g_{h A A} &=& \frac{-1}{v} \left( s_{\beta-\alpha} (2 m_A^2 - m_h^2) +
 \frac{c_{\alpha+\beta}}{s_{2\beta}} (2 m_h^2 - v^2 \Lambda_5)
\right) \label{eq:gH1AA} \\
g_{HAG^0} &=& \frac{- s_{\beta-\alpha}}{v} ( m_A^2 -
m_H^2) \label{eq:gH2AG} \\
g_{hAG^0} &=& \frac{c_{\beta-\alpha}}{v} ( m_A^2 - m_h^2) \;.
\label{eq:gH1AG}
\eeq
By using the explicit form of the tadpole counterterm $\delta T_{G^0
  G^0}$ given in Eq.~(\ref{eq:tadpoleg0g0}) the vanishing Goldstone
boson mass receives the shift contribution $\Delta D_{G^0 G^0}$ 
\begin{equation}
\begin{split}
\Delta D _{G^0G^0} &= \delta T_{G^0G^0} = i \frac{-i c _{\beta -
    \alpha}}{v} m_{H}^2 \frac{-i}{m_{H}^2} i \delta T_{H} + i
\frac{-i s _{\beta - \alpha}}{v} m_{h}^2 \frac{-i}{m_{h}^2}
i \delta T_{h} \\
&=i \, (i g_{HG^0 G^0}) \, \left(\frac{-i}{m_H^2}\right) \,  (i\delta T_H) +
     i \, (i g_{hG^0 G^0}) \, \left(\frac{-i}{m_H^2}\right) \,  (i\delta T_h)  \\
&= \mathord{ i \left(\rule{0cm}{27px}\right. \vcenter{
    \hbox{ \includegraphics[height=57px , trim = 10.4mm 13mm 9.25mm
      12mm, clip]{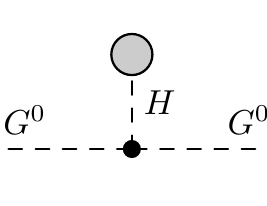}
    } } \left.\rule{0cm}{27px}\right) + i
  \left(\rule{0cm}{27px}\right. \vcenter{
    \hbox{ \includegraphics[height=57px , trim = 10.4mm 13mm 9.25mm
      12mm, clip]{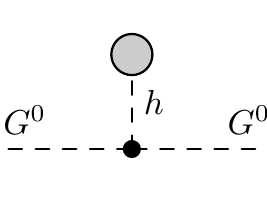}
    } } \left.\rule{0cm}{27px}\right) } ~.
\end{split}
\end{equation}
In the second line we have used Eqs.~(\ref{eq:gH2GG}) and
(\ref{eq:gH1GG}). The last line is the diagrammatic representation of $\Delta T_{G^0
  G^0}$. It is given by two tadpole contributions from the CP-even
Higgs bosons to the neutral Goldstone boson self-energy. Analogously
we find for $\Delta D_{AA}$ by using 
  Eqs.~(\ref{eq:tadpoleAA}), (\ref{eq:shiftbasesrelation}),
  (\ref{eq:gH2AA}) and (\ref{eq:gH1AA}),
\begin{equation}
\begin{split}
\Delta D _{AA} &= \delta T_{AA} - \frac{\Lambda _5
  v}{s_{2\beta}} \left( s_\beta \delta v_1 +
  c _\beta v_2 \right) + \frac{2m_{A^0}^2}{v} \left(
  c_\beta \delta v_1 + s_\beta \delta v_2 \right) \\
&= i \frac{-i}{v} \bigg( c_{\beta -\alpha} \left( 2m_{A}^2 -
  m_{H}^2 \right) + \frac{s_{\alpha + \beta
  }}{s_{2\beta}} \left( 2m_{H}^2 - v^2 \Lambda _5 \right)
\bigg) \frac{-i}{m_{H}^2} i\delta T_{H} \\
&\hspace*{0.37cm} + i \frac{-i}{v} \bigg( s_{\beta -\alpha}
\left( 2m_{A}^2 - m_{h}^2 \right) + \frac{c_{\alpha +
    \beta }}{s_{2\beta}} \left( 2m_{h}^2 - v^2 \Lambda _5
\right) \bigg) \frac{-i}{m_{h}^2} i\delta T_{h} \\
&= i\, (ig_{HAA}) \, \left( \frac{-i}{m_H^2} \right) \, (i \delta T_H) 
+ i\, (ig_{hAA}) \, \left( \frac{-i}{m_h^2} \right) \, (i \delta T_h) 
\\
&= \mathord{ i \left(\rule{0cm}{27px}\right. \vcenter{
    \hbox{ \includegraphics[height=57px , trim = 10.4mm 13mm 9.25mm
      12mm, clip]{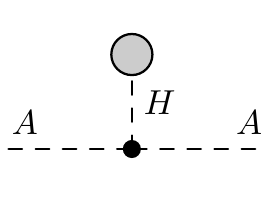}
    } } \left.\rule{0cm}{27px}\right) + i
  \left(\rule{0cm}{27px}\right. \vcenter{
    \hbox{ \includegraphics[height=57px , trim = 10.4mm 13mm 9.25mm
      12mm, clip]{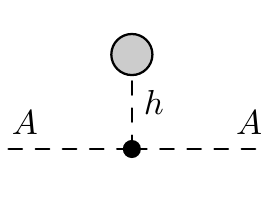}
    } } \left.\rule{0cm}{27px}\right) } ~. 
\end{split}
\end{equation}
The last line again reproduces the diagrammatic representation of the
shift. The shift is hence given by two CP-even tadpole contributions
to the $A$ boson self-energy. The off-diagonal shift $\Delta D_{G^0A}$
finally can be cast into the form by applying
Eqs.~(\ref{eq:tadpoleg0A}), (\ref{eq:shiftbasesrelation}),
(\ref{eq:gH2AG}) and (\ref{eq:gH1AG}),
\begin{equation}
\begin{split}
\Delta D _{G^0 A} &= \delta T_{G^0 A} + \frac{m_{A}^2}{v} \left(
  s_\beta \delta v_1 - c_\beta \delta v_2 \right) \\
&= i \frac{-i s_{\beta - \alpha}}{v} \left( m_{A}^2 - m_{H}^2
\right) \frac{-i}{m_{H}^2} i\delta T_{H} + i \frac{i c_{\beta
    - \alpha}}{v} \left( m_{A}^2 - m_{h}^2 \right)
\frac{-i}{m_{h}^2} i\delta T_{h} \\
& = i \, (ig_{HAG^0}) \, \left( \frac{-i}{m_H^2} \right) \, (i\delta T_H
)
+ i \, (ig_{hAG^0}) \, \left( \frac{-i}{m_h^2} \right) \, (i\delta T_h) \\
&= \mathord{ i \left(\rule{0cm}{27px}\right. \vcenter{
    \hbox{ \includegraphics[height=57px , trim = 10.4mm 13mm 9.25mm
      12mm, clip]{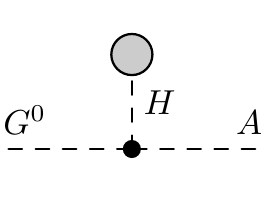}
    } } \left.\rule{0cm}{27px}\right) + i
  \left(\rule{0cm}{27px}\right. \vcenter{
    \hbox{ \includegraphics[height=57px , trim = 10.4mm 13mm 9.25mm
      12mm, clip]{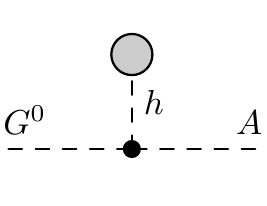}
    } } \left.\rule{0cm}{27px}\right) } ~.
\end{split}
\end{equation}
The diagrammatic representation in the last line reveals that the
shift $\Delta G_{GA}$ consists of two CP-even tadpole contributions to
the off-diagonal $G^0 A$ self-energy. It is straightforward to derive
the remaining shifts for the whole scalar sector. The total shift
of the mass matrices, which is given by the shifts $\Delta D$ induced
by the NLO shifts of the VEVs and by the mass counterterms, then reads
\begin{equation}
\delta D_\phi ^2 = \begin{pmatrix} \delta m_{\phi _1}^2 && 0 \\ 0 &&
  \delta m_{\phi _2} ^2 \end{pmatrix} + \begin{pmatrix} \Delta D
  _{\phi _1 \phi _1} && \Delta D _{\phi _1 \phi _2} \\ \Delta D _{\phi
    _1 \phi _2} && \Delta D _{\phi _2 \phi _2} \end{pmatrix} ~,
\label{eq:finalmassct}
\end{equation}
with the explicit form of the additional mass shifts ($i=1,2$)
\begin{equation}
\Delta D _{\phi _i \phi _j} = \mathord{ i
  \left(\rule{0cm}{27px}\right. \vcenter{
    \hbox{ \includegraphics[height=57px , trim = 10.4mm 13mm 9.25mm
      12mm,
      clip]{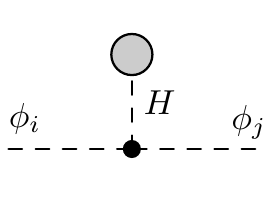} } }
  \left.\rule{0cm}{27px}\right) + i
  \left(\rule{0cm}{27px}\right. \vcenter{
    \hbox{ \includegraphics[height=57px , trim = 10.4mm 13mm 9.25mm
      12mm,
      clip]{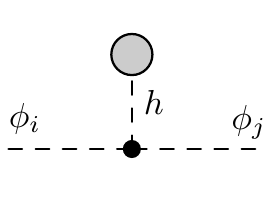} } }
  \left.\rule{0cm}{27px}\right) } ~, 
\end{equation}
where $(\phi_1, \phi_2)$ refers to the pairs $(H,h)$, $(G^0,A)$ and
$(G^\pm, H^\pm)$, respectively. Equation (\ref{eq:finalmassct})
makes evident that in the tadpole scheme the tadpole counterterms
$\delta T_1$ and $\delta T_2$ induced through the VEV shifts in
Eqs.~(\ref{eq:delt1shift}) and (\ref{eq:delt2shift}) have become part
of the shift parameters $\Delta D_{\phi_i \phi_j}$ of the physical
mass matrices of the scalar sector. They do not
appear explicitly as counterterms, in contrast to the standard scheme
where $\delta T_1$ and $\delta T_2$ were considered as counterterms
being explicitly part of $\delta D_\phi^2$, {\it
  cf.}~Eq.~(\ref{eq:massmatrixct}). Therefore, in the tadpole scheme,
the tadpole counterterms
Eqs.~(\ref{eq:tadpoleH2H2})-(\ref{eq:tadpoleHpHm}) do not belong to
the definition of the mass counterterms and wave function
renormalization constants. With the redefinition of the 1PI
self-energy as
\beq
i \Sigma^{\text{tad}}_{\phi_i \phi_j} (p^2) \equiv i \Sigma_{\phi_i
  \phi_j} (p^2) - i \Delta D_{\phi_i \phi_j}
\eeq
we obtain by inserting  Eq.~(\ref{eq:finalmassct}) in
Eq.~(\ref{eq:renselfenergy}) the following form of the renormalized
self-energy,
\begin{equation}
\hat{\Sigma } _\phi (p^2) \equiv \Sigma ^\textrm{tad} _\phi (p^2)
- \begin{pmatrix} \delta m_{\phi _1}^2 && 0 \\ 0 && \delta m_{\phi
    _2}^2 \end{pmatrix} + \frac{\delta Z _\phi ^\dagger }{2} \left(
  p^2 \mathbbm{1} _{2\times 2} - D_\phi ^2 \right) + \left( p^2
  \mathbbm{1} _{2\times 2} - D_\phi ^2 \right) \frac{\delta Z _\phi
}{2} ~~ . 
\end{equation}
And finally the counterterms and wave function renormalization
constants in the tadpole scheme read
\beq
\delta m_{\phi_i}^2 &=& \mbox{Re} \left[ \Sigma^{\text{tad}}_{\phi_i \phi_i}
  (m_{\phi_i}^2) \right] 
\label{eq:tadpolect1} \\
\delta Z_{\phi_i \phi_i} &=& - \mbox{Re} \left[ \frac{\partial
    \Sigma^{\text{tad}}_{\phi_i \phi_i} (p^2)}{\partial p^2}
\right]_{p^2=m_{\phi_i}^2} 
\label{eq:tadpolect2} \\
\delta Z_{\phi_i \phi_j} &=& \frac{2}{m_{\phi_i}^2 - m_{\phi_j}^2}
\mbox{Re} \left[ \Sigma^{\text{tad}}_{\phi_i \phi_j} (m_{\phi_j}^2) \right]  \;, \qquad\hspace*{0.05cm} i\ne j \;. \label{eq:tadpolect3}
\eeq
These results can be generalized to the gauge boson and fermion
sectors. The application of the tadpole scheme hence requires a
redefinition of the self-energies as depicted diagrammatically in
Fig.~\ref{fig:tadpoleselfen}. In the gauge and fermion sectors this
implies that the tadpole diagrams of the scalar Higgs
bosons that couple to the gauge boson and fermion, respectively, have to be
included in their self-energy. Furthermore, in the scalar sector the
tadpole counterterms drop out of the definition of the wave function
renormalization constants and mass
counterterms.\footnote{In the gauge and fermion
    sectors they do not appear anyway as the mass matrices do not
    depend on $m_{11}^2$ and $m_{22}^2$ that are traded for the tadpoles.} \s

The VEV shifts introduced in Eq.~(\ref{eq:vevshift}) also have
implications for the coupling constants of the vertices. Let us
consider the example of the Higgs $H$ coupling to a pair of $Z^\mu
Z^\nu$ bosons. Defining the needed coupling constants through the
Feynman rules
\beq
HZ^\mu Z^\nu : & ig_{HZZ} g^{\mu\nu} \\
HH Z^\mu Z^\nu: & i g_{HHZZ} g^{\mu\nu} \;,
\eeq
we have 
\beq
g_{HZZ} &=& \frac{g^2 v c_{\beta-\alpha}}{2 c_W^2} = \frac{g^2}{2
  c_W^2} (c_\alpha v_1 + s_\alpha v_2)  \\
g_{HHZZ} &=& \frac{g^2}{2 c_W^2} \;. \label{eq:quartic}
\eeq
The shifts Eq.~(\ref{eq:vevshift}) introduce a shift in the coupling
constants. In order to perform this shift consistently, the coupling
constants must be expressed in terms of the VEVs $v_1$ and $v_2$. When
doing so, care has to be taken, to differentiate between the
angles $\alpha$ and $\beta$ in the sense of mixing angles and $\beta$
in the sense of the ratio of the VEVs, {\it
  cf.}~Eq.~(\ref{eq:tanbetadef}), and $\alpha$ in the sense of the ratio of the
2HDM parameters\footnote{Note that in all couplings but the trilinear
  and quartic Higgs self-couplings $\alpha$ has the role of a mixing
  angle. Only in the Higgs self-couplings $\alpha$ partly appears in
  the sense of the ratio of 2HDM potential parameters.} given in
Eq.~(\ref{eq:alphadef}). The VEV shifts only affect the latter two. \s

The quartic coupling obviously does not receive any shift, while
$g_{HZZ}$ contains $\beta$ as ratio of the VEVs so that it receives a
shift. The angle $\alpha$ is a mixing angle here. At NLO we therefore
have to make the replacement
\begin{equation}
\begin{split}
i g_{HZZ} ~ &\rightarrow ~ i g_{HZZ} + \frac{i g^2}{2c_W^2} (c_\alpha \delta
v_1 + s_\alpha \delta v_2) \\
&= i g_{HZZ} + \frac{ig^2}{2 c_W^2} \left[ (c_\alpha^2 +
  s_\alpha^2) \frac{\delta T_H}{m_H^2} + (s_\alpha c_\alpha - s_\alpha
  c_\alpha) \frac{\delta T_h}{m_h^2} \right]\\
&= i g_{HZZ} + i g_{HHZZ} \, \left( \frac{-i}{m_H^2} \right) \, i
\delta T_H \\
&= i g_{HZZ} +  \mathord{ \left(\rule{0cm}{30px}\right. \vcenter{
    \hbox{ \includegraphics[height=56px , trim = 7.3mm 8mm 3mm 8mm,
      clip]{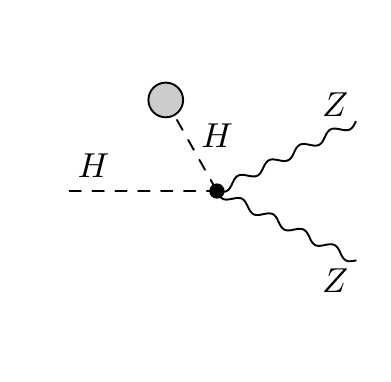} } }
  \left.\rule{0cm}{30px}\right)_\textrm{trunc}} \\
& \equiv i g_{HZZ}^{\textrm{tad}} \;.
\end{split}
\label{eq:vertexchange}
\end{equation}
The subscript 'trunc' means that all Lorentz structure of the
vector bosons as well as the Lorentz structure of the coupling has
been suppressed here for simplicity. In the derivation of this
equation we have used Eq.~(\ref{eq:shiftbasesrelation}) and the
explicit form of the quartic coupling constant,
Eq.~(\ref{eq:quartic}). The Feynman rule for the $HZZ$ vertex in the
tadpole scheme is then given by
\beq
i g_{HZZ}^{\textrm{tad}} \, g^{\mu\nu} \;.
\eeq

The above result can be generalized to the whole 2HDM. In the tadpole
scheme additional virtual vertex corrections have to be taken into
account that manifest themselves in form of tadpdole vertex
diagrams. The rule to be applied here is, that every 2HDM trilinear
vertex receives corrections if the resulting quartic coupling
constant exists, that connects the original trilinear vertex to the CP-even
Higgs, that is attached to the tadpole. In the case above the vertex
$g_{HHZZ}$ exists, so that the vertex acquires a tadpole contribution
with $H$, but not with $h$, as the vertex $g_{hHZZ}$ does not
exist. \s

As last example we look at the coupling between $W^\pm_\mu$, $H^\pm (p')$ and
$h (p)$, where $p' (p)$ denotes the outgoing (incoming) momentum of
$H^\pm$ ($h$). The Feynman rule for the coupling is given by
\beq
W^\pm_\mu H^\pm h : \quad \mp i \, g_{W^\pm H^\pm h} \, (p+p')_\mu \,,
\eeq
with the coupling constant
\beq
g_{W^\pm H^\pm h} = \frac{g c_{\beta-\alpha}}{2} \;.
\eeq
Both angles in this coupling are true mixing angles, so that no VEV
shift has to be applied. Therefore the vertex does not change in the
tadpole scheme. This is in accordance with the rule given above: There
exists no vertex $g_{W^\pm H^\pm hh}$ nor $g_{W^\pm H^\pm hH}$ that could
connect a tadpole with $h$ or $H$ to the trilinear vertex. \s

\subsection{Rules for the Tadpole Scheme in the 2HDM \label{app:tadpolerules}}

In this Appendix we summarize all rules of the tadpole scheme for the
2HDM at NLO. The general rules are: \s

\noindent
{\bf Self-energies:} The self-energies in the wave function
renormalization constants and counterterms change such that they
contain additional tadpole contributions: $\Sigma (p^2) \to
\Sigma^{\text{tad}} (p^2)$. \s

\noindent 
{\bf Tadpole counterterms:} The tadpole counterterms in the scalar
sector vanish: $\delta T_{\phi_i \phi_j} \to 0$ ($i,j=1,2$). \s

\noindent
{\bf Vertex corrections:} In the virtual vertex corrections
additional tadpole contributions have to be taken into account if the
resulting coupling exists in the 2HDM. \s

\noindent Explicitly, this means that the
following counterterms are the same in the standard and the alternative tadpole scheme:\\[0.3cm]
\noindent
{\it Counterterms independent of the choice of the tadpole scheme:}
\beq
\begin{array}{ll}
\hspace*{-0.5cm} \mbox{Tadpoles:} & \; \delta T_H, \delta T_h \\[0.1cm]
\hspace*{-0.5cm} \mbox{Gauge sector:} & \;\delta Z_e, \delta g, \delta Z_{WW},
\delta Z_{ZZ}, \delta Z_{Z\gamma}, \delta Z_{\gamma Z} \\[0.1cm]
\hspace*{-0.5cm} \mbox{Fermion sector:} & \; \delta Z^L_{FF}, \delta Z^R_{FF}
\hspace*{-0.5cm} \\[0.1cm]
\hspace*{-0.5cm} \mbox{Scalar sector:} & \; \delta Z_{\phi_i \phi_i}\\[0.1cm]
\hspace*{-0.5cm} \mbox{Vertices:} &\; \lambda_{FFS}, \lambda_{FFV},
\lambda_{SSV}, \lambda_{SUU}, \lambda_{UUV}, \lambda_{VVV}, \lambda_{VVVV} 
\end{array}
\eeq
for all possible combinations of fermions $F$, gauge bosons $V$,
ghosts $U$, scalars $S$ and $\phi_{i,j} \equiv H,h,G^0, A, G^\pm,
H^\pm$ within the 2HDM. \\[0.3cm]
The following counterterms and wave function renormalization constants
depend on the choice of the tadpole scheme. We give the relations
between the standard tadpole scheme, denoted by the superscript
'stand', and the alternative tadpole scheme denoted by the superscript
'tad'. The subscript 'trunc' means, that all spinors, all Lorentz structure of the
vector bosons and the Lorentz structure of the coupling has been
suppressed where applicable. \\[0.3cm]
\noindent
{\it Tadpole-scheme-dependent counterterms:} \\[0.1cm]
\noindent Gauge sector:
\begin{align}
( \delta m_W^2 )^\textrm{tad} &= ( \delta m_W^2 )^\textrm{stand} +i
\mathord{ \left(\rule{0cm}{26px}\right. \vcenter{
    \hbox{ \includegraphics[height=41px , trim = 10mm 20mm 9mm 10.4mm,
      clip]{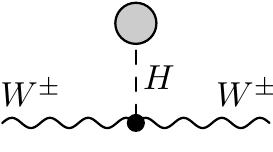} } }
  \left.\rule{0cm}{26px}\right)_\textrm{trunc} + i
  \left(\rule{0cm}{26px}\right. \vcenter{
    \hbox{ \includegraphics[height=41px , trim = 10mm 20mm 9mm 10.4mm,
      clip]{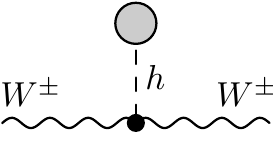} } }
  \left.\rule{0cm}{26px}\right)_\textrm{trunc} }  \\ 
( \delta m_Z^2 )^\textrm{tad} &= ( \delta m_Z^2 )^\textrm{stand} +i
\mathord{ \left(\rule{0cm}{26px}\right. \vcenter{
    \hbox{ \includegraphics[height=41px , trim = 10mm 20mm 9mm 10.4mm,
      clip]{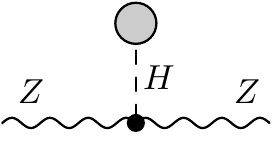} } }
  \left.\rule{0cm}{26px}\right)_\textrm{trunc} + i
  \left(\rule{0cm}{26px}\right. \vcenter{
    \hbox{ \includegraphics[height=41px , trim = 10mm 20mm 9mm 10.4mm,
      clip]{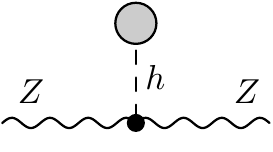} } }
  \left.\rule{0cm}{26px}\right)_\textrm{trunc} }  
\end{align}	

\noindent Fermion sector: 
\begin{align}
( \delta m_F  )^\textrm{tad} &= ( \delta m_F  ) ^\textrm{stand}
-i  \mathord{ \left(\rule{0cm}{26px}\right. \vcenter{
\hbox{ \includegraphics[height=41px , trim = 10mm 20mm 9mm 10.4mm,
      clip]{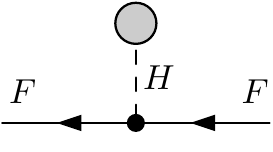} } }
\left.\rule{0cm}{26px}\right)_\textrm{trunc} - i
\left(\rule{0cm}{26px}\right. \vcenter{
    \hbox{ \includegraphics[height=41px , trim = 10mm 20mm 9mm 10.4mm,
      clip]{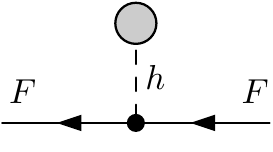} } }
  \left.\rule{0cm}{26px}\right)_\textrm{trunc} } 
\end{align}		
			
\noindent Scalar sector: 
\begin{align}
( \delta m_{\phi _i}  )^\textrm{tad} &= ( \delta m_{\phi
  _i})^\textrm{stand} + \delta T_{\phi _i \phi _i} -i  \mathord{  
  \left(\rule{0cm}{26px}\right. \vcenter{
    \hbox{ \includegraphics[height=61px , trim = 10mm 14mm 9mm 10.4mm,
      clip]{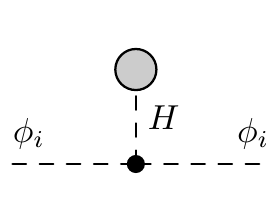} } }
  \left.\rule{0cm}{26px}\right)_\textrm{trunc} - i
  \left(\rule{0cm}{26px}\right. \vcenter{
    \hbox{ \includegraphics[height=61px , trim = 10mm 14mm 9mm 10.4mm,
      clip]{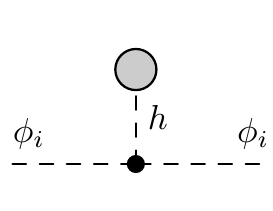} } }
  \left.\rule{0cm}{26px}\right)_\textrm{trunc} } 
\end{align}	
for all possible combinations of $\phi_{i,j} \equiv H , h , A , H^\pm
$. 
\begin{align}
( \delta Z_{\phi _i \phi _j}  )^\textrm{tad} &= ( \delta Z_{\phi _i
  \phi _j}  )^\textrm{stand} \\ \nonumber 
&+ \frac{2}{m_{\phi _i}^2 - m_{\phi _j}
  ^2} \left[ \delta T_{\phi _i \phi _j} -i  \mathord{
    \left(\rule{0cm}{26px}\right. \vcenter{
      \hbox{ \includegraphics[height=61px , trim = 10mm 14mm 9mm
        10.4mm, clip]{VEVShiftResultPhiPhioffHH.pdf} } }
    \left.\rule{0cm}{26px}\right)_\textrm{trunc} - i
    \left(\rule{0cm}{26px}\right. \vcenter{
      \hbox{ \includegraphics[height=61px , trim = 10mm 14mm 9mm
        10.4mm, clip]{VEVShiftResultPhiPhioffh0.pdf} } }
    \left.\rule{0cm}{26px}\right)_\textrm{trunc} } \right] ~, 
\end{align}	
where $\phi _i \neq \phi _j$. \s

We encounter additional contributions to the vertices when changing
from the standard to the tadpole scheme. 
Here below, the $g$ denote the coupling constants, {\it i.e.}~we have
suppressed the Lorentz structure of the vertex where applicable. \s

\noindent Triple scalar vertices: 
\begin{equation}
i g_{\phi _i \phi _j \phi _k} ~ \rightarrow ~ i g_{\phi _i
  \phi _j \phi _k} + \mathord{ \left(\rule{0cm}{30px}\right. \vcenter{
    \hbox{ \includegraphics[height=56px , trim = 7.3mm 8mm 3mm 8mm,
      clip]{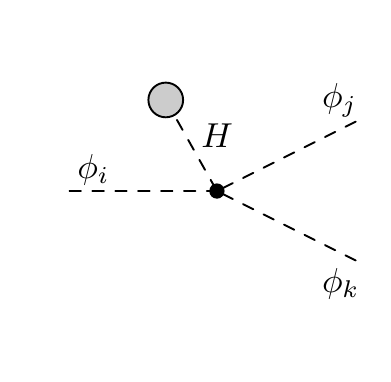} } }
  \left.\rule{0cm}{30px}\right)_\textrm{trunc} +
  \left(\rule{0cm}{30px}\right. \vcenter{
    \hbox{ \includegraphics[height=56px , trim = 7.3mm 8mm 3mm 8mm,
      clip]{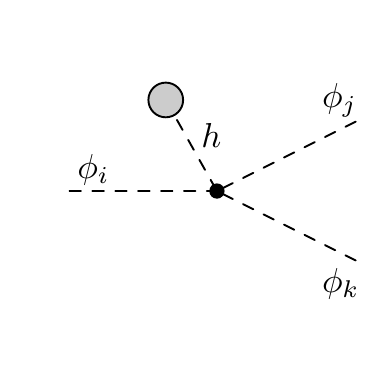} } }
  \left.\rule{0cm}{30px}\right)_\textrm{trunc} }
\end{equation}
for all scalars $\phi_{i,j,k} \equiv H , h ,
G^0 , A , G^\pm , H^\pm$, 
wherever the resulting quartic couplings $\lambda _{\phi _i \phi _j
  \phi _k h}$ and $\lambda _{\phi _i \phi _j \phi _k H}$ exist in the
2HDM. \s
			
\noindent Scalar-vector-vector vertices: 
\begin{equation}
i g_{\phi _i V _j V _k} ~ \rightarrow ~ i g_{\phi _i V _j V
  _k} + \mathord{ \left(\rule{0cm}{30px}\right. \vcenter{
    \hbox{ \includegraphics[height=56px , trim = 7.3mm 8mm 3mm 8mm,
      clip]{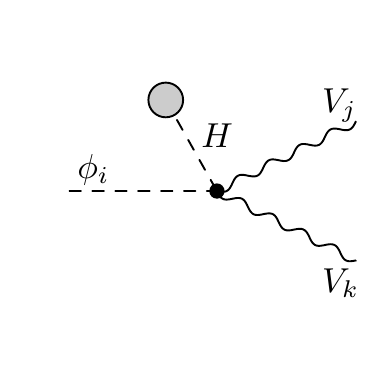} } }
  \left.\rule{0cm}{30px}\right)_\textrm{trunc} +
  \left(\rule{0cm}{30px}\right. \vcenter{
    \hbox{ \includegraphics[height=56px , trim = 7.3mm 8mm 3mm 8mm,
      clip]{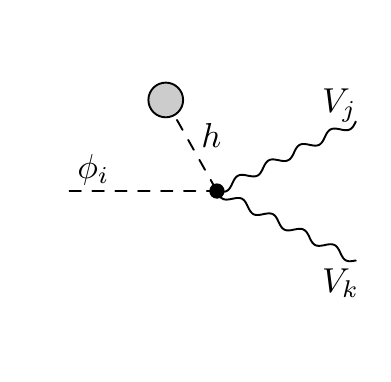} } }
  \left.\rule{0cm}{30px}\right)_\textrm{trunc} } 
\end{equation}
for all scalars $\phi_{i,j,k} \equiv H , h , G^0 , A , G^\pm , H^\pm$,
and gauge bosons $V_{j,k} \equiv \gamma , Z , W^\pm $, wherever the
resulting quartic couplings $\lambda _{\phi _i V _j V _k h}$ and
$\lambda _{\phi _i V _j V _k H}$ exist in the 2HDM. 
\end{appendix}

\clearpage
\newpage
\vspace*{1cm}
\bibliographystyle{h-physrev}

\end{document}